\documentclass[letterpaper,aps,prl,twocolumn,floatfix,amsfonts,amssymb,notitlepage]{revtex4-2}
\pdfpageattr {/Group << /S /Transparency /I true /CS /DeviceRGB>>}

\usepackage{amsmath}
\usepackage{amsthm}
\usepackage{amssymb}
\usepackage{graphicx}
\usepackage{tabularx}
\usepackage{color}

\usepackage[colorlinks=true]{hyperref}
\usepackage{xcolor}
\usepackage[normalem]{ulem}

\newcommand{\figfour}[0]{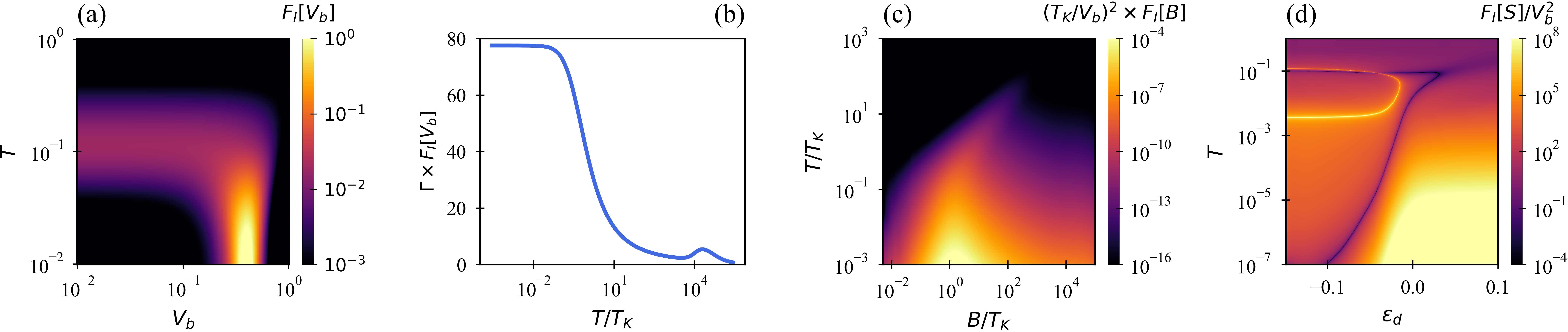}
\newcommand{\rev}[1]{{\color{black}#1}}

\begin{document}
\title{Quantum Sensing with Nanoelectronics:\\Fisher Information for an \rev{Applied} Perturbation}

\author{George Mihailescu}
\affiliation{School of Physics, University College Dublin, Belfield, Dublin 4, Ireland}
\affiliation{Centre for Quantum Engineering, Science, and Technology, University College Dublin, Ireland}
\author{Anthony Kiely}
\affiliation{School of Physics, University College Dublin, Belfield, Dublin 4, Ireland}
\affiliation{Centre for Quantum Engineering, Science, and Technology, University College Dublin, Ireland}
%\affiliation{School of Physics, University College Cork, College Road, Cork, Ireland}
\author{Andrew K. Mitchell}
\affiliation{School of Physics, University College Dublin, Belfield, Dublin 4, Ireland}
\affiliation{Centre for Quantum Engineering, Science, and Technology, University College Dublin, Ireland}

\begin{abstract}
\noindent Quantum systems used for metrology can offer enhanced precision over their classical counterparts. 
The design of quantum sensors can be optimized by maximizing the quantum Fisher information (QFI), which characterizes the precision of parameter estimation for an ideal measurement.
Here we consider the response of a quantum system as a means to estimate the strength of a weak external perturbation. General expressions for the QFI in the nonequilibrium steady-state are derived, which hold for arbitrary interacting many-body systems %in the thermodynamic limit 
at finite or zero temperature, and can be related to susceptibilities or linear-response transport coefficients. For quantum dot nanoelectronics devices, we show that electron interactions can lead to \textit{exponential} scaling of the QFI with system size, highlighting that quantum resources can be utilized in the full Fock space. The precision estimation of voltages and fields can also be achieved by practical measurements. In particular, we show that current-based metrology in quantum circuits can leverage many-body effects for enhanced sensing. 
\end{abstract}

\maketitle
%%%%%%%%%%%%%%%%%%%%%%%%%%%%%%%%%%%%%%%%%%%%%%%%%%%%%%%%%%%%%%%%%%%%%%%
%%%%%%%%%%%%%%%%%%%%%%%%%%%%%%%%%%%%%%%%%%%%%%%%%%%%%%%%%%%%%%%%%%%%%%%

Parameter estimation is a key part of any experiment, both in terms of calibration and readout; and the technological applications of devices for precision sensing are diverse. Scientific discoveries in the modern era have long been driven by our ability to make ever more precise measurements, from the Michelson–Morley interferometer to LIGO \cite{martynov2016sensitivity}. In the context of new quantum technologies such as NISQ devices \cite{preskill2018}, or other experiments on quantum systems, parameter estimation techniques may themselves be inherently quantum \cite{Paris_Quantum_Estimation,toth}. Indeed, the use of quantum systems as sensors can provide an advantage in terms of precision over classical counterparts \cite{degen2017,Giovannetti2011,yang2022super,aslam2023,abiuso2025fundamental}. Atomic, molecular and optical systems have been widely studied in this regard \cite{degen2017,Ye2024,PhysRevA.107.033318,PhysRevLett.130.090802,DiCandia2023,e24081015,mag1,mag2,mag3}.

In this Letter we explore nanoelectronics devices \cite{goldhaber1998kondo,*van2000kondo,perrin2015single,*chen2024quantum,nowack2007coherent,*barthelemy2013quantum} as an alternative and promising platform for quantum sensing. The charge or electrical current through a contacted nanostructure can be accurately measured in an external circuit, and can depend sensitively on voltages, fields, and temperature \cite{ihn2009semiconductor}. Entanglement and quantum many-body effects arising from electron interactions such as Coulomb blockade \cite{meirav1990single,park2002coulomb}, Kondo effect \cite{goldhaber1998kondo,*van2000kondo,liang2002kondo,keller2014emergent,piquard2023observing} and quantum criticality \cite{potok2007observation,roch2008quantum,iftikhar2018tunable,pouse2023quantum,*karki2023z} observed in such devices may be a useful resource for sensing. They are also appealing from a practical perspective, since sophisticated quantum devices can now be fabricated in commercial-process semiconductor technologies \cite{chanrion2020charge,xue2021cmos,ruffino2022cryo,petropoulos2024nanoscale}.

\begin{figure}[b]
\centering
\includegraphics[width=\linewidth]{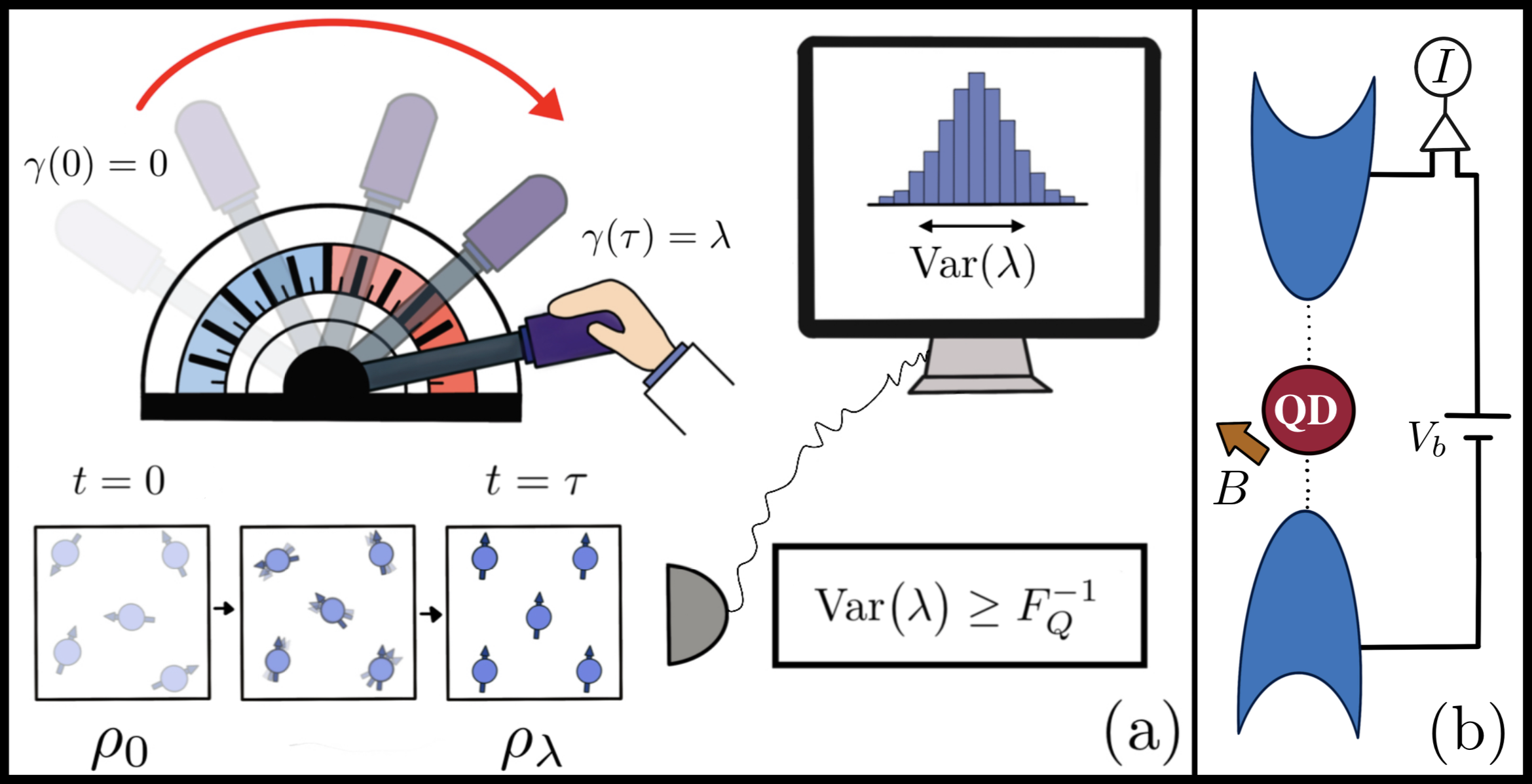}
\caption{\rev{Quantum sensing of a perturbation. (a) A system is subject to a weak perturbation $\lambda$, switched on in time $\tau$.} Measurements of the system response yield statistics on the estimator for $\lambda$ whose variance is constrained by the QFI Eq.~\ref{eq:QFIpert} via the CRB. (b) A typical nanoelectronics setup in which an electrical current $I$ flows through a quantum dot due to a bias voltage $V_b$ between source and drain leads, in a field $B$.} 
\label{fig:schematic}
\end{figure}

We consider a general scenario in which the response of a quantum system to an external perturbation is used to estimate the strength of this perturbation. \rev{We focus on the case of a weak perturbation, which allows exact results to be obtained via linear response (LR) theory, even for highly complex systems. We compute the quantum Fisher information (QFI) \cite{Paris_Quantum_Estimation,toth}, which characterizes the maximum precision of parameter estimation attainable by making optimal measurements on the system, see Fig.~\ref{fig:schematic}(a), and relate this to steady-state transport coefficients in the adiabatic regime.} This is applied to many-body models of quantum dot (QD) nanoelectronics circuits \cite{goldhaber1998kondo,*van2000kondo}, Fig.~\ref{fig:schematic}(b), and compared with the precision attainable when using a \textit{practical} measurement of the electrical current in such devices.

%%%%%%%%%%%%%%%%%%%%%%%%%%
%%%%%%%%%%%%%%%%%%%%%%%%%%

Intuitively, the more sensitive the state of a system $\hat{\rho}_{\lambda}$ is to changes in the external perturbation $\lambda$, the more information about the perturbation that can be gained from measurements on the system. For a given number $N$ of independent measurements, the precision of estimation for $\lambda$ is characterized by its statistical variance ${\rm Var}(\lambda)=\mathbb{E}[(\lambda_{\rm est}-\lambda)^2]$, which for unbiased estimators is controlled by the QFI $F_Q[\lambda]$ through the Cram\'er-Rao bound (CRB) \cite{braunstein1994statistical} $N {\rm Var}(\lambda) \ge 1/F_Q[\lambda]$. The QFI corresponds to an optimal measurement and is therefore the best-case scenario against which any practical measurement scheme should be compared. The design of advanced quantum sensors therefore necessitates the characterization and optimization of the QFI.

We take an arbitrary system $\hat{H}_0$, initially at thermal equilibrium, to which a small perturbation $\lambda\hat{A}$ is introduced \rev{over a time $\tau$ via the ramp protocol $\gamma(t)$.} Our main result is an expression for the nonequilibrium QFI $F_Q[\lambda]$ for estimation of parameter $\lambda$ at the end of the ramp. In the adiabatic limit where the perturbation is switched on slowly, the QFI reduces to, 
\begin{equation}
  F_Q[\lambda] = 2\sum_{n\ne m}\frac{(p_n^0-p_m^0)^2}{p_n^0+p_m^0}\frac{|\langle n_0|\hat{A}|m_0\rangle |^2}{(E_n^0-E_m^0)^2} \;, \label{eq:QFIpert}  
\end{equation}
where $\hat{H}_0 |n_0\rangle = E^0_n |n_0\rangle$ and the density matrix $\hat{\rho}_0 = \sum_n p_n^0 |n_0\rangle \langle n_0|$ characterizes the unperturbed state of the system at inverse temperature $\beta\equiv 1/T$, with $p_n^0=\exp(-\beta E_n^0)/Z_0$ and $Z_0=\sum_m\exp(-\beta E_m^0)$. \rev{The QFI in Eq.~\ref{eq:QFIpert} can be expressed in terms of properties of the unperturbed system $\hat{H}_0$ because the perturbation is weak and the system remains close to equilibrium. Details of the drive $\gamma(t)$ do not enter in the adiabatic limit, but the result is straightforwardly generalized to finite ramp time $\tau$ as shown in the Supplemental Material (SM)~\cite{sm}.}

%##################
%##################

\noindent{\textit{Derivation via adiabatic gauge potential.--}}
The perturbation is switched on smoothly via $\hat{H}_1(t)=\gamma(t)\hat{A}$, \rev{where $\gamma(0)=0$ and $\gamma(\tau)=\lambda$}. We label $\hat{H}_{\gamma}=\hat{H}_0+\gamma(t) \hat{A}$ by the running value of $\gamma$ at time $t$. The instantaneous spectral decomposition then reads $\hat{H}_{\gamma}=\sum_n E_n^{\gamma}|n_{\gamma}\rangle\langle n_{\gamma}|$. For adiabatic evolution $\tau\to \infty$ \cite{sm}, states stay in these instantaneous eigenstates and so,
\begin{eqnarray}\label{eq:rhof}
    \hat{\rho}_{\gamma}=\sum_n p_n^0 |n_{\gamma}\rangle\langle n_\gamma| =\hat{U}_{\gamma}^{\phantom{\dagger}} \hat{\rho}_0 \hat{U}^{\dagger}_{\gamma} \;.
\end{eqnarray}
The second equality follows from the relation,
\begin{equation}
    \partial_{\gamma}|n_{\gamma} \rangle= i\hat{G}_{\gamma}|n_{\gamma}\rangle \;,
\end{equation}
where the generator $\hat{G}_{\gamma}$ of the adiabatic evolution is known as the adiabatic gauge potential \cite{berry2009transitionless,kolodrubetz2017geometry}.
Since $|n_{\gamma+d\gamma}\rangle=e^{id\gamma \hat{G}_{\gamma}} |n_{\gamma}\rangle$ it follows that $|n_{\gamma}\rangle \simeq \hat{U}_{\gamma}|n_0\rangle$ with $\hat{U}_{\gamma}=e^{i\gamma \hat{G}_0}$ when $\gamma$ is small. The QFI for estimating $\lambda$ from a state transformed by Eq.~\ref{eq:rhof} is then \cite{toth,sm},
\begin{equation}\label{eq:qfi}
    F_Q[\lambda] = 2\sum_{n \ne m}\frac{(p_n^0-p_m^0)^2}{p_n^0+p_m^0} |\langle m_0|\hat{G}_0 | n_0\rangle |^2 \;.
\end{equation}
To find the relation between $\hat{G}_0$ and the physical perturbation $\hat{A}$, we note that $\partial_{\gamma} \langle m_\gamma | \hat{H}_{\gamma}|n_\gamma\rangle = i\left ( E_m^{\gamma} - E_n^{\gamma}\right ) \langle  m_\gamma|\hat{G}_{\gamma} | n_\gamma \rangle+ \langle  m_\gamma| \partial_\gamma\hat{H}_{\gamma} | n_\gamma \rangle$.
Since $\partial_\gamma \hat{H}_{\gamma}= \hat{A}$ and $\langle m_\gamma | \hat{H}_{\gamma}|n_\gamma\rangle =0$ for $n\ne m$, we thus find,
\begin{equation}
    \langle  m_\gamma|\hat{G}_\gamma | n_\gamma \rangle = i\frac{\langle  m_\gamma| \hat{A} | n_\gamma \rangle }{E_m^{\gamma} - E_n^{\gamma}} \qquad :~ n\ne m
\end{equation}
which yields Eq.~\ref{eq:QFIpert} when inserted into Eq.~\ref{eq:qfi}.

%####################
%####################

\noindent{\textit{Susceptibilities and transport coefficients.--}}
Our results for the QFI can be expressed in terms of the system's dynamical susceptibility $K(\omega)$ to the perturbation $\hat{A}$. For a weak perturbation switched on adiabatically, the Kubo formula then yields an alternative formulation in terms of the LR transport coefficient $\chi(\omega)$.

\rev{First, we introduce the retarded correlation function, $K(\omega)=-i\int_0^\infty dt\:\exp(i\omega t){\rm Tr}\{\hat{\rho}_0  [\hat{A}(0),\hat{A}(t)]\}$ evaluated in the  equilibrium thermal state of $\hat{H}_0$,} where $\hat{\Omega}(t)=e^{i\hat{H}_0t} \:\hat{\Omega}\: e^{-i\hat{H}_0t}$. Its Lehmann representation reads, 
${\rm Im} K(\omega)= \pi\sum_{n,m} (p_n^0-p_m^0)\:|\langle n_0|\hat{A}|m_0\rangle|^2\:\delta(\omega-E_m^0+E_n^0)$. Together with the identity $\int d\omega~\omega^{-2}\tanh(\omega/2T)\delta(\omega-E_m^0+E_n^0) = \frac{p_n^0-p_m^0}{p_n^0+p_m^0}\times(E_n^0-E_m^0)^{-2}$ we may express Eq.~\ref{eq:QFIpert},
\begin{eqnarray}\label{eq:qfi_susc}
F_Q[\lambda] = \frac{2}{\pi}\int d\omega~\frac{\tanh(\omega/2T)}{\omega^2}\: {\rm Im} K(\omega) \;.
\end{eqnarray}
This exact expression holds equally for interacting quantum many-body systems in the thermodynamic limit, as well as for finite open or closed systems.

\rev{In a different context, the QFI following unitary evolution was obtained in terms of a susceptibility in Ref.~\cite{zoller}.  
However, for quantum sensing one must explicitly connect the QFI to the \textit{physical evolution} induced by the perturbation $\lambda$, which results in the additional (and crucial) excitation energy denominator appearing in Eq.~\ref{eq:qfi_susc}.}

\rev{In general, applying the perturbation $\lambda \hat{A}$ will induce a current $\dot{A}$, where $\dot{A}=\tfrac{d}{dt}\hat{A}$. In LR (for a weak perturbation $\lambda$ switched on adiabatically) we have $\dot{A}\simeq \chi^{dc} \lambda$, with $\chi^{dc}$ the steady-state (dc) transport coefficient. This can be generalized to an ac perturbation 
$\lambda\cos(\omega t)\hat{A}$ of frequency $\omega$, which then endows the transport coefficient with a frequency dependence. The Kubo formula~\cite{kubo1957statistical, minarelli2022linear} relates this dynamical conductance to the dynamical susceptibility, $\chi(\omega)=-\omega\: {\rm Im} K(\omega)$, and hence,}
\begin{eqnarray}\label{eq:qfi_cond}
F_Q[\lambda] = -\frac{2}{\pi}\int d\omega~\frac{\tanh(\omega/2T)}{\omega^3}\: \chi(\omega) \;.
\end{eqnarray}
This immediately implies that optimal measurements on a quantum system yield \textit{perfect} metrological sensitivity due to a diverging QFI whenever the $\omega\to 0$ dc conductance is finite. 
Eqs.~\ref{eq:qfi_susc} and \ref{eq:qfi_cond} make a concrete connection between the maximum achievable metrological precision and standard physical observables.

%####################
%####################

\noindent{\textit{Nanoelectronic quantum sensors.--}}
We turn now to the application and implications of this for quantum nanoelectronic devices. 
We focus on the simplest model for a single semiconductor QD with local Coulomb interaction, coupled to source and drain leads -- the celebrated Anderson impurity model (AIM) \cite{hewson1997Kondo,goldhaber1998kondo,*van2000kondo,pustilnik2004kondo},
\begin{align}\label{eq:aim}
&\hat{H}_0=(\epsilon_d^{\phantom{\dagger}}+\tfrac{1}{2}B)  d_{\uparrow}^{\dagger}d_{\uparrow}^{\phantom{\dagger}} +(\epsilon_d^{\phantom{\dagger}}-\tfrac{1}{2}B) d_{\downarrow}^{\dagger}d_{\downarrow}^{\phantom{\dagger}} + U_d^{\phantom{\dagger}} \left (d_{\uparrow}^{\dagger}d_{\uparrow}^{\phantom{\dagger}}d_{\downarrow}^{\dagger}d_{\downarrow}^{\phantom{\dagger}} \right ) \nonumber\\ 
&+ t\sum_{j=1}^L \sum_{\alpha,\sigma} \left (c_{\alpha j \sigma}^{\dagger} c_{\alpha j+1 \sigma}^{\phantom{\dagger}} + {\rm H.c.} \right ) + V\sum_{\alpha,\sigma}  \left ( d_{\sigma}^{\dagger}c_{\alpha 1 \sigma}^{\phantom{\dagger}} + {\rm H.c.}  \right ) %\nonumber
\end{align}
where $\alpha=s,d$ for source and drain leads, $\sigma=\uparrow,\downarrow$ for up and down spin, $c_{\alpha j \sigma}^{(\dagger)}$ are annihilation (creation) operators for the conduction electrons, and $d_{\sigma}^{(\dagger)}$ are QD operators. Here we have given the Hamiltonian for each of the leads in the form of a 1d nanowire comprising $L$ sites. This allows us to study the scaling of the QFI with system size. The thermodynamic limit, where quantum transport can be meaningfully considered, corresponds to $L\to \infty$. The noninteracting ($U_d=0$) limit of this model is the resonant level model (RLM), which can be solved exactly using Green's function methods. For finite $L$ up to around $L=8$, the interacting AIM can be solved directly using exact diagonalization (ED). The full AIM with $L\to \infty$ can also be solved with sophisticated many-body techniques such as the numerical renormalization group (NRG) \cite{wilson1975renormalization,*bulla2008numerical}. We combine these methods below to study the metrological capability of the AIM.

%########################
\begin{figure}[t]
 	\centering
	\includegraphics[width=\linewidth]{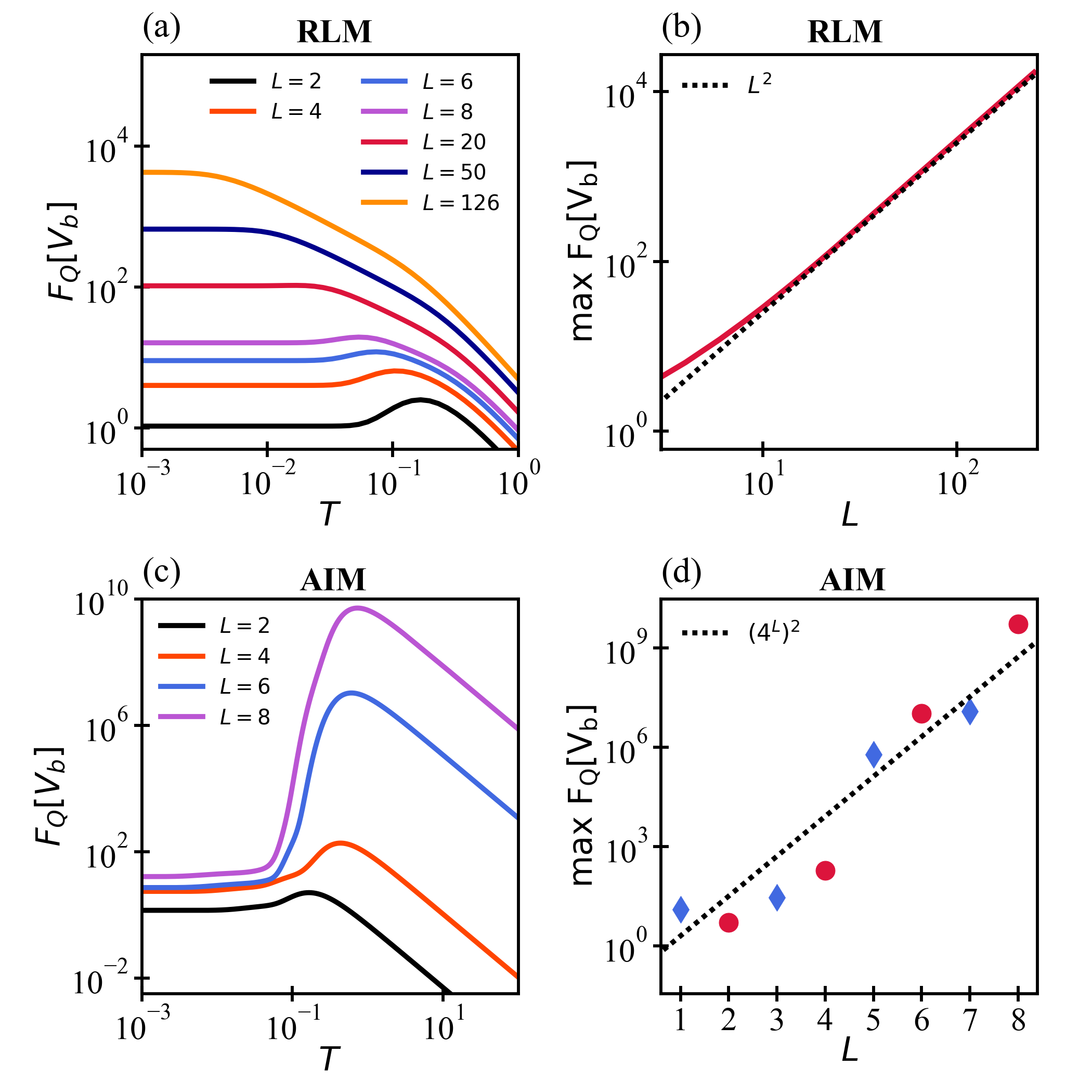}
\caption{QFI for a voltage bias perturbation in the noninteracting RLM with $U_d=0$ in (a,b) compared with interacting AIM for $U_d=0.2$ in (c,d).  $F_Q[V_b]$ plotted vs temperature $T$ for finite leads of $L$ sites in (a,c) and the scaling of the maximum QFI with $L$ shown in (b,d). Asymptotes as dotted lines. Parameters: $t=0.4$, $V=0.2$, $\epsilon_d=-U_d/2$ and $B=0$.} 
	\label{fig:QFI_volt}
\end{figure}
%########################

\noindent{\textit{Voltometry.--}}
We consider first a bias voltage perturbation switched on adiabatically, with $\lambda=-eV_b$ and $\hat{A}=\tfrac{1}{2}(\hat{N}_s-\hat{N}_d)$, where 
$\hat{N}_{\alpha}=\sum_{j=1}^L \hat{n}_{\alpha j}$
is the total number operator for lead $\alpha$ and $\hat{n}_{\alpha j}= \sum_{\sigma}c_{\alpha j \sigma}^{\dagger} c_{\alpha j \sigma}^{\phantom{\dagger}}$. This perturbation can be defined for finite or infinite $L$.
For infinite leads in the quantum transport context, the natural observable is the (average) electrical current into the drain lead, $I\equiv\langle \hat{I} \rangle$ with current operator $\hat{I}=-e\dot{N}_d$. In LR, $I  = \chi^{dc} V_b$, where $\chi^{dc}=\lim_{\omega\to 0}\chi(\omega)$ is the dc conductance. $\chi^{dc}$ is typically finite in the nanoelectronics context, and so the QFI for estimating the bias voltage diverges according to Eq.~\ref{eq:qfi_cond}.
Fundamentally this is because optimal measurements can be arbitrarily complex and leverage the full (in this case infinite) Fock space. 

It is however instructive to examine \textit{how} the QFI diverges with system size $L$. As shown in the SM \cite{sm}, the adiabatic voltage QFI for the AIM can be cast exactly~as 
\begin{eqnarray}\label{eq:QFIVb_poles}
    F_Q[V_b] = 4V^2 \sum_{k,p} \frac{\tanh \left (\frac{\xi_k-\epsilon_p}{2T}\right )}{(\xi_k-\epsilon_p)^4}\Big [f(\epsilon_p)  - f(\xi_k)  \Big ] \: |a_k  b_p|^2 \qquad~~
\end{eqnarray}
where $f(\omega)$ is the equilibrium Fermi function. $a_k$ and $\xi_k$ are the pole weights and energies of the uncoupled leads with spectrum $\rho_0(\omega)=\sum_{k=1}^L |a_k|^2 \delta(\omega-\xi_k)$ whereas
$b_p$ and $\epsilon_p$ are pole weights and energies of the lead-coupled QD spectral function $A_{QD}(\omega)= \sum_p |b_p|^2 \delta(\omega-\epsilon_p)$.\\   \rev{Contributions to the QFI in Eq.~\ref{eq:QFIVb_poles} are dominated by the minimum excitation energy gap $\Delta E={\rm min}(\xi_k-\epsilon_p)$, and for small gaps the QFI scales as $F_Q[V_b]\sim 1/\Delta E^2$.}

For the noninteracting RLM, the sum on $p$ runs over all $2L+1$ poles of the full system, $\hat{H}_0$. In this case $\epsilon_p$ are the single-particle energies in the diagonal representation $\hat{H}_0=\sum_{p\sigma} \epsilon_p f_{p\sigma}^{\dagger}f_{p\sigma}$ and $b_p$ is the weight of eigenstate $p$ on the QD orbital. 
Fig.~\ref{fig:QFI_volt}(a) shows the exact QFI for representative model parameters as a function of temperature $T$ for different system sizes. 
We see a saturation of $F_Q[V_b]$ for $T\ll t/L$ due to the existence of a minimum excitation gap $\Delta E$ in the finite system, and we generally see better sensitivity at lower temperatures. Fig.~\ref{fig:QFI_volt}(b) shows Heisenberg-type scaling of the QFI with the system size \cite{gietka2021adiabatic,demkowicz2012elusive,zwierz2012ultimate,zwierz2010general,rams2018limits}, ${\rm max}~F_Q[V_b]\sim L^2$. \rev{This is because the typical excitation gap is $\Delta E\sim 1/L$ for generic  noninteracting systems. 
Importantly, the quantum resources being utilized here are essentially the \textit{single-particle states}.}

For the interacting AIM, Figs.~\ref{fig:QFI_volt}(c,d), the story is quite different. The QD spectral function $A_{QD}(\omega)$ now involves exponentially-many terms, corresponding to the proliferation of many-particle excitations \cite{weichselbaum2007sum}. We now see  \textit{exponential scaling} of the QFI, ${\rm max}~F_Q[V_b]\sim 4^{2L}$. \rev{This is consistent with an exponential growth of the Fock space with increasing system size $L$, and the typical excitation gap $\Delta E\sim 1/4^L$. This highlights that optimal global measurements may exploit the full Fock space in strongly correlated systems~\cite{boixo2007generalized,sm}. The gap scale is a key ingredient in understanding the QFI~\cite{abiuso2025fundamental} -- and the gap has an implicit dependence on system size in many-body systems, which permits super-Heisenberg scaling.}

%######################
\begin{figure*}[t!]
\begin{center}
\includegraphics[width=\linewidth]{\figfour}
\end{center}
\caption{Quantum metrology with current measurements. (a) Precision of voltage estimation $F_I[V_b]$ in the nonequilibrium RLM as a function of $V_b$ and $T$ for $\epsilon_d=-0.2$, $\Gamma=0.05$, $B=0$. (b,c,d) NRG results for interacting AIM with $U_d=0.3$,  $\Gamma=0.013$ in LR. (b) $F_I[V_b]$ vs  $T/T_K$ in the universal Kondo regime for $\epsilon_d=-0.15$, $B=0$. (c) Magnetometry $F_I[B]$ vs $T/T_K$ and $B/T_K$ with $\epsilon_d=-0.15$. (d)  $F_I[S]$ vs $T$ and $\epsilon_d$ for estimation of the QD entropy $S$ from a current measurement. See \cite{sm} for details.}
	\label{fig:currentFI}
\end{figure*}
%######################

\rev{Here the voltage bias perturbation acts globally, and information about it is encoded non-redundantly across the many-body states of the system. Therefore the QFI $F_Q[V_b]$ diverges in the thermodynamic limit. However, perturbations that act locally (for example a magnetic field acting only on the QD) are expected to yield a QFI that saturates to a finite value as $L\to \infty$ because spatially distant degrees of freedom carry limited information about the perturbation. %, depending on the correlation length. 
This is discussed further for the case of QD magnetometry in the \textit{End Matter}.}

%#####################
%#####################

\noindent{\textit{Quantum sensing from current measurements.--}}
While the QFI considered above provides a bound on the best possible sensitivity, the optimal global measurement in a many-body system is typically impractical. For nanoelectronics devices with leads in the thermodynamic limit, the standard experimental measurement~\cite{goldhaber1998kondo,*van2000kondo} is the dc electrical current $I$ due to a voltage perturbation $V_b$. The parameter $\lambda$ to be estimated can be the perturbation strength $V_b$, some other parameter of $\hat{H}_0$, or indeed a non-Hamiltonian parameter such as temperature.  
The error-propagation formula \cite{toth} gives the precision $F_I[\lambda]$ for estimating parameter $\lambda$ from a current measurement,
\begin{eqnarray}\label{eq:errprop}
    F_I[\lambda]\equiv 1/{\rm Var}(\lambda)=|\partial_{\lambda} \langle \hat{I}\rangle|^2/{\rm Var}(I) \;,
\end{eqnarray}
and $F_Q[\lambda]\ge F_I[\lambda]$. The precision is equal to the classical Fisher information when current measurements are Gaussian distributed \cite{ding2022enhanced}. %Here we take $L\to \infty$.
\rev{High precision is obtained when the current $\langle \hat{I}\rangle$ is sensitive to changes in the parameter $\lambda$, and when current fluctuations ${\rm Var}(I)$ are small.}

For a noninteracting QD modelled by the RLM, the current is given by the Landauer-B\"uttiker (LB) formula \cite{buttiker1985generalized,minarelli2022linear},
$\langle \hat{I}\rangle = \frac{e}{h} \int d\omega\: \mathcal{T}(\omega)\times [f(\omega-\mu_s)-f(\omega-\mu_d)] $ which holds at finite bias $eV_b=\mu_s-\mu_d$, with $\mu_{\alpha}$ the chemical potential of lead $\alpha$. 
We assume for simplicity a flat lead density of states $\rho_0(\omega)=\rho_0\Theta(D-|\omega|)$ in a band of halfwidth $D\equiv 1$, and hybridization $\Gamma=\pi\rho_0 V^2$. For the RLM the transmission function is then given by $\mathcal{T}(\omega)= 4\pi\Gamma A_{QD}(\omega)$ in terms of the QD spectral function, which takes a Lorentzian form \cite{sm}. We consider instantaneous current measurements and so ${\rm Var}(I)=\tfrac{1}{2\pi}\int d\omega S(\omega)$ in terms of the nonequilibrium noise spectrum $S(\omega)$ \cite{sm}, which can be obtained in closed form from the transmission function via the Lesovik formula \cite{lesovik,sm}. In Fig.~\ref{fig:currentFI}(a) we use this machinery to compute $F_I[V_b]$ for voltage estimation, and find that resonances in the transmission function enhance the precision at sweet-spot values of $T$ and $V_b$.
Unlike the QFI, the precision $F_I[V_b]$ is finite.

\rev{Real nanoelectronics devices are however typically strongly interacting and non-Markovian, exhibiting Coulomb blockade and Kondo effects at low temperatures~\cite{goldhaber1998kondo,*van2000kondo}. Quantum transport properties in the many-body regime are not simply governed by a transmission function $\mathcal{T}(\omega)$ and the LB formalism no longer applies~\cite{meir1992landauer}. Instead we restrict to LR, and obtain the current $\langle \hat{I}\rangle$ directly from the Kubo formula~\cite{kubo1957statistical,minarelli2022linear}. 
In LR we can also utilize the fluctuation-dissipation relation to obtain an expression for ${\rm Var}(I)$. For an interacting QD modelled by the AIM, we therefore find that the precision $F_I[\lambda]$ in Eq.~\ref{eq:errprop} can be formulated entirely in terms of a dynamical susceptibility, as with the QFI in Eq.~\ref{eq:qfi_susc}, see \textit{End Matter} and SM \cite{sm}. The precision can therefore be computed numerically-exactly using NRG~\cite{wilson1975renormalization,*bulla2008numerical,weichselbaum2007sum}.}

Fig.~\ref{fig:currentFI}(b) shows current-based metrology for the applied voltage, $F_I[V_b]$ vs $T/T_K$ for the interacting AIM. We present universal results, with axes rescaled in terms of the emergent Kondo temperature $T_K$~\cite{hewson1997Kondo}. We see a dramatic enhancement of the precision at low temperatures due to the Kondo effect, since the current changes rapidly with $V_b$ when $T\sim T_K$. In panel (c) we consider magnetometry in the AIM, showing a nontrivial precision profile, with strongly enhanced performance for sensing the magnetic field $B$ at low $T\ll T_K$ for fields $B\sim T_K$. 

Finally, we briefly examine the capability of estimating thermodynamic properties using current measurements. Recent interest in measuring the entropy $S$ of a QD system \cite{hartman2018direct,han2022fractional,child2022entropy,campbell2025roadmap} motivates us to compute the precision $F_I[S]$ at $B=0$ in Fig.~\ref{fig:currentFI}(d).
We do this from Eq.~\ref{eq:errprop} by rewriting $\partial_S \langle \hat{I}\rangle = \partial_{\epsilon_d} \langle \hat{I}\rangle / \partial_{\epsilon_d} S$ and then exploiting the Maxwell relation $\partial_{\epsilon_d} S=-\partial_T \langle \hat{n}_d\rangle$ in terms of the QD charge $\hat{n}_d=\sum_{\sigma}d_{\sigma}^{\dagger}d_{\sigma}$. The required derivatives are obtained from NRG results for the evolution of $\langle \hat{I}\rangle$ and $\langle \hat{n}_d\rangle$ with $T$ and $\epsilon_d$.
Fig.~\ref{fig:currentFI}(d) shows highly nontrivial behavior, with distinct regions of the precision phase diagram corresponding to the different renormalization group fixed points of the AIM \cite{hewson1997Kondo}. Further details in the SM include the case of an integrated current measurement \cite{sm}.

%####################

\noindent{\textit{Conclusion.--}}
The metrological performance of a quantum system can be optimized by maximizing its QFI and measurement-specific precision.
Here we developed a general strategy for computing the QFI for a perturbation applied to a quantum many-body system, focusing on the linear response regime. We demonstrated the formalism for models of nanoelectronics devices, comparing the QFI for the optimal measurement with the precision for a more practical current-based measurement.

For generic extensive perturbations we expect the QFI for the optimal global measurement to grow exponentially in the number of degrees of freedom for strongly correlated systems, since the full Fock space can in principle be utilized. In particular, when a perturbation induces finite dc conductivity, we show that the QFI is non-redundantly encoded across the full Fock space, leading to a finite QFI density for the ensemble in the thermodynamic limit. 
Effective single-particle states in noninteracting systems appear to be a much weaker quantum resource for metrology. Local perturbations will typically yield a finite QFI even in the thermodynamic limit. 

The precision for practical measurements may be far from the QFI ideal -- as with the electrical current in QD devices. This presents ample opportunity to optimize the design and measurement protocol for nanoelectronics devices beyond the single QD paradigm. Many-body quantum effects \cite{goldhaber1998kondo,*van2000kondo,mitchell2017kondo,*sen2023many} and quantum criticality \cite{iftikhar2018tunable,pouse2023quantum,*karki2023z} may provide a route to enhanced sensing \cite{frerot2018quantum,PhysRevA.107.042614,mihailescu2023multiparameter,SciPostPhys.13.4.077}.

%The adiabatic gauge potential \cite{Takahashi2024,morawetz2024} has recently been computed using the Lanczos algorithm and might provide a route to the QFI beyond linear response.
Our results have potential impact beyond metrology, since the QFI plays many roles across quantum science and technology e.g.~witness for entanglement \cite{laurell2024}, measure of non-Markovianity \cite{Liu2020,scandi2023}, resource quantifier in quantum thermodynamics \cite{marvian2022} and fidelity susceptibility in quantum control \cite{Poggi2024}, quantum speed limits \cite{Deffner2017}, optimisation of variational quantum algorithms \cite{meyer2021}, and for continuous measurement currents \cite{Gammelmark2013,Gammelmark2014,radaelli2024,PRXQuantum.3.010354,boeyens2307probe}. \\
%\vfill

%%%%%%%%%%%%%%%%%%%%
%%%%%%%%%%%%%%%%%%%%

\begin{acknowledgments}
\noindent{\textit{Note added in proof.--}}  Recently, Ref.~\cite{mark-landi} studied current-based metrology using Eq.~\ref{eq:errprop} in the non-equilibrium regime of non-interacting nanoelectronics devices, uncovering ideal transmission function profiles for sensing.\\

\noindent{\textit{Acknowledgments.--}} We thank Steve Campbell, Mark Mitchison, Abolfazl Bayat, Tomohiro Shitara and Adeline Cr\'{e}pieux for insightful discussions. We acknowledge financial support from Science Foundation Ireland through Grant 21/RP-2TF/10019 (AKM) and Equal1 Laboratories Ireland Limited (GM).\\
%\vfill
%\clearpage
\end{acknowledgments}

%###################
%###################

\section*{End Matter}

\textit{Quantum transport in linear response.--}
For a system $\hat{H}_0$ initially at thermal equilibrium, consider an ac perturbation $\hat{H}_1=\lambda \cos(\omega t)\hat{A}$, switched on slowly. In LR the induced ac current is $\langle\dot{A}\rangle\simeq\chi(\omega)\lambda$. The Kubo formula~\cite{kubo1957statistical,galpin2014conductance} expresses the transport coefficient $\chi(\omega)$ in terms of the current-current correlation function $\bar{K}(\omega)=-i\int_0^\infty dt\:\exp(i\omega t){\rm Tr}\{\hat{\rho}_0  [\dot{A}(0),\dot{A}(t)]\}$ as $\chi(\omega)={\rm Im}\bar{K}(\omega)/\omega$. The dc steady-state then corresponds to the $\omega\to 0$ limit. As shown in Ref.~\cite{minarelli2022linear} one can write ${\rm Im}\bar{K}(\omega)=-\omega^2{\rm Im}K(\omega)$ in terms of the dynamical susceptibility $K(\omega)$. For an adiabatic perturbation, this provides the link between Eqs.~\ref{eq:qfi_susc} and \ref{eq:qfi_cond}. 

For the electrical current in a quantum circuit $I=-e\langle \dot{N}_d\rangle$, we have $\lambda=-eV_b$ and $\hat{A}=\tfrac{1}{2}(\hat{N}_s-\hat{N}_d)$. The variance of instantaneous current measurements is ${\rm Var}(I)=\langle \delta \hat{I}^2 \rangle$ with $\delta \hat{I}=\hat{I}-\langle\hat{I}\rangle$. Defining the noise spectrum $S(\omega)=\int dt\: e^{i\omega t}\langle \delta\hat{I}(0)\delta\hat{I}(t)\rangle$ we can then write 
${\rm Var}(I)=\tfrac{1}{2\pi}\int d\omega\:S(\omega)$. Current fluctuations in LR are therefore to leading order controlled by the equilibrium noise spectrum, itself obtained in terms of the retarded current autocorrelator $\bar{K}(\omega)$ via the fluctuation-dissipation relation $\pi S(\omega)=e^2n_B(\omega)\:{\rm Im}\bar{K}(\omega)$, with $n_B(\omega)$ the Bose-Einstein distribution~\cite{sm}.

Quantum transport properties in LR can therefore be computed from either $K(\omega)$ or $\bar{K}(\omega)$~\cite{minarelli2022linear}. Note that this formalism holds for arbitrary (interacting, non-Markovian) open quantum many-body systems. For interacting quantum impurity models of nanoelectronics circuits, NRG~\cite{wilson1975renormalization,*bulla2008numerical} can be used to evaluate the Lehmann representation of these correlation functions~\cite{weichselbaum2007sum}. The precision of current-based metrology for interacting nanoelectronics devices can then be assessed from Eq.~\ref{eq:errprop}.

%###############

\textit{Non-interacting limit.--}
The situation for non-interacting systems is far simpler because the physics is controlled by single-particle (product) states and exact solutions for any $L$ can be found using Green's function methods~\cite{sm}. For the RLM model of a non-interacting QD ($U_d=B=0$ limit of the AIM, Eq.~\ref{eq:aim}) the QD Green's function $G_{QD}(\omega)=-i\int_0^\infty dt\:\exp(i\omega t){\rm Tr}[\hat{\rho}_0  \{d(0),d^{\dagger}(t)\}]$ is given exactly by $G_{QD}(\omega)=[\omega+i0^+-\epsilon_d-2\Delta(\omega)]^{-1}$, where the dot-lead hybridization $\Delta(\omega)=V^2G_{\rm lead}(\omega)$ is given in terms of the free lead Green's function, which for infinite 1d nanowires (Eq.~\ref{eq:aim} with $L\to \infty$) is given by the solution to the self-consistent equation $G_{\rm lead}(\omega)=[\omega-t^2G_{\rm lead}(\omega)]^{-1}$. The uncoupled lead density of states then follows as $\rho_0(\omega)=-\tfrac{1}{\pi}{\rm Im}G_{\rm lead}(\omega)$ and the lead-coupled QD spectral function is $A_{QD}(\omega)=-\tfrac{1}{\pi}{\rm Im}G_{QD}(\omega)$. In the non-interacting limit, quantum transport is controlled by the transmission function $\mathcal{T}(\omega)= 4\pi\Gamma A_{QD}(\omega)$. For current-based metrology Eq.~\ref{eq:errprop}, the non-equilibrium current $\langle \hat{I}\rangle$ is given by the LB formula~\cite{buttiker1985generalized} and the non-equilibrium noise spectrum $S(\omega)$, upon which ${\rm Var}(I)$ depends, is given by the Lesovik formula~\cite{lesovik,sm}.

For the RLM with finite $L$ (see Fig.~\ref{fig:QFI_volt}a,b) one can numerically diagonalize the Hamiltonian operator using a canonical transformation, $\hat{H}_0=\sum_{p\sigma} \epsilon_p f_{p\sigma}^{\dagger}f_{p\sigma}$ where $f_{p\sigma}=b_p d_{\sigma}+\sum_j (\alpha_j c_{sj\sigma} + \bar{\alpha}_j c_{dj\sigma})$. The QD Green's function then immediately follows as $G_{QD}(\omega)=\sum_p |b_p|^2/(\omega^+-\epsilon_p)$ and $A_{QD}(\omega)=\sum_p |b_p|^2\delta(\omega-\epsilon_p)$.

%################

\textit{Derivation and behavior of Eq.~\ref{eq:QFIVb_poles}.--} For QD-lead hybridization in proportionate coupling as per Eq.~\ref{eq:aim}, one can decouple the current-current correlator~\cite{minarelli2022linear,sm} viz,
${\rm Im}\; \bar{K}(\omega)=\pi V^2\int d\omega' A_{QD}(\omega')  [ \rho_0(\omega'-\omega) \{ f(\omega'-\omega) - f(\omega')\} - \rho_0(\omega'+\omega)\{ f(\omega'+\omega) - f(\omega')\}  ]$.
%\begin{eqnarray}
%\frac{{\rm Im}\; \bar{K}(\omega)}{\pi V^2}= \int d\omega' A_{QD}(\omega')  \Big [& \rho_0&(\omega'-\omega) \{ f(\omega'-\omega) - f(\omega')\} \nonumber \\-& \rho_0&(\omega'+\omega)\{ f(\omega'+\omega) - f(\omega')\} \Big ] \;.\nonumber
%\end{eqnarray}
With Eq.~\ref{eq:qfi_cond} and $\chi(\omega)=\bar{K}(\omega)/\omega$ one may then express the voltometry QFI for the AIM as,
\begin{equation}\label{eq:voltintegral}
   \frac{F_Q[V_b]}{4V^2}= \iint d\omega d\omega'\frac{\tanh(\frac{\omega-\omega'}{2T})}{(\omega-\omega')^4} \left [  f(\omega') - f(\omega) \right]  \rho_0(\omega)  A_{QD}(\omega') \;.
\end{equation}
This expression holds in LR for interacting or non-interacting systems. The density of states of the uncoupled leads can be resolved in terms of single-particle excitations, $\rho_0(\omega)=\sum_k |a_k|^2 \delta(\omega-\xi_k)$. For interacting systems, the lead-coupled QD spectral function can be Lehmann-resolved,  $A_{QD}(\omega)=Z^{-1}\sum_{j,n}|\langle \psi_j|d_{\sigma}^{\dagger}|\psi_n\rangle|^2 (e^{-E_n/T}+e^{-E_j/T})\delta(\omega-E_j+E_n)$ where $\hat{H}_0|\psi_n\rangle=E_n|\psi_n\rangle$ and $Z=\sum_n e^{-E_n/T}$. This again takes the form of a sum over poles $A_{QD}(\omega)= \sum_p |b_p|^2 \delta(\omega-\epsilon_p)$ but where $\epsilon_p$ are now \textit{many-particle excitations}.
Inserting these expressions into Eq.~\ref{eq:voltintegral} yields Eq.~\ref{eq:QFIVb_poles}. We evaluate $A_{QD}(\omega)$ for the interacting AIM using exact diagonalization for $L\le 8$ and in the thermodynamic limit $L\to \infty$ using NRG.

The scaling behavior of Eq.~\ref{eq:QFIVb_poles} can be understood by noting that contributions to the QFI are dominated by the term in the sum with the minimum excitation energy gap $\Delta E={\rm min}(\xi_k-\epsilon_p)$ when $\Delta E$ is itself small. Approximating $\tanh(\Delta E/2T)\approx \Delta E/2T$ and $f(\epsilon_p)-f(\xi_k)\approx \Delta E \times f(\epsilon_p)[1-f(\epsilon_p)]$ when $\Delta E\ll 2T$, we find
\begin{eqnarray}
    F_Q[V_b] \sim 1/\Delta E^2 \;.
\end{eqnarray}

%############

\textit{Numerical Renormalization Group for the AIM.--}
Generalized quantum impurity models describing two-terminal quantum circuits exhibit notoriously complex many-body physics, due to the strong nanostructure-leads coupling which generates highly non-Markovian dynamics (rendering many standard open-systems techniques inapplicable) and strong electron correlations which generate extensive multipartite entanglement (rendering brute-force methods inapplicable). The gold-standard method of choice for solving impurity models such as the AIM in equilibrium is Wilson's numerical renormalization group (NRG) \cite{wilson1975renormalization,*bulla2008numerical,hewson1997Kondo}, which applies directly in the thermodynamic limit. The lead density of states $\rho_0(\omega)$ is discretized logarithmically and mapped to 1d Wilson chains. With the impurity at one end, the chain is constructed iteratively by adding one Wilson orbital at a time. At each step, the system is diagonalized and the Fock space is truncated by discarding high-energy states. As the chain is built up, the physics on progressively lower energy scales is revealed. Dynamical quantities such as $A_{QD}(\omega)$ can be computed via the Lehmann sum in the complete Anders-Schiller basis~\cite{weichselbaum2007sum}. %The method allows numerically-exact results to be obtained for thermodynamics and real-frequency dynamical quantities at any temperature down to $T=0$ at equilibrium, and applies directly in the thermodynamic limit $L\to \infty$.

%##############

%#####################
\begin{figure}[t]
    \centering
    \includegraphics[width=\linewidth]{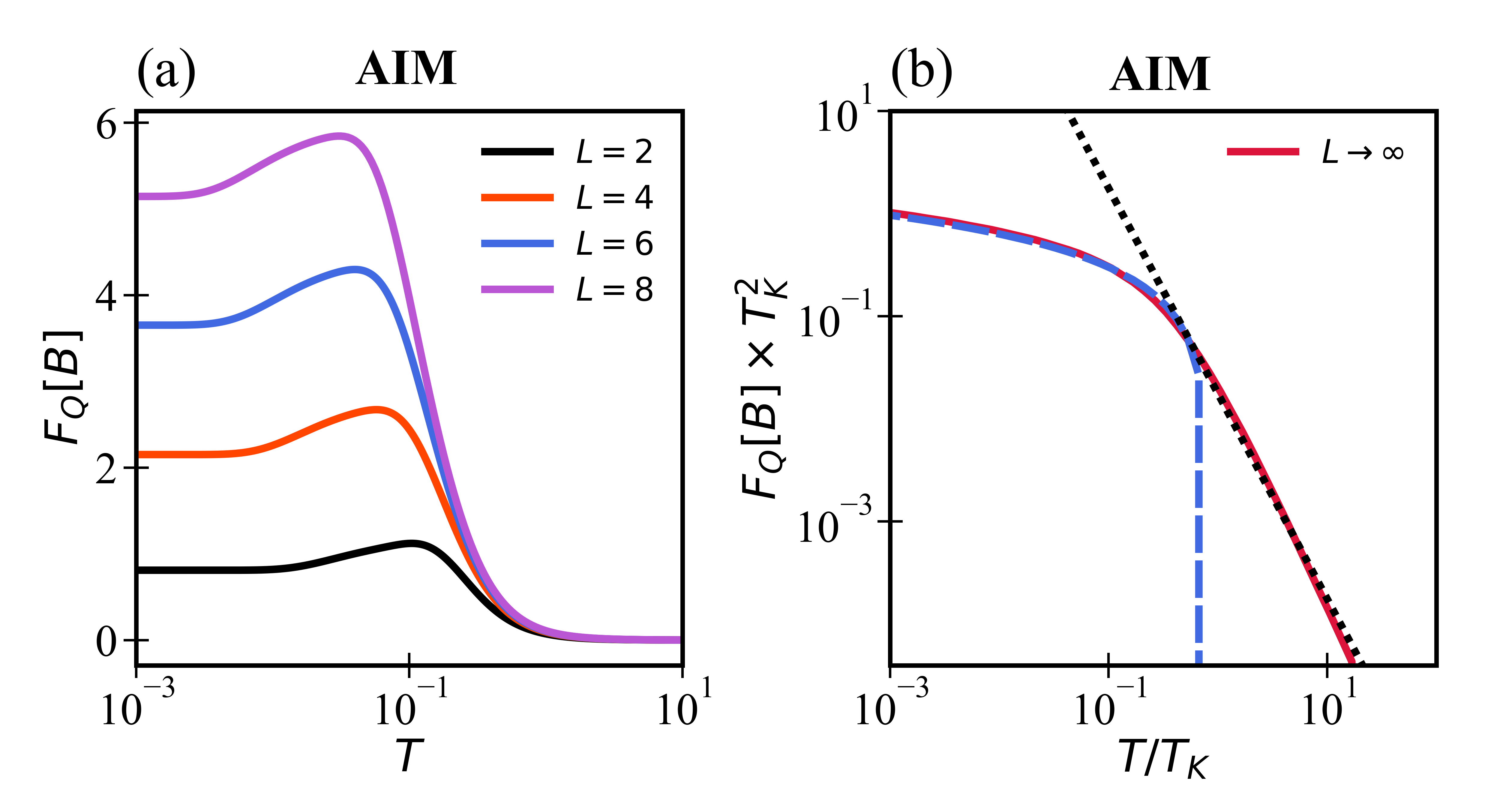}
    \caption{QFI for a local magnetic field $B$ applied adiabatically in the AIM. (a) ED results for $F_Q[B]$ vs $T$ in systems with finite leads using the same parameters as Fig.~\ref{fig:QFI_volt}c. (b) NRG results for the AIM in the thermodynamic limit of infinite leads, scaled in terms of the Kondo temperature $T_K$, for $U_d=0.2$, $\epsilon_d=-0.1$, $V=0.07$, $t=0.5$ in the universal regime. Dotted/dashed lines are the asymptotes discussed in the text. }
    \label{fig:Magnetometry}
\end{figure}
%#####################

\textit{Magnetometry QFI.--}
In the main text we considered an \textit{extensive} bias voltage perturbation applied to the full system. Here we contrast those results to the case of a \textit{local} magnetic field perturbation applied just to the QD site in Eq.~\ref{eq:aim}, $\lambda \hat{A} \equiv B\hat{S}_d^z$, where $\hat{S}_d^z=\tfrac{1}{2}(d^{\dagger}_{\uparrow} d^{\phantom{\dagger}}_{\uparrow}-d^{\dagger}_{\downarrow} d^{\phantom{\dagger}}_{\downarrow})$. We focus here on the interacting AIM since $\sigma=\uparrow,\downarrow$ are decoupled for $U_d=0$ and so the RLM has no spin dynamics. For a weak applied field $B$ we can study the QFI for magnetometry using Eq.~\ref{eq:qfi_susc}. In this case $K(\omega)=\langle\langle \hat{S}_d^z ; \hat{S}_d^z\rangle\rangle$ is the QD dynamical spin susceptibility~\cite{hanl2014local}, obtained for small finite systems directly using ED in Fig.~\ref{fig:Magnetometry}(a) and via NRG \cite{weichselbaum2007sum} for $L\to \infty$ in Fig.~\ref{fig:Magnetometry}(b).

The ED results for $F_Q[B]$ vs $T$ with finite leads show only a modest increase of sensitivity with $L$. Even though the underlying Fock space dimension increases exponentially with $L$, information on the local perturbation is seemingly not strongly imprinted on all many-particle states (unlike for the extensive voltage perturbation). Jumping to the case with infinite leads in Fig.~\ref{fig:Magnetometry}(b), we in fact find a finite $F_Q[B]$ for all finite $T$. To understand this behavior of the QFI one must understand the physics of the AIM and its spin dynamics. At half-filling, the interacting QD hosts a spin-$\tfrac{1}{2}$ local moment, which becomes dynamically screened by conduction electrons at low temperatures $T\ll T_K$ due to the formation of a many-body singlet state through a process known as the Kondo effect \cite{hewson1997Kondo}. The Kondo temperature $T_K$ is an emergent low-energy scale in terms of which physical properties exhibit universal scaling \cite{wilson1975renormalization,*bulla2008numerical}. While an asymptotically-free local moment is polarized by an infinitesimal field, giving a diverging dynamical spin susceptibility, the QD local moment in the AIM only starts to become polarized for $B\sim T_K$ due to formation of the Kondo singlet \cite{costi2000kondo}. On the lowest energy scales, $K(\omega)\sim \omega$ 
 \cite{hanl2014local} and thus Eq.~\ref{eq:qfi_susc} is finite. Up to log corrections, our NRG results are consistent with the ansatz ${\rm Im}\:K(\omega) \sim \frac{\omega/T_K^2}{1+(T/T_K)^2+(\omega/T_K)^2}$ in the universal regime $|\omega|,T \ll U_d$, from which we extract the universal asymptotes $T_K^2 F_Q[B]\sim (T_K/T)^2$ for $T\gg T_K$ (Fig.~\ref{fig:Magnetometry}(b),  dotted line) and $\sim \log[T_K/T]$ for $T\ll T_K$ (dashed line), in agreement with NRG data.
Our results are consistent with Eq.~\ref{eq:qfi_cond} since %the local field cannot induce a persistent dc spin current, and so 
$\chi(0)= 0$.

%############

\textit{Discussion of quantum sensing performance.--} 
%The QFI for voltometry shows strong scaling with system size $L$ and diverges as $L\to \infty$, see Fig.~\ref{fig:QFI_volt}. In the context of nanoelectronics circuits where the leads have a continuous electronic density of states and the thermodynamic limit $L\to \infty$ pertains, the QFI is therefore infinite. However, a practical current measurement yields finite precision for voltometry, see Figs.~\ref{fig:currentFI}(a,b). 
For nanoelectronics circuits, the practical measurement is the electrical current (and the thermodynamic limit $L\to \infty$ is implicit). For voltometry, note that the quantity $\partial_{V_b}\langle \hat{I}\rangle$ appearing in Eq.~\ref{eq:errprop} is precisely the conductance $G_c$, which is finite, taking a quantized maximum value of $2e^2/h$ per channel (spin-summed). We present results in units where $e\equiv 1$, $k_{\rm B}\equiv 1$ and $\hbar\equiv 1$. For a single non-interacting QD [Fig.~\ref{fig:currentFI}(a)] we find $F_I[V_b]$ has a maximum of order 1 for the typical parameters used. For the interacting QD case in LR [Fig.~\ref{fig:currentFI}(b)] we show the \textit{universal} curves scaled in terms of the Kondo temperature $T_K$ (which for these parameters is $\sim 10^{-6}D$). The precision, scaled in terms of the hybridization $\Gamma\times F_I[V_b]$, is strongly enhanced at low temperatures $T\ll T_K$ because the current is boosted by the many-body physics of the Kondo effect~\cite{goldhaber1998kondo,*van2000kondo, hewson1997Kondo}. On the other hand, instantaneous current fluctuations are controlled by the hybridization, ${\rm Var}(I)=\Gamma D/2\pi^2$ \cite{sm} and are minimized by a weak QD-lead coupling. The precision therefore takes a maximum \textit{universal} value $F_I[V_b]=8\pi^2/\Gamma D$. 

For magnetometry [Fig.~\ref{fig:currentFI}(c)] the situation is somewhat different, since the parameter to be estimated (the magnetic field $B$) is not the parameter inducing the current (which is the voltage $V_b$). In LR the precision $F_I[B]$ is therefore proportional to  $V_b^2$ which is small by construction. On the other hand, the current is extremely sensitive to the field $B$ on the scale of $T_K$ due to the Kondo effect~\cite{hewson1997Kondo}. Indeed, the precision takes a fully \textit{universal} form, scaling as $(V_b/T_K)^2$ when plotted vs $B/T_K$ and $T/T_K$. Since $T_K\sim D \exp(-\pi U_d/8\Gamma)$, the precision of magnetometry in the Kondo regime is exponentially boosted when the QD-lead hybridization $\Gamma$ is small and correlations $U_d$ are strong. In practice, $T_K$ can vary widely in real devices. 
As an example, in the QD device of Ref.~\cite{piquard2023observing}, the effective bandwidth is $D\sim 450$mK, $eV_b\sim 0.1D$, and a representative value of $T_K$ is $0.01D$. Using $\Gamma=0.013D$ as per Fig.~\ref{fig:currentFI}(c) we have max $F_I[B]\approx 0.05$ (in SI units of ${\rm K}^{-2} \:{\rm T}^{-2}$). 

For a review of metrological sensitivity scaling in other platforms, see e.g.~Fig.~2 of Ref.~\cite{Pezze2018}. For magnetometry, NV centers in diamond~\cite{barry2016optical,barry2020sensitivity,jensen2016magnetometry} and SQUIDs~\cite{drung2007highly} provide exceptional sensitivity, with the smallest detectable change in magnetic field with a measurement bandwidth of 1Hz being about 1pT/$\sqrt{\rm Hz}$ and 1fT/$\sqrt{\rm Hz}$ respectively. \\

%\vfill 
%\clearpage
%\bibliographystyle{unsrt}
%\bibliography{bibo}

\begin{thebibliography}{91}%
\makeatletter
\providecommand \@ifxundefined [1]{%
 \@ifx{#1\undefined}
}%
\providecommand \@ifnum [1]{%
 \ifnum #1\expandafter \@firstoftwo
 \else \expandafter \@secondoftwo
 \fi
}%
\providecommand \@ifx [1]{%
 \ifx #1\expandafter \@firstoftwo
 \else \expandafter \@secondoftwo
 \fi
}%
\providecommand \natexlab [1]{#1}%
\providecommand \enquote  [1]{``#1''}%
\providecommand \bibnamefont  [1]{#1}%
\providecommand \bibfnamefont [1]{#1}%
\providecommand \citenamefont [1]{#1}%
\providecommand \href@noop [0]{\@secondoftwo}%
\providecommand \href [0]{\begingroup \@sanitize@url \@href}%
\providecommand \@href[1]{\@@startlink{#1}\@@href}%
\providecommand \@@href[1]{\endgroup#1\@@endlink}%
\providecommand \@sanitize@url [0]{\catcode `\\12\catcode `\$12\catcode `\&12\catcode `\#12\catcode `\^12\catcode `\_12\catcode `\%12\relax}%
\providecommand \@@startlink[1]{}%
\providecommand \@@endlink[0]{}%
\providecommand \url  [0]{\begingroup\@sanitize@url \@url }%
\providecommand \@url [1]{\endgroup\@href {#1}{\urlprefix }}%
\providecommand \urlprefix  [0]{URL }%
\providecommand \Eprint [0]{\href }%
\providecommand \doibase [0]{https://doi.org/}%
\providecommand \selectlanguage [0]{\@gobble}%
\providecommand \bibinfo  [0]{\@secondoftwo}%
\providecommand \bibfield  [0]{\@secondoftwo}%
\providecommand \translation [1]{[#1]}%
\providecommand \BibitemOpen [0]{}%
\providecommand \bibitemStop [0]{}%
\providecommand \bibitemNoStop [0]{.\EOS\space}%
\providecommand \EOS [0]{\spacefactor3000\relax}%
\providecommand \BibitemShut  [1]{\csname bibitem#1\endcsname}%
\let\auto@bib@innerbib\@empty
%</preamble>
\bibitem [{\citenamefont {Martynov}\ \emph {et~al.}(2016)\citenamefont {Martynov}, \citenamefont {Hall}, \citenamefont {Abbott}, \citenamefont {Abbott}, \citenamefont {Abbott}, \citenamefont {Adams}, \citenamefont {Adhikari}, \citenamefont {Anderson}, \citenamefont {Anderson}, \citenamefont {Arai},\ and\ \citenamefont {et~al}}]{martynov2016sensitivity}%
  \BibitemOpen
  \bibfield  {author} {\bibinfo {author} {\bibfnamefont {D.~V.}\ \bibnamefont {Martynov}}, \bibinfo {author} {\bibfnamefont {E.~D.}\ \bibnamefont {Hall}}, \bibinfo {author} {\bibfnamefont {B.~P.}\ \bibnamefont {Abbott}}, \bibinfo {author} {\bibfnamefont {R.}~\bibnamefont {Abbott}}, \bibinfo {author} {\bibfnamefont {T.~D.}\ \bibnamefont {Abbott}}, \bibinfo {author} {\bibfnamefont {C.}~\bibnamefont {Adams}}, \bibinfo {author} {\bibfnamefont {R.~X.}\ \bibnamefont {Adhikari}}, \bibinfo {author} {\bibfnamefont {R.~A.}\ \bibnamefont {Anderson}}, \bibinfo {author} {\bibfnamefont {S.~B.}\ \bibnamefont {Anderson}}, \bibinfo {author} {\bibfnamefont {K.}~\bibnamefont {Arai}},\ and\ \bibinfo {author} {\bibnamefont {et~al}},\ }\bibfield  {title} {\bibinfo {title} {Sensitivity of the advanced ligo detectors at the beginning of gravitational wave astronomy},\ }\href {https://doi.org/10.1103/PhysRevD.93.112004} {\bibfield  {journal} {\bibinfo  {journal} {Phys. Rev. D}\ }\textbf {\bibinfo {volume} {93}},\ \bibinfo {pages}
  {112004} (\bibinfo {year} {2016})}\BibitemShut {NoStop}%
\bibitem [{\citenamefont {Preskill}(2018)}]{preskill2018}%
  \BibitemOpen
  \bibfield  {author} {\bibinfo {author} {\bibfnamefont {J.}~\bibnamefont {Preskill}},\ }\bibfield  {title} {\bibinfo {title} {Quantum {C}omputing in the {NISQ} era and beyond},\ }\href {https://doi.org/10.22331/q-2018-08-06-79} {\bibfield  {journal} {\bibinfo  {journal} {Quantum}\ }\textbf {\bibinfo {volume} {2}},\ \bibinfo {pages} {79} (\bibinfo {year} {2018})}\BibitemShut {NoStop}%
\bibitem [{\citenamefont {Paris}(2009)}]{Paris_Quantum_Estimation}%
  \BibitemOpen
  \bibfield  {author} {\bibinfo {author} {\bibfnamefont {M.~G.~A.}\ \bibnamefont {Paris}},\ }\bibfield  {title} {\bibinfo {title} {Quantum estimation for quantum technology},\ }\href {https://doi.org/10.1142/S0219749909004839} {\bibfield  {journal} {\bibinfo  {journal} {{I}nt. {J}. {Q}uantum {I}nf.}\ }\textbf {\bibinfo {volume} {07}},\ \bibinfo {pages} {125} (\bibinfo {year} {2009})}\BibitemShut {NoStop}%
\bibitem [{\citenamefont {Tóth}\ and\ \citenamefont {Apellaniz}(2014)}]{toth}%
  \BibitemOpen
  \bibfield  {author} {\bibinfo {author} {\bibfnamefont {G.}~\bibnamefont {Tóth}}\ and\ \bibinfo {author} {\bibfnamefont {I.}~\bibnamefont {Apellaniz}},\ }\bibfield  {title} {\bibinfo {title} {Quantum metrology from a quantum information science perspective},\ }\href {https://doi.org/10.1088/1751-8113/47/42/424006} {\bibfield  {journal} {\bibinfo  {journal} {J. Phys. A: Math. Theor}\ }\textbf {\bibinfo {volume} {47}},\ \bibinfo {pages} {424006} (\bibinfo {year} {2014})}\BibitemShut {NoStop}%
\bibitem [{\citenamefont {Degen}\ \emph {et~al.}(2017)\citenamefont {Degen}, \citenamefont {Reinhard},\ and\ \citenamefont {Cappellaro}}]{degen2017}%
  \BibitemOpen
  \bibfield  {author} {\bibinfo {author} {\bibfnamefont {C.~L.}\ \bibnamefont {Degen}}, \bibinfo {author} {\bibfnamefont {F.}~\bibnamefont {Reinhard}},\ and\ \bibinfo {author} {\bibfnamefont {P.}~\bibnamefont {Cappellaro}},\ }\bibfield  {title} {\bibinfo {title} {Quantum sensing},\ }\href {https://doi.org/10.1103/RevModPhys.89.035002} {\bibfield  {journal} {\bibinfo  {journal} {Rev. Mod. Phys.}\ }\textbf {\bibinfo {volume} {89}},\ \bibinfo {pages} {035002} (\bibinfo {year} {2017})}\BibitemShut {NoStop}%
\bibitem [{\citenamefont {Giovannetti}\ \emph {et~al.}(2011)\citenamefont {Giovannetti}, \citenamefont {Lloyd},\ and\ \citenamefont {Maccone}}]{Giovannetti2011}%
  \BibitemOpen
  \bibfield  {author} {\bibinfo {author} {\bibfnamefont {V.}~\bibnamefont {Giovannetti}}, \bibinfo {author} {\bibfnamefont {S.}~\bibnamefont {Lloyd}},\ and\ \bibinfo {author} {\bibfnamefont {L.}~\bibnamefont {Maccone}},\ }\bibfield  {title} {\bibinfo {title} {Advances in quantum metrology},\ }\href {https://doi.org/10.1038/nphoton.2011.35} {\bibfield  {journal} {\bibinfo  {journal} {Nature {P}hotonics}\ }\textbf {\bibinfo {volume} {5}},\ \bibinfo {pages} {222} (\bibinfo {year} {2011})}\BibitemShut {NoStop}%
\bibitem [{\citenamefont {Yang}\ \emph {et~al.}(2022)\citenamefont {Yang}, \citenamefont {Pang}, \citenamefont {del Campo},\ and\ \citenamefont {Jordan}}]{yang2022super}%
  \BibitemOpen
  \bibfield  {author} {\bibinfo {author} {\bibfnamefont {J.}~\bibnamefont {Yang}}, \bibinfo {author} {\bibfnamefont {S.}~\bibnamefont {Pang}}, \bibinfo {author} {\bibfnamefont {A.}~\bibnamefont {del Campo}},\ and\ \bibinfo {author} {\bibfnamefont {A.~N.}\ \bibnamefont {Jordan}},\ }\bibfield  {title} {\bibinfo {title} {Super-heisenberg scaling in hamiltonian parameter estimation in the long-range kitaev chain},\ }\href {https://doi.org/10.1103/PhysRevResearch.4.013133} {\bibfield  {journal} {\bibinfo  {journal} {Phys. Rev. Res.}\ }\textbf {\bibinfo {volume} {4}},\ \bibinfo {pages} {013133} (\bibinfo {year} {2022})}\BibitemShut {NoStop}%
\bibitem [{\citenamefont {Aslam}\ \emph {et~al.}(2023)\citenamefont {Aslam}, \citenamefont {Zhou}, \citenamefont {Urbach}, \citenamefont {Turner}, \citenamefont {Walsworth}, \citenamefont {Lukin},\ and\ \citenamefont {Park}}]{aslam2023}%
  \BibitemOpen
  \bibfield  {author} {\bibinfo {author} {\bibfnamefont {N.}~\bibnamefont {Aslam}}, \bibinfo {author} {\bibfnamefont {H.}~\bibnamefont {Zhou}}, \bibinfo {author} {\bibfnamefont {E.~K.}\ \bibnamefont {Urbach}}, \bibinfo {author} {\bibfnamefont {M.~J.}\ \bibnamefont {Turner}}, \bibinfo {author} {\bibfnamefont {R.~L.}\ \bibnamefont {Walsworth}}, \bibinfo {author} {\bibfnamefont {M.~D.}\ \bibnamefont {Lukin}},\ and\ \bibinfo {author} {\bibfnamefont {H.}~\bibnamefont {Park}},\ }\bibfield  {title} {\bibinfo {title} {Quantum sensors for biomedical applications},\ }\href {https://doi.org/10.1038/s42254-023-00558-3} {\bibfield  {journal} {\bibinfo  {journal} {Nat. Rev. Phys.}\ }\textbf {\bibinfo {volume} {5}},\ \bibinfo {pages} {157} (\bibinfo {year} {2023})}\BibitemShut {NoStop}%
\bibitem [{\citenamefont {Abiuso}\ \emph {et~al.}(2025)\citenamefont {Abiuso}, \citenamefont {Sekatski}, \citenamefont {Calsamiglia},\ and\ \citenamefont {Perarnau-Llobet}}]{abiuso2025fundamental}%
  \BibitemOpen
  \bibfield  {author} {\bibinfo {author} {\bibfnamefont {P.}~\bibnamefont {Abiuso}}, \bibinfo {author} {\bibfnamefont {P.}~\bibnamefont {Sekatski}}, \bibinfo {author} {\bibfnamefont {J.}~\bibnamefont {Calsamiglia}},\ and\ \bibinfo {author} {\bibfnamefont {M.}~\bibnamefont {Perarnau-Llobet}},\ }\bibfield  {title} {\bibinfo {title} {Fundamental limits of metrology at thermal equilibrium},\ }\href {https://doi.org/10.1103/PhysRevLett.134.010801} {\bibfield  {journal} {\bibinfo  {journal} {Phys. Rev. Lett.}\ }\textbf {\bibinfo {volume} {134}},\ \bibinfo {pages} {010801} (\bibinfo {year} {2025})}\BibitemShut {NoStop}%
\bibitem [{\citenamefont {Ye}\ and\ \citenamefont {Zoller}(2024)}]{Ye2024}%
  \BibitemOpen
  \bibfield  {author} {\bibinfo {author} {\bibfnamefont {J.}~\bibnamefont {Ye}}\ and\ \bibinfo {author} {\bibfnamefont {P.}~\bibnamefont {Zoller}},\ }\bibfield  {title} {\bibinfo {title} {Essay: Quantum sensing with atomic, molecular, and optical platforms for fundamental physics},\ }\href {https://doi.org/10.1103/PhysRevLett.132.190001} {\bibfield  {journal} {\bibinfo  {journal} {Phys. Rev. Lett.}\ }\textbf {\bibinfo {volume} {132}},\ \bibinfo {pages} {190001} (\bibinfo {year} {2024})}\BibitemShut {NoStop}%
\bibitem [{\citenamefont {Pelayo}\ \emph {et~al.}(2023)\citenamefont {Pelayo}, \citenamefont {Gietka},\ and\ \citenamefont {Busch}}]{PhysRevA.107.033318}%
  \BibitemOpen
  \bibfield  {author} {\bibinfo {author} {\bibfnamefont {J.~C.}\ \bibnamefont {Pelayo}}, \bibinfo {author} {\bibfnamefont {K.}~\bibnamefont {Gietka}},\ and\ \bibinfo {author} {\bibfnamefont {T.}~\bibnamefont {Busch}},\ }\bibfield  {title} {\bibinfo {title} {Distributed quantum sensing with optical lattices},\ }\href {https://doi.org/10.1103/PhysRevA.107.033318} {\bibfield  {journal} {\bibinfo  {journal} {Phys. Rev. A}\ }\textbf {\bibinfo {volume} {107}},\ \bibinfo {pages} {033318} (\bibinfo {year} {2023})}\BibitemShut {NoStop}%
\bibitem [{\citenamefont {Gietka}\ and\ \citenamefont {Ritsch}(2023)}]{PhysRevLett.130.090802}%
  \BibitemOpen
  \bibfield  {author} {\bibinfo {author} {\bibfnamefont {K.}~\bibnamefont {Gietka}}\ and\ \bibinfo {author} {\bibfnamefont {H.}~\bibnamefont {Ritsch}},\ }\bibfield  {title} {\bibinfo {title} {Squeezing and overcoming the heisenberg scaling with spin-orbit coupled quantum gases},\ }\href {https://doi.org/10.1103/PhysRevLett.130.090802} {\bibfield  {journal} {\bibinfo  {journal} {Phys. Rev. Lett.}\ }\textbf {\bibinfo {volume} {130}},\ \bibinfo {pages} {090802} (\bibinfo {year} {2023})}\BibitemShut {NoStop}%
\bibitem [{\citenamefont {Di~Candia}\ \emph {et~al.}(2023)\citenamefont {Di~Candia}, \citenamefont {Minganti}, \citenamefont {Petrovnin}, \citenamefont {Paraoanu},\ and\ \citenamefont {Felicetti}}]{DiCandia2023}%
  \BibitemOpen
  \bibfield  {author} {\bibinfo {author} {\bibfnamefont {R.}~\bibnamefont {Di~Candia}}, \bibinfo {author} {\bibfnamefont {F.}~\bibnamefont {Minganti}}, \bibinfo {author} {\bibfnamefont {K.}~\bibnamefont {Petrovnin}}, \bibinfo {author} {\bibfnamefont {G.}~\bibnamefont {Paraoanu}},\ and\ \bibinfo {author} {\bibfnamefont {S.}~\bibnamefont {Felicetti}},\ }\bibfield  {title} {\bibinfo {title} {Critical parametric quantum sensing},\ }\href {https://doi.org/10.1038/s41534-023-00690-z} {\bibfield  {journal} {\bibinfo  {journal} {npj Quantum Information}\ }\textbf {\bibinfo {volume} {9}},\ \bibinfo {pages} {23} (\bibinfo {year} {2023})}\BibitemShut {NoStop}%
\bibitem [{\citenamefont {Ying}\ \emph {et~al.}(2022)\citenamefont {Ying}, \citenamefont {Felicetti}, \citenamefont {Liu},\ and\ \citenamefont {Braak}}]{e24081015}%
  \BibitemOpen
  \bibfield  {author} {\bibinfo {author} {\bibfnamefont {Z.-J.}\ \bibnamefont {Ying}}, \bibinfo {author} {\bibfnamefont {S.}~\bibnamefont {Felicetti}}, \bibinfo {author} {\bibfnamefont {G.}~\bibnamefont {Liu}},\ and\ \bibinfo {author} {\bibfnamefont {D.}~\bibnamefont {Braak}},\ }\bibfield  {title} {\bibinfo {title} {Critical quantum metrology in the non-linear quantum rabi model},\ }\href {https://www.mdpi.com/1099-4300/24/8/1015} {\bibfield  {journal} {\bibinfo  {journal} {Entropy}\ }\textbf {\bibinfo {volume} {24}},\ \bibinfo {pages} {1015} (\bibinfo {year} {2022})}\BibitemShut {NoStop}%
\bibitem [{\citenamefont {Montenegro}\ \emph {et~al.}(2022)\citenamefont {Montenegro}, \citenamefont {Jones}, \citenamefont {Bose},\ and\ \citenamefont {Bayat}}]{mag1}%
  \BibitemOpen
  \bibfield  {author} {\bibinfo {author} {\bibfnamefont {V.}~\bibnamefont {Montenegro}}, \bibinfo {author} {\bibfnamefont {G.~S.}\ \bibnamefont {Jones}}, \bibinfo {author} {\bibfnamefont {S.}~\bibnamefont {Bose}},\ and\ \bibinfo {author} {\bibfnamefont {A.}~\bibnamefont {Bayat}},\ }\bibfield  {title} {\bibinfo {title} {Sequential measurements for quantum-enhanced magnetometry in spin chain probes},\ }\href {https://doi.org/10.1103/PhysRevLett.129.120503} {\bibfield  {journal} {\bibinfo  {journal} {Phys. {R}ev. {L}ett.}\ }\textbf {\bibinfo {volume} {129}},\ \bibinfo {pages} {120503} (\bibinfo {year} {2022})}\BibitemShut {NoStop}%
\bibitem [{\citenamefont {Troiani}\ and\ \citenamefont {Paris}(2018)}]{mag2}%
  \BibitemOpen
  \bibfield  {author} {\bibinfo {author} {\bibfnamefont {F.}~\bibnamefont {Troiani}}\ and\ \bibinfo {author} {\bibfnamefont {M.~G.~A.}\ \bibnamefont {Paris}},\ }\bibfield  {title} {\bibinfo {title} {Universal quantum magnetometry with spin states at equilibrium},\ }\href {https://doi.org/10.1103/PhysRevLett.120.260503} {\bibfield  {journal} {\bibinfo  {journal} {Phys. {R}ev. {L}ett.}\ }\textbf {\bibinfo {volume} {120}},\ \bibinfo {pages} {260503} (\bibinfo {year} {2018})}\BibitemShut {NoStop}%
\bibitem [{\citenamefont {Albarelli}\ \emph {et~al.}(2017)\citenamefont {Albarelli}, \citenamefont {Rossi}, \citenamefont {Paris},\ and\ \citenamefont {Genoni}}]{mag3}%
  \BibitemOpen
  \bibfield  {author} {\bibinfo {author} {\bibfnamefont {F.}~\bibnamefont {Albarelli}}, \bibinfo {author} {\bibfnamefont {M.}~\bibnamefont {Rossi}}, \bibinfo {author} {\bibfnamefont {M.}~\bibnamefont {Paris}},\ and\ \bibinfo {author} {\bibfnamefont {M.}~\bibnamefont {Genoni}},\ }\bibfield  {title} {\bibinfo {title} {Ultimate limits for quantum magnetometry via time-continuous measurements},\ }\href {https://doi.org/10.1088/1367-2630/aa9840} {\bibfield  {journal} {\bibinfo  {journal} {New {J}. {P}hys.}\ }\textbf {\bibinfo {volume} {19}},\ \bibinfo {pages} {123011} (\bibinfo {year} {2017})}\BibitemShut {NoStop}%
\bibitem [{\citenamefont {Goldhaber-Gordon}\ \emph {et~al.}(1998)\citenamefont {Goldhaber-Gordon}, \citenamefont {Shtrikman}, \citenamefont {Mahalu}, \citenamefont {Abusch-Magder}, \citenamefont {Meirav},\ and\ \citenamefont {Kastner}}]{goldhaber1998kondo}%
  \BibitemOpen
  \bibfield  {author} {\bibinfo {author} {\bibfnamefont {D.}~\bibnamefont {Goldhaber-Gordon}}, \bibinfo {author} {\bibfnamefont {H.}~\bibnamefont {Shtrikman}}, \bibinfo {author} {\bibfnamefont {D.}~\bibnamefont {Mahalu}}, \bibinfo {author} {\bibfnamefont {D.}~\bibnamefont {Abusch-Magder}}, \bibinfo {author} {\bibfnamefont {U.}~\bibnamefont {Meirav}},\ and\ \bibinfo {author} {\bibfnamefont {M.~A.}\ \bibnamefont {Kastner}},\ }\bibfield  {title} {\bibinfo {title} {Kondo effect in a single-electron transistor},\ }\href {https://doi.org/10.1038/34373} {\bibfield  {journal} {\bibinfo  {journal} {Nature}\ }\textbf {\bibinfo {volume} {391}},\ \bibinfo {pages} {156} (\bibinfo {year} {1998})}\BibitemShut {NoStop}%
\bibitem [{\citenamefont {Van~der Wiel}\ \emph {et~al.}(2000)\citenamefont {Van~der Wiel}, \citenamefont {Franceschi}, \citenamefont {Fujisawa}, \citenamefont {Elzerman}, \citenamefont {Tarucha},\ and\ \citenamefont {Kouwenhoven}}]{van2000kondo}%
  \BibitemOpen
  \bibfield  {author} {\bibinfo {author} {\bibfnamefont {W.}~\bibnamefont {Van~der Wiel}}, \bibinfo {author} {\bibfnamefont {S.~D.}\ \bibnamefont {Franceschi}}, \bibinfo {author} {\bibfnamefont {T.}~\bibnamefont {Fujisawa}}, \bibinfo {author} {\bibfnamefont {J.}~\bibnamefont {Elzerman}}, \bibinfo {author} {\bibfnamefont {S.}~\bibnamefont {Tarucha}},\ and\ \bibinfo {author} {\bibfnamefont {L.}~\bibnamefont {Kouwenhoven}},\ }\bibfield  {title} {\bibinfo {title} {The kondo effect in the unitary limit},\ }\href {https://doi.org/10.1126/science.289.5487.2105} {\bibfield  {journal} {\bibinfo  {journal} {Science}\ }\textbf {\bibinfo {volume} {289}},\ \bibinfo {pages} {2105} (\bibinfo {year} {2000})}\BibitemShut {NoStop}%
\bibitem [{\citenamefont {Perrin}\ \emph {et~al.}(2015)\citenamefont {Perrin}, \citenamefont {Burzurí},\ and\ \citenamefont {van~der Zant}}]{perrin2015single}%
  \BibitemOpen
  \bibfield  {author} {\bibinfo {author} {\bibfnamefont {M.~L.}\ \bibnamefont {Perrin}}, \bibinfo {author} {\bibfnamefont {E.}~\bibnamefont {Burzurí}},\ and\ \bibinfo {author} {\bibfnamefont {H.~S.~J.}\ \bibnamefont {van~der Zant}},\ }\bibfield  {title} {\bibinfo {title} {Single-molecule transistors},\ }\href {https://doi.org/10.1039/C4CS00231H} {\bibfield  {journal} {\bibinfo  {journal} {Chem. Soc. Rev.}\ }\textbf {\bibinfo {volume} {44}},\ \bibinfo {pages} {902} (\bibinfo {year} {2015})}\BibitemShut {NoStop}%
\bibitem [{\citenamefont {Chen}\ \emph {et~al.}(2024)\citenamefont {Chen}, \citenamefont {Grace}, \citenamefont {Woltering}, \citenamefont {Chen}, \citenamefont {Gee}, \citenamefont {Baugh}, \citenamefont {Briggs}, \citenamefont {Bogani}, \citenamefont {Mol}, \citenamefont {Lambert}, \citenamefont {Anderson},\ and\ \citenamefont {Thomas}}]{chen2024quantum}%
  \BibitemOpen
  \bibfield  {author} {\bibinfo {author} {\bibfnamefont {Z.}~\bibnamefont {Chen}}, \bibinfo {author} {\bibfnamefont {I.~M.}\ \bibnamefont {Grace}}, \bibinfo {author} {\bibfnamefont {S.~L.}\ \bibnamefont {Woltering}}, \bibinfo {author} {\bibfnamefont {L.}~\bibnamefont {Chen}}, \bibinfo {author} {\bibfnamefont {A.}~\bibnamefont {Gee}}, \bibinfo {author} {\bibfnamefont {J.}~\bibnamefont {Baugh}}, \bibinfo {author} {\bibfnamefont {G.~A.~D.}\ \bibnamefont {Briggs}}, \bibinfo {author} {\bibfnamefont {L.}~\bibnamefont {Bogani}}, \bibinfo {author} {\bibfnamefont {J.~A.}\ \bibnamefont {Mol}}, \bibinfo {author} {\bibfnamefont {C.~J.}\ \bibnamefont {Lambert}}, \bibinfo {author} {\bibfnamefont {H.~L.}\ \bibnamefont {Anderson}},\ and\ \bibinfo {author} {\bibfnamefont {J.~O.}\ \bibnamefont {Thomas}},\ }\bibfield  {title} {\bibinfo {title} {Quantum interference enhances the performance of single-molecule transistors},\ }\href {https://doi.org/10.1038/s41565-024-01633-1} {\bibfield  {journal} {\bibinfo  {journal} {Nat.
  Nanotechnol.}\ ,\ \bibinfo {pages} {1}} (\bibinfo {year} {2024})}\BibitemShut {NoStop}%
\bibitem [{\citenamefont {Nowack}\ \emph {et~al.}(2007)\citenamefont {Nowack}, \citenamefont {Koppens}, \citenamefont {Nazarov},\ and\ \citenamefont {Vandersypen}}]{nowack2007coherent}%
  \BibitemOpen
  \bibfield  {author} {\bibinfo {author} {\bibfnamefont {K.}~\bibnamefont {Nowack}}, \bibinfo {author} {\bibfnamefont {F.~H.~L.}\ \bibnamefont {Koppens}}, \bibinfo {author} {\bibfnamefont {Y.~V.}\ \bibnamefont {Nazarov}},\ and\ \bibinfo {author} {\bibfnamefont {L.~M.~K.}\ \bibnamefont {Vandersypen}},\ }\bibfield  {title} {\bibinfo {title} {Coherent control of a single electron spin with electric fields},\ }\href {https://doi.org/10.1126/science.1148092} {\bibfield  {journal} {\bibinfo  {journal} {Science}\ }\textbf {\bibinfo {volume} {318}},\ \bibinfo {pages} {1430} (\bibinfo {year} {2007})}\BibitemShut {NoStop}%
\bibitem [{\citenamefont {Barthelemy}\ and\ \citenamefont {Vandersypen}(2013)}]{barthelemy2013quantum}%
  \BibitemOpen
  \bibfield  {author} {\bibinfo {author} {\bibfnamefont {P.}~\bibnamefont {Barthelemy}}\ and\ \bibinfo {author} {\bibfnamefont {L.~M.~K.}\ \bibnamefont {Vandersypen}},\ }\bibfield  {title} {\bibinfo {title} {Quantum dot systems: a versatile platform for quantum simulations},\ }\href {https://doi.org/https://doi.org/10.1002/andp.201300124} {\bibfield  {journal} {\bibinfo  {journal} {Annalen der Physik}\ }\textbf {\bibinfo {volume} {525}},\ \bibinfo {pages} {808} (\bibinfo {year} {2013})}\BibitemShut {NoStop}%
\bibitem [{\citenamefont {Ihn}(2009)}]{ihn2009semiconductor}%
  \BibitemOpen
  \bibfield  {author} {\bibinfo {author} {\bibfnamefont {T.}~\bibnamefont {Ihn}},\ }\href@noop {} {\emph {\bibinfo {title} {Semiconductor Nanostructures: Quantum states and electronic transport}}}\ (\bibinfo  {publisher} {OUP Oxford},\ \bibinfo {year} {2009})\BibitemShut {NoStop}%
\bibitem [{\citenamefont {Meirav}\ \emph {et~al.}(1990)\citenamefont {Meirav}, \citenamefont {Kastner},\ and\ \citenamefont {Wind}}]{meirav1990single}%
  \BibitemOpen
  \bibfield  {author} {\bibinfo {author} {\bibfnamefont {U.}~\bibnamefont {Meirav}}, \bibinfo {author} {\bibfnamefont {M.~A.}\ \bibnamefont {Kastner}},\ and\ \bibinfo {author} {\bibfnamefont {S.~J.}\ \bibnamefont {Wind}},\ }\bibfield  {title} {\bibinfo {title} {Single-electron charging and periodic conductance resonances in gaas nanostructures},\ }\href {https://doi.org/10.1103/PhysRevLett.65.771} {\bibfield  {journal} {\bibinfo  {journal} {Phys. Rev. Lett.}\ }\textbf {\bibinfo {volume} {65}},\ \bibinfo {pages} {771} (\bibinfo {year} {1990})}\BibitemShut {NoStop}%
\bibitem [{\citenamefont {Park}\ \emph {et~al.}(2002)\citenamefont {Park}, \citenamefont {Pasupathy}, \citenamefont {Goldsmith}, \citenamefont {Chang}, \citenamefont {Yaish}, \citenamefont {Petta}, \citenamefont {Rinkoski}, \citenamefont {Sethna}, \citenamefont {Abru{\~n}a}, \citenamefont {McEuen},\ and\ \citenamefont {Ralph}}]{park2002coulomb}%
  \BibitemOpen
  \bibfield  {author} {\bibinfo {author} {\bibfnamefont {J.}~\bibnamefont {Park}}, \bibinfo {author} {\bibfnamefont {A.}~\bibnamefont {Pasupathy}}, \bibinfo {author} {\bibfnamefont {J.}~\bibnamefont {Goldsmith}}, \bibinfo {author} {\bibfnamefont {C.}~\bibnamefont {Chang}}, \bibinfo {author} {\bibfnamefont {Y.}~\bibnamefont {Yaish}}, \bibinfo {author} {\bibfnamefont {J.}~\bibnamefont {Petta}}, \bibinfo {author} {\bibfnamefont {M.}~\bibnamefont {Rinkoski}}, \bibinfo {author} {\bibfnamefont {J.}~\bibnamefont {Sethna}}, \bibinfo {author} {\bibfnamefont {H.}~\bibnamefont {Abru{\~n}a}}, \bibinfo {author} {\bibfnamefont {P.}~\bibnamefont {McEuen}},\ and\ \bibinfo {author} {\bibfnamefont {D.}~\bibnamefont {Ralph}},\ }\bibfield  {title} {\bibinfo {title} {Coulomb blockade and the kondo effect in single-atom transistors},\ }\href {https://doi.org/10.1038/nature00791} {\bibfield  {journal} {\bibinfo  {journal} {Nature}\ }\textbf {\bibinfo {volume} {417}},\ \bibinfo {pages} {722} (\bibinfo {year} {2002})}\BibitemShut
  {NoStop}%
\bibitem [{\citenamefont {Liang}\ \emph {et~al.}(2002)\citenamefont {Liang}, \citenamefont {Shores}, \citenamefont {Bockrath}, \citenamefont {Long},\ and\ \citenamefont {Park}}]{liang2002kondo}%
  \BibitemOpen
  \bibfield  {author} {\bibinfo {author} {\bibfnamefont {W.}~\bibnamefont {Liang}}, \bibinfo {author} {\bibfnamefont {M.}~\bibnamefont {Shores}}, \bibinfo {author} {\bibfnamefont {M.}~\bibnamefont {Bockrath}}, \bibinfo {author} {\bibfnamefont {J.}~\bibnamefont {Long}},\ and\ \bibinfo {author} {\bibfnamefont {H.}~\bibnamefont {Park}},\ }\bibfield  {title} {\bibinfo {title} {Kondo resonance in a single-molecule transistor},\ }\href {https://doi.org/10.1038/nature00790} {\bibfield  {journal} {\bibinfo  {journal} {Nature}\ }\textbf {\bibinfo {volume} {417}},\ \bibinfo {pages} {725} (\bibinfo {year} {2002})}\BibitemShut {NoStop}%
\bibitem [{\citenamefont {Keller}\ \emph {et~al.}(2014)\citenamefont {Keller}, \citenamefont {Amasha}, \citenamefont {Weymann}, \citenamefont {Moca}, \citenamefont {Rau}, \citenamefont {Katine}, \citenamefont {Shtrikman}, \citenamefont {Zar{\'a}nd},\ and\ \citenamefont {Goldhaber-Gordon}}]{keller2014emergent}%
  \BibitemOpen
  \bibfield  {author} {\bibinfo {author} {\bibfnamefont {A.}~\bibnamefont {Keller}}, \bibinfo {author} {\bibfnamefont {S.}~\bibnamefont {Amasha}}, \bibinfo {author} {\bibfnamefont {I.}~\bibnamefont {Weymann}}, \bibinfo {author} {\bibfnamefont {C.}~\bibnamefont {Moca}}, \bibinfo {author} {\bibfnamefont {I.}~\bibnamefont {Rau}}, \bibinfo {author} {\bibfnamefont {J.}~\bibnamefont {Katine}}, \bibinfo {author} {\bibfnamefont {D.}~\bibnamefont {Shtrikman}}, \bibinfo {author} {\bibfnamefont {G.}~\bibnamefont {Zar{\'a}nd}},\ and\ \bibinfo {author} {\bibfnamefont {D.}~\bibnamefont {Goldhaber-Gordon}},\ }\bibfield  {title} {\bibinfo {title} {Emergent su (4) kondo physics in a spin--charge-entangled double quantum dot},\ }\href {https://doi.org/10.1038/nphys2844} {\bibfield  {journal} {\bibinfo  {journal} {Nat.Phys.}\ }\textbf {\bibinfo {volume} {10}},\ \bibinfo {pages} {145} (\bibinfo {year} {2014})}\BibitemShut {NoStop}%
\bibitem [{\citenamefont {Piquard}\ \emph {et~al.}(2023)\citenamefont {Piquard}, \citenamefont {Glidic}, \citenamefont {Han}, \citenamefont {Aassime}, \citenamefont {Cavanna}, \citenamefont {Gennser}, \citenamefont {Meir}, \citenamefont {Sela}, \citenamefont {Anthore},\ and\ \citenamefont {Pierre}}]{piquard2023observing}%
  \BibitemOpen
  \bibfield  {author} {\bibinfo {author} {\bibfnamefont {C.}~\bibnamefont {Piquard}}, \bibinfo {author} {\bibfnamefont {P.}~\bibnamefont {Glidic}}, \bibinfo {author} {\bibfnamefont {C.}~\bibnamefont {Han}}, \bibinfo {author} {\bibfnamefont {A.}~\bibnamefont {Aassime}}, \bibinfo {author} {\bibfnamefont {A.}~\bibnamefont {Cavanna}}, \bibinfo {author} {\bibfnamefont {U.}~\bibnamefont {Gennser}}, \bibinfo {author} {\bibfnamefont {Y.}~\bibnamefont {Meir}}, \bibinfo {author} {\bibfnamefont {E.}~\bibnamefont {Sela}}, \bibinfo {author} {\bibfnamefont {A.}~\bibnamefont {Anthore}},\ and\ \bibinfo {author} {\bibfnamefont {F.}~\bibnamefont {Pierre}},\ }\bibfield  {title} {\bibinfo {title} {Observing the universal screening of a kondo impurity},\ }\href {https://doi.org/10.1038/s41467-023-42857-4} {\bibfield  {journal} {\bibinfo  {journal} {Nat. Commun.}\ }\textbf {\bibinfo {volume} {14}},\ \bibinfo {pages} {7263} (\bibinfo {year} {2023})}\BibitemShut {NoStop}%
\bibitem [{\citenamefont {Potok}\ \emph {et~al.}(2007)\citenamefont {Potok}, \citenamefont {Rau}, \citenamefont {Shtrikman}, \citenamefont {Oreg},\ and\ \citenamefont {Goldhaber-Gordon}}]{potok2007observation}%
  \BibitemOpen
  \bibfield  {author} {\bibinfo {author} {\bibfnamefont {R.}~\bibnamefont {Potok}}, \bibinfo {author} {\bibfnamefont {I.}~\bibnamefont {Rau}}, \bibinfo {author} {\bibfnamefont {H.}~\bibnamefont {Shtrikman}}, \bibinfo {author} {\bibfnamefont {Y.}~\bibnamefont {Oreg}},\ and\ \bibinfo {author} {\bibfnamefont {D.}~\bibnamefont {Goldhaber-Gordon}},\ }\bibfield  {title} {\bibinfo {title} {Observation of the two-channel kondo effect},\ }\href {https://doi.org/10.1038/nature05556} {\bibfield  {journal} {\bibinfo  {journal} {Nature}\ }\textbf {\bibinfo {volume} {446}},\ \bibinfo {pages} {167} (\bibinfo {year} {2007})}\BibitemShut {NoStop}%
\bibitem [{\citenamefont {Roch}\ \emph {et~al.}(2008)\citenamefont {Roch}, \citenamefont {Florens}, \citenamefont {Bouchiat}, \citenamefont {Wernsdorfer},\ and\ \citenamefont {Balestro}}]{roch2008quantum}%
  \BibitemOpen
  \bibfield  {author} {\bibinfo {author} {\bibfnamefont {N.}~\bibnamefont {Roch}}, \bibinfo {author} {\bibfnamefont {S.}~\bibnamefont {Florens}}, \bibinfo {author} {\bibfnamefont {B.}~\bibnamefont {Bouchiat}}, \bibinfo {author} {\bibfnamefont {W.}~\bibnamefont {Wernsdorfer}},\ and\ \bibinfo {author} {\bibfnamefont {F.}~\bibnamefont {Balestro}},\ }\bibfield  {title} {\bibinfo {title} {Quantum phase transition in a single-molecule quantum dot},\ }\href {https://doi.org/10.1038/nature06930} {\bibfield  {journal} {\bibinfo  {journal} {Nature}\ }\textbf {\bibinfo {volume} {453}},\ \bibinfo {pages} {633} (\bibinfo {year} {2008})}\BibitemShut {NoStop}%
\bibitem [{\citenamefont {Iftikhar}\ \emph {et~al.}(2018)\citenamefont {Iftikhar}, \citenamefont {Anthore}, \citenamefont {Mitchell}, \citenamefont {Parmentier}, \citenamefont {Gennser}, \citenamefont {Ouerghi}, \citenamefont {Cavanna}, \citenamefont {Mora}, \citenamefont {Simon},\ and\ \citenamefont {Pierre}}]{iftikhar2018tunable}%
  \BibitemOpen
  \bibfield  {author} {\bibinfo {author} {\bibfnamefont {Z.}~\bibnamefont {Iftikhar}}, \bibinfo {author} {\bibfnamefont {A.}~\bibnamefont {Anthore}}, \bibinfo {author} {\bibfnamefont {A.~K.}\ \bibnamefont {Mitchell}}, \bibinfo {author} {\bibfnamefont {F.~D.}\ \bibnamefont {Parmentier}}, \bibinfo {author} {\bibfnamefont {U.}~\bibnamefont {Gennser}}, \bibinfo {author} {\bibfnamefont {A.}~\bibnamefont {Ouerghi}}, \bibinfo {author} {\bibfnamefont {A.}~\bibnamefont {Cavanna}}, \bibinfo {author} {\bibfnamefont {C.}~\bibnamefont {Mora}}, \bibinfo {author} {\bibfnamefont {P.}~\bibnamefont {Simon}},\ and\ \bibinfo {author} {\bibfnamefont {F.}~\bibnamefont {Pierre}},\ }\bibfield  {title} {\bibinfo {title} {Tunable quantum criticality and super-ballistic transport in a “charge” {K}ondo circuit},\ }\href {https://doi.org/10.1126/science.aan5592} {\bibfield  {journal} {\bibinfo  {journal} {Science}\ }\textbf {\bibinfo {volume} {360}},\ \bibinfo {pages} {1315} (\bibinfo {year} {2018})}\BibitemShut {NoStop}%
\bibitem [{\citenamefont {Pouse}\ \emph {et~al.}(2023)\citenamefont {Pouse}, \citenamefont {Peeters}, \citenamefont {Hsueh}, \citenamefont {Gennser}, \citenamefont {Cavanna}, \citenamefont {Kastner}, \citenamefont {Mitchell},\ and\ \citenamefont {Goldhaber-Gordon}}]{pouse2023quantum}%
  \BibitemOpen
  \bibfield  {author} {\bibinfo {author} {\bibfnamefont {W.}~\bibnamefont {Pouse}}, \bibinfo {author} {\bibfnamefont {L.}~\bibnamefont {Peeters}}, \bibinfo {author} {\bibfnamefont {C.~L.}\ \bibnamefont {Hsueh}}, \bibinfo {author} {\bibfnamefont {U.}~\bibnamefont {Gennser}}, \bibinfo {author} {\bibfnamefont {A.}~\bibnamefont {Cavanna}}, \bibinfo {author} {\bibfnamefont {M.~A.}\ \bibnamefont {Kastner}}, \bibinfo {author} {\bibfnamefont {A.~K.}\ \bibnamefont {Mitchell}},\ and\ \bibinfo {author} {\bibfnamefont {D.}~\bibnamefont {Goldhaber-Gordon}},\ }\bibfield  {title} {\bibinfo {title} {Quantum simulation of an exotic quantum critical point in a two-site charge kondo circuit},\ }\href {https://doi.org/10.1038/s41567-022-01905-4} {\bibfield  {journal} {\bibinfo  {journal} {Nat. Phys.}\ }\textbf {\bibinfo {volume} {19}},\ \bibinfo {pages} {492} (\bibinfo {year} {2023})}\BibitemShut {NoStop}%
\bibitem [{\citenamefont {Karki}\ \emph {et~al.}(2023)\citenamefont {Karki}, \citenamefont {Boulat}, \citenamefont {Pouse}, \citenamefont {Goldhaber-Gordon}, \citenamefont {Mitchell},\ and\ \citenamefont {Mora}}]{karki2023z}%
  \BibitemOpen
  \bibfield  {author} {\bibinfo {author} {\bibfnamefont {D.~B.}\ \bibnamefont {Karki}}, \bibinfo {author} {\bibfnamefont {E.}~\bibnamefont {Boulat}}, \bibinfo {author} {\bibfnamefont {W.}~\bibnamefont {Pouse}}, \bibinfo {author} {\bibfnamefont {D.}~\bibnamefont {Goldhaber-Gordon}}, \bibinfo {author} {\bibfnamefont {A.~K.}\ \bibnamefont {Mitchell}},\ and\ \bibinfo {author} {\bibfnamefont {C.}~\bibnamefont {Mora}},\ }\bibfield  {title} {\bibinfo {title} {${\mathbb{z}}_{3}$ parafermion in the double charge kondo model},\ }\href {https://doi.org/10.1103/PhysRevLett.130.146201} {\bibfield  {journal} {\bibinfo  {journal} {Phys. Rev. Lett.}\ }\textbf {\bibinfo {volume} {130}},\ \bibinfo {pages} {146201} (\bibinfo {year} {2023})}\BibitemShut {NoStop}%
\bibitem [{\citenamefont {Chanrion}\ \emph {et~al.}(2020)\citenamefont {Chanrion}, \citenamefont {Niegemann}, \citenamefont {Bertrand}, \citenamefont {Spence}, \citenamefont {Jadot}, \citenamefont {Li}, \citenamefont {Mortemousque}, \citenamefont {Hutin}, \citenamefont {Maurand}, \citenamefont {Jehl}, \citenamefont {Sanquer}, \citenamefont {De~Franceschi}, \citenamefont {B\"auerle}, \citenamefont {Balestro}, \citenamefont {Niquet}, \citenamefont {Vinet}, \citenamefont {Meunier},\ and\ \citenamefont {Urdampilleta}}]{chanrion2020charge}%
  \BibitemOpen
  \bibfield  {author} {\bibinfo {author} {\bibfnamefont {E.}~\bibnamefont {Chanrion}}, \bibinfo {author} {\bibfnamefont {D.~J.}\ \bibnamefont {Niegemann}}, \bibinfo {author} {\bibfnamefont {B.}~\bibnamefont {Bertrand}}, \bibinfo {author} {\bibfnamefont {C.}~\bibnamefont {Spence}}, \bibinfo {author} {\bibfnamefont {B.}~\bibnamefont {Jadot}}, \bibinfo {author} {\bibfnamefont {J.}~\bibnamefont {Li}}, \bibinfo {author} {\bibfnamefont {P.}~\bibnamefont {Mortemousque}}, \bibinfo {author} {\bibfnamefont {L.}~\bibnamefont {Hutin}}, \bibinfo {author} {\bibfnamefont {R.}~\bibnamefont {Maurand}}, \bibinfo {author} {\bibfnamefont {X.}~\bibnamefont {Jehl}}, \bibinfo {author} {\bibfnamefont {M.}~\bibnamefont {Sanquer}}, \bibinfo {author} {\bibfnamefont {S.}~\bibnamefont {De~Franceschi}}, \bibinfo {author} {\bibfnamefont {C.}~\bibnamefont {B\"auerle}}, \bibinfo {author} {\bibfnamefont {F.}~\bibnamefont {Balestro}}, \bibinfo {author} {\bibfnamefont {Y.}~\bibnamefont {Niquet}}, \bibinfo {author} {\bibfnamefont
  {M.}~\bibnamefont {Vinet}}, \bibinfo {author} {\bibfnamefont {T.}~\bibnamefont {Meunier}},\ and\ \bibinfo {author} {\bibfnamefont {M.}~\bibnamefont {Urdampilleta}},\ }\bibfield  {title} {\bibinfo {title} {Charge detection in an array of cmos quantum dots},\ }\href {https://doi.org/10.1103/PhysRevApplied.14.024066} {\bibfield  {journal} {\bibinfo  {journal} {Phys. Rev. Appl.}\ }\textbf {\bibinfo {volume} {14}},\ \bibinfo {pages} {024066} (\bibinfo {year} {2020})}\BibitemShut {NoStop}%
\bibitem [{\citenamefont {Xue}\ \emph {et~al.}(2021)\citenamefont {Xue}, \citenamefont {Patra}, \citenamefont {van Dijk}, \citenamefont {Samkharadze}, \citenamefont {Subramanian}, \citenamefont {Corna}, \citenamefont {Paquelet~Wuetz}, \citenamefont {Jeon}, \citenamefont {Sheikh}, \citenamefont {Juarez-Hernandez}, \citenamefont {Esparza}, \citenamefont {Rampurawala}, \citenamefont {Carlton}, \citenamefont {Ravikumar}, \citenamefont {Nieva}, \citenamefont {Kim}, \citenamefont {Lee}, \citenamefont {Sammak}, \citenamefont {Scappucci}, \citenamefont {Veldhorst}, \citenamefont {Sebastiano}, \citenamefont {Babaie}, \citenamefont {Pellerano}, \citenamefont {Charbon},\ and\ \citenamefont {Vandersypen}}]{xue2021cmos}%
  \BibitemOpen
  \bibfield  {author} {\bibinfo {author} {\bibfnamefont {X.}~\bibnamefont {Xue}}, \bibinfo {author} {\bibfnamefont {B.}~\bibnamefont {Patra}}, \bibinfo {author} {\bibfnamefont {J.~P.~G.}\ \bibnamefont {van Dijk}}, \bibinfo {author} {\bibfnamefont {N.}~\bibnamefont {Samkharadze}}, \bibinfo {author} {\bibfnamefont {S.}~\bibnamefont {Subramanian}}, \bibinfo {author} {\bibfnamefont {A.}~\bibnamefont {Corna}}, \bibinfo {author} {\bibfnamefont {B.}~\bibnamefont {Paquelet~Wuetz}}, \bibinfo {author} {\bibfnamefont {C.}~\bibnamefont {Jeon}}, \bibinfo {author} {\bibfnamefont {F.}~\bibnamefont {Sheikh}}, \bibinfo {author} {\bibfnamefont {E.}~\bibnamefont {Juarez-Hernandez}}, \bibinfo {author} {\bibfnamefont {B.}~\bibnamefont {Esparza}}, \bibinfo {author} {\bibfnamefont {H.}~\bibnamefont {Rampurawala}}, \bibinfo {author} {\bibfnamefont {B.}~\bibnamefont {Carlton}}, \bibinfo {author} {\bibfnamefont {S.}~\bibnamefont {Ravikumar}}, \bibinfo {author} {\bibfnamefont {C.}~\bibnamefont {Nieva}}, \bibinfo {author} {\bibfnamefont
  {S.}~\bibnamefont {Kim}}, \bibinfo {author} {\bibfnamefont {H.}~\bibnamefont {Lee}}, \bibinfo {author} {\bibfnamefont {A.}~\bibnamefont {Sammak}}, \bibinfo {author} {\bibfnamefont {G.}~\bibnamefont {Scappucci}}, \bibinfo {author} {\bibfnamefont {M.}~\bibnamefont {Veldhorst}}, \bibinfo {author} {\bibfnamefont {F.}~\bibnamefont {Sebastiano}}, \bibinfo {author} {\bibfnamefont {M.}~\bibnamefont {Babaie}}, \bibinfo {author} {\bibfnamefont {S.}~\bibnamefont {Pellerano}}, \bibinfo {author} {\bibfnamefont {E.}~\bibnamefont {Charbon}},\ and\ \bibinfo {author} {\bibfnamefont {L.~M.~K.}\ \bibnamefont {Vandersypen}},\ }\bibfield  {title} {\bibinfo {title} {Cmos-based cryogenic control of silicon quantum circuits},\ }\href {https://doi.org/10.1038/s41586-021-03469-4} {\bibfield  {journal} {\bibinfo  {journal} {Nature}\ }\textbf {\bibinfo {volume} {593}},\ \bibinfo {pages} {205} (\bibinfo {year} {2021})}\BibitemShut {NoStop}%
\bibitem [{\citenamefont {Ruffino}\ \emph {et~al.}(2022)\citenamefont {Ruffino}, \citenamefont {Yang}, \citenamefont {Michniewicz}, \citenamefont {Peng}, \citenamefont {Charbon},\ and\ \citenamefont {Gonzalez-Zalba}}]{ruffino2022cryo}%
  \BibitemOpen
  \bibfield  {author} {\bibinfo {author} {\bibfnamefont {A.}~\bibnamefont {Ruffino}}, \bibinfo {author} {\bibfnamefont {T.}~\bibnamefont {Yang}}, \bibinfo {author} {\bibfnamefont {J.}~\bibnamefont {Michniewicz}}, \bibinfo {author} {\bibfnamefont {Y.}~\bibnamefont {Peng}}, \bibinfo {author} {\bibfnamefont {E.}~\bibnamefont {Charbon}},\ and\ \bibinfo {author} {\bibfnamefont {M.}~\bibnamefont {Gonzalez-Zalba}},\ }\bibfield  {title} {\bibinfo {title} {A cryo-cmos chip that integrates silicon quantum dots and multiplexed dispersive readout electronics},\ }\href {https://doi.org/10.1038/s41928-021-00687-6} {\bibfield  {journal} {\bibinfo  {journal} {Nature Electronics}\ }\textbf {\bibinfo {volume} {5}},\ \bibinfo {pages} {53} (\bibinfo {year} {2022})}\BibitemShut {NoStop}%
\bibitem [{\citenamefont {Petropoulos}\ \emph {et~al.}(2024)\citenamefont {Petropoulos}, \citenamefont {Wu}, \citenamefont {Sokolov}, \citenamefont {Giounanlis}, \citenamefont {Bashir}, \citenamefont {Mitchell}, \citenamefont {Asker}, \citenamefont {Leipold}, \citenamefont {Staszewski},\ and\ \citenamefont {Blokhina}}]{petropoulos2024nanoscale}%
  \BibitemOpen
  \bibfield  {author} {\bibinfo {author} {\bibfnamefont {N.}~\bibnamefont {Petropoulos}}, \bibinfo {author} {\bibfnamefont {X.}~\bibnamefont {Wu}}, \bibinfo {author} {\bibfnamefont {A.}~\bibnamefont {Sokolov}}, \bibinfo {author} {\bibfnamefont {P.}~\bibnamefont {Giounanlis}}, \bibinfo {author} {\bibfnamefont {I.}~\bibnamefont {Bashir}}, \bibinfo {author} {\bibfnamefont {A.~K.}\ \bibnamefont {Mitchell}}, \bibinfo {author} {\bibfnamefont {M.}~\bibnamefont {Asker}}, \bibinfo {author} {\bibfnamefont {D.}~\bibnamefont {Leipold}}, \bibinfo {author} {\bibfnamefont {R.~B.}\ \bibnamefont {Staszewski}},\ and\ \bibinfo {author} {\bibfnamefont {E.}~\bibnamefont {Blokhina}},\ }\bibfield  {title} {\bibinfo {title} {{Nanoscale single-electron box with a floating lead for quantum sensing: Modeling and device characterization}},\ }\href {https://doi.org/10.1063/5.0203421} {\bibfield  {journal} {\bibinfo  {journal} {Appl. Phys. Lett.}\ }\textbf {\bibinfo {volume} {124}},\ \bibinfo {pages} {173503} (\bibinfo {year}
  {2024})}\BibitemShut {NoStop}%
\bibitem [{\citenamefont {Braunstein}\ and\ \citenamefont {Caves}(1994)}]{braunstein1994statistical}%
  \BibitemOpen
  \bibfield  {author} {\bibinfo {author} {\bibfnamefont {S.~L.}\ \bibnamefont {Braunstein}}\ and\ \bibinfo {author} {\bibfnamefont {C.~M.}\ \bibnamefont {Caves}},\ }\bibfield  {title} {\bibinfo {title} {Statistical distance and the geometry of quantum states},\ }\href {https://doi.org/10.1103/PhysRevLett.72.3439} {\bibfield  {journal} {\bibinfo  {journal} {Phys. Rev. Lett.}\ }\textbf {\bibinfo {volume} {72}},\ \bibinfo {pages} {3439} (\bibinfo {year} {1994})}\BibitemShut {NoStop}%
\bibitem [{sm()}]{sm}%
  \BibitemOpen
  \href@noop {} {\bibinfo {title} {{S}upplementary {M}aterial}}\BibitemShut {NoStop}%
\bibitem [{\citenamefont {Berry}(2009)}]{berry2009transitionless}%
  \BibitemOpen
  \bibfield  {author} {\bibinfo {author} {\bibfnamefont {M.}~\bibnamefont {Berry}},\ }\bibfield  {title} {\bibinfo {title} {Transitionless quantum driving},\ }\href {https://doi.org/10.1088/1751-8113/42/36/365303} {\bibfield  {journal} {\bibinfo  {journal} {J. Phys. A: Math. Theor.}\ }\textbf {\bibinfo {volume} {42}},\ \bibinfo {pages} {365303} (\bibinfo {year} {2009})}\BibitemShut {NoStop}%
\bibitem [{\citenamefont {Kolodrubetz}\ \emph {et~al.}(2017)\citenamefont {Kolodrubetz}, \citenamefont {Sels}, \citenamefont {Mehta},\ and\ \citenamefont {Polkovnikov}}]{kolodrubetz2017geometry}%
  \BibitemOpen
  \bibfield  {author} {\bibinfo {author} {\bibfnamefont {M.}~\bibnamefont {Kolodrubetz}}, \bibinfo {author} {\bibfnamefont {D.}~\bibnamefont {Sels}}, \bibinfo {author} {\bibfnamefont {P.}~\bibnamefont {Mehta}},\ and\ \bibinfo {author} {\bibfnamefont {A.}~\bibnamefont {Polkovnikov}},\ }\bibfield  {title} {\bibinfo {title} {Geometry and non-adiabatic response in quantum and classical systems},\ }\href {https://doi.org/https://doi.org/10.1016/j.physrep.2017.07.001} {\bibfield  {journal} {\bibinfo  {journal} {Phys. Rep.}\ }\textbf {\bibinfo {volume} {697}},\ \bibinfo {pages} {1} (\bibinfo {year} {2017})}\BibitemShut {NoStop}%
\bibitem [{\citenamefont {Hauke}\ \emph {et~al.}(2016)\citenamefont {Hauke}, \citenamefont {Heyl}, \citenamefont {Tagliacozzo},\ and\ \citenamefont {Zoller}}]{zoller}%
  \BibitemOpen
  \bibfield  {author} {\bibinfo {author} {\bibfnamefont {P.}~\bibnamefont {Hauke}}, \bibinfo {author} {\bibfnamefont {M.}~\bibnamefont {Heyl}}, \bibinfo {author} {\bibfnamefont {L.}~\bibnamefont {Tagliacozzo}},\ and\ \bibinfo {author} {\bibfnamefont {P.}~\bibnamefont {Zoller}},\ }\bibfield  {title} {\bibinfo {title} {Measuring multipartite entanglement through dynamic susceptibilities},\ }\href {https://doi.org/10.1038/nphys3700} {\bibfield  {journal} {\bibinfo  {journal} {Nat. Phys.}\ }\textbf {\bibinfo {volume} {12}},\ \bibinfo {pages} {778} (\bibinfo {year} {2016})}\BibitemShut {NoStop}%
\bibitem [{\citenamefont {Kubo}(1957)}]{kubo1957statistical}%
  \BibitemOpen
  \bibfield  {author} {\bibinfo {author} {\bibfnamefont {R.}~\bibnamefont {Kubo}},\ }\bibfield  {title} {\bibinfo {title} {Statistical-mechanical theory of irreversible processes. i. general theory and simple applications to magnetic and conduction problems},\ }\href {https://doi.org/10.1143/JPSJ.12.570} {\bibfield  {journal} {\bibinfo  {journal} {J. Phys. Soc. Jpn.}\ }\textbf {\bibinfo {volume} {12}},\ \bibinfo {pages} {570} (\bibinfo {year} {1957})}\BibitemShut {NoStop}%
\bibitem [{\citenamefont {Minarelli}\ \emph {et~al.}(2022)\citenamefont {Minarelli}, \citenamefont {Rigo},\ and\ \citenamefont {Mitchell}}]{minarelli2022linear}%
  \BibitemOpen
  \bibfield  {author} {\bibinfo {author} {\bibfnamefont {E.~L.}\ \bibnamefont {Minarelli}}, \bibinfo {author} {\bibfnamefont {J.~B.}\ \bibnamefont {Rigo}},\ and\ \bibinfo {author} {\bibfnamefont {A.~K.}\ \bibnamefont {Mitchell}},\ }\href {https://arxiv.org/abs/2209.01208} {\bibinfo {title} {Linear response quantum transport through interacting multi-orbital nanostructures}} (\bibinfo {year} {2022}),\ \Eprint {https://arxiv.org/abs/2209.01208} {2209.01208} \BibitemShut {NoStop}%
\bibitem [{\citenamefont {Hewson}(1993)}]{hewson1997Kondo}%
  \BibitemOpen
  \bibfield  {author} {\bibinfo {author} {\bibfnamefont {A.}~\bibnamefont {Hewson}},\ }\href {https://doi.org/10.1017/CBO9780511470752} {\emph {\bibinfo {title} {The {K}ondo problem to heavy fermions}}}\ (\bibinfo  {publisher} {Cambridge Studies in Magnetism, CUP},\ \bibinfo {year} {1993})\BibitemShut {NoStop}%
\bibitem [{\citenamefont {Pustilnik}\ and\ \citenamefont {Glazman}(2004)}]{pustilnik2004kondo}%
  \BibitemOpen
  \bibfield  {author} {\bibinfo {author} {\bibfnamefont {M.}~\bibnamefont {Pustilnik}}\ and\ \bibinfo {author} {\bibfnamefont {L.}~\bibnamefont {Glazman}},\ }\bibfield  {title} {\bibinfo {title} {Kondo effect in quantum dots},\ }\href {https://doi.org/10.1088/0953-8984/16/16/R01} {\bibfield  {journal} {\bibinfo  {journal} {J. Phys. Condens. Matter}\ }\textbf {\bibinfo {volume} {16}},\ \bibinfo {pages} {R513} (\bibinfo {year} {2004})}\BibitemShut {NoStop}%
\bibitem [{\citenamefont {Wilson}(1975)}]{wilson1975renormalization}%
  \BibitemOpen
  \bibfield  {author} {\bibinfo {author} {\bibfnamefont {K.~G.}\ \bibnamefont {Wilson}},\ }\bibfield  {title} {\bibinfo {title} {The renormalization group: Critical phenomena and the {K}ondo problem},\ }\href@noop {} {\bibfield  {journal} {\bibinfo  {journal} {Reviews of {M}odern {P}hysics}\ }\textbf {\bibinfo {volume} {47}},\ \bibinfo {pages} {773} (\bibinfo {year} {1975})}\BibitemShut {NoStop}%
\bibitem [{\citenamefont {Bulla}\ \emph {et~al.}(2008)\citenamefont {Bulla}, \citenamefont {Costi},\ and\ \citenamefont {Pruschke}}]{bulla2008numerical}%
  \BibitemOpen
  \bibfield  {author} {\bibinfo {author} {\bibfnamefont {R.}~\bibnamefont {Bulla}}, \bibinfo {author} {\bibfnamefont {T.~A.}\ \bibnamefont {Costi}},\ and\ \bibinfo {author} {\bibfnamefont {T.}~\bibnamefont {Pruschke}},\ }\bibfield  {title} {\bibinfo {title} {Numerical renormalization group method for quantum impurity systems},\ }\href {https://doi.org/10.1103/RevModPhys.80.395} {\bibfield  {journal} {\bibinfo  {journal} {Rev. Mod. Phys.}\ }\textbf {\bibinfo {volume} {80}},\ \bibinfo {pages} {395} (\bibinfo {year} {2008})}\BibitemShut {NoStop}%
\bibitem [{\citenamefont {Gietka}\ \emph {et~al.}(2021)\citenamefont {Gietka}, \citenamefont {Metz}, \citenamefont {Keller},\ and\ \citenamefont {Li}}]{gietka2021adiabatic}%
  \BibitemOpen
  \bibfield  {author} {\bibinfo {author} {\bibfnamefont {K.}~\bibnamefont {Gietka}}, \bibinfo {author} {\bibfnamefont {F.}~\bibnamefont {Metz}}, \bibinfo {author} {\bibfnamefont {T.}~\bibnamefont {Keller}},\ and\ \bibinfo {author} {\bibfnamefont {J.}~\bibnamefont {Li}},\ }\bibfield  {title} {\bibinfo {title} {Adiabatic critical quantum metrology cannot reach the {H}eisenberg limit even when shortcuts to adiabaticity are applied},\ }\href {https://doi.org/10.22331/q-2021-07-01-489} {\bibfield  {journal} {\bibinfo  {journal} {{Quantum}}\ }\textbf {\bibinfo {volume} {5}},\ \bibinfo {pages} {489} (\bibinfo {year} {2021})}\BibitemShut {NoStop}%
\bibitem [{\citenamefont {Demkowicz-Dobrza{\'n}ski}\ \emph {et~al.}(2012)\citenamefont {Demkowicz-Dobrza{\'n}ski}, \citenamefont {Ko{\l}ody{\'n}ski},\ and\ \citenamefont {Gu{\c{t}}{\u{a}}}}]{demkowicz2012elusive}%
  \BibitemOpen
  \bibfield  {author} {\bibinfo {author} {\bibfnamefont {R.}~\bibnamefont {Demkowicz-Dobrza{\'n}ski}}, \bibinfo {author} {\bibfnamefont {J.}~\bibnamefont {Ko{\l}ody{\'n}ski}},\ and\ \bibinfo {author} {\bibfnamefont {M.}~\bibnamefont {Gu{\c{t}}{\u{a}}}},\ }\bibfield  {title} {\bibinfo {title} {The elusive heisenberg limit in quantum-enhanced metrology},\ }\href {https://doi.org/10.1038/ncomms2067} {\bibfield  {journal} {\bibinfo  {journal} {Nat. Commun.}\ }\textbf {\bibinfo {volume} {3}},\ \bibinfo {pages} {1063} (\bibinfo {year} {2012})}\BibitemShut {NoStop}%
\bibitem [{\citenamefont {Zwierz}\ \emph {et~al.}(2012)\citenamefont {Zwierz}, \citenamefont {P\'erez-Delgado},\ and\ \citenamefont {Kok}}]{zwierz2012ultimate}%
  \BibitemOpen
  \bibfield  {author} {\bibinfo {author} {\bibfnamefont {M.}~\bibnamefont {Zwierz}}, \bibinfo {author} {\bibfnamefont {C.~A.}\ \bibnamefont {P\'erez-Delgado}},\ and\ \bibinfo {author} {\bibfnamefont {P.}~\bibnamefont {Kok}},\ }\bibfield  {title} {\bibinfo {title} {Ultimate limits to quantum metrology and the meaning of the heisenberg limit},\ }\href {https://doi.org/10.1103/PhysRevA.85.042112} {\bibfield  {journal} {\bibinfo  {journal} {Phys. Rev. A}\ }\textbf {\bibinfo {volume} {85}},\ \bibinfo {pages} {042112} (\bibinfo {year} {2012})}\BibitemShut {NoStop}%
\bibitem [{\citenamefont {Zwierz}\ \emph {et~al.}(2010)\citenamefont {Zwierz}, \citenamefont {P\'erez-Delgado},\ and\ \citenamefont {Kok}}]{zwierz2010general}%
  \BibitemOpen
  \bibfield  {author} {\bibinfo {author} {\bibfnamefont {M.}~\bibnamefont {Zwierz}}, \bibinfo {author} {\bibfnamefont {C.~A.}\ \bibnamefont {P\'erez-Delgado}},\ and\ \bibinfo {author} {\bibfnamefont {P.}~\bibnamefont {Kok}},\ }\bibfield  {title} {\bibinfo {title} {General optimality of the heisenberg limit for quantum metrology},\ }\href {https://doi.org/10.1103/PhysRevLett.105.180402} {\bibfield  {journal} {\bibinfo  {journal} {Phys. Rev. Lett.}\ }\textbf {\bibinfo {volume} {105}},\ \bibinfo {pages} {180402} (\bibinfo {year} {2010})}\BibitemShut {NoStop}%
\bibitem [{\citenamefont {Rams}\ \emph {et~al.}(2018)\citenamefont {Rams}, \citenamefont {Sierant}, \citenamefont {Dutta}, \citenamefont {Horodecki},\ and\ \citenamefont {Zakrzewski}}]{rams2018limits}%
  \BibitemOpen
  \bibfield  {author} {\bibinfo {author} {\bibfnamefont {M.~M.}\ \bibnamefont {Rams}}, \bibinfo {author} {\bibfnamefont {P.}~\bibnamefont {Sierant}}, \bibinfo {author} {\bibfnamefont {O.}~\bibnamefont {Dutta}}, \bibinfo {author} {\bibfnamefont {P.}~\bibnamefont {Horodecki}},\ and\ \bibinfo {author} {\bibfnamefont {J.}~\bibnamefont {Zakrzewski}},\ }\bibfield  {title} {\bibinfo {title} {At the limits of criticality-based quantum metrology: Apparent super-heisenberg scaling revisited},\ }\href {https://doi.org/10.1103/PhysRevX.8.021022} {\bibfield  {journal} {\bibinfo  {journal} {Phys. Rev. X}\ }\textbf {\bibinfo {volume} {8}},\ \bibinfo {pages} {021022} (\bibinfo {year} {2018})}\BibitemShut {NoStop}%
\bibitem [{\citenamefont {Weichselbaum}\ and\ \citenamefont {von Delft}(2007)}]{weichselbaum2007sum}%
  \BibitemOpen
  \bibfield  {author} {\bibinfo {author} {\bibfnamefont {A.}~\bibnamefont {Weichselbaum}}\ and\ \bibinfo {author} {\bibfnamefont {J.}~\bibnamefont {von Delft}},\ }\bibfield  {title} {\bibinfo {title} {Sum-rule conserving spectral functions from the numerical renormalization group},\ }\href {https://doi.org/10.1103/PhysRevLett.99.076402} {\bibfield  {journal} {\bibinfo  {journal} {Phys. Rev. Lett.}\ }\textbf {\bibinfo {volume} {99}},\ \bibinfo {pages} {076402} (\bibinfo {year} {2007})}\BibitemShut {NoStop}%
\bibitem [{\citenamefont {Boixo}\ \emph {et~al.}(2007)\citenamefont {Boixo}, \citenamefont {Flammia}, \citenamefont {Caves},\ and\ \citenamefont {Geremia}}]{boixo2007generalized}%
  \BibitemOpen
  \bibfield  {author} {\bibinfo {author} {\bibfnamefont {S.}~\bibnamefont {Boixo}}, \bibinfo {author} {\bibfnamefont {S.~T.}\ \bibnamefont {Flammia}}, \bibinfo {author} {\bibfnamefont {C.~M.}\ \bibnamefont {Caves}},\ and\ \bibinfo {author} {\bibfnamefont {J.~M.}\ \bibnamefont {Geremia}},\ }\bibfield  {title} {\bibinfo {title} {Generalized limits for single-parameter quantum estimation},\ }\href {https://doi.org/10.1103/PhysRevLett.98.090401} {\bibfield  {journal} {\bibinfo  {journal} {Physical {R}eview {L}etters}\ }\textbf {\bibinfo {volume} {98}},\ \bibinfo {pages} {090401} (\bibinfo {year} {2007})}\BibitemShut {NoStop}%
\bibitem [{\citenamefont {Ding}\ \emph {et~al.}(2022)\citenamefont {Ding}, \citenamefont {Liu}, \citenamefont {Shi}, \citenamefont {Guo}, \citenamefont {M{\o}lmer},\ and\ \citenamefont {Adams}}]{ding2022enhanced}%
  \BibitemOpen
  \bibfield  {author} {\bibinfo {author} {\bibfnamefont {D.}~\bibnamefont {Ding}}, \bibinfo {author} {\bibfnamefont {Z.}~\bibnamefont {Liu}}, \bibinfo {author} {\bibfnamefont {B.}~\bibnamefont {Shi}}, \bibinfo {author} {\bibfnamefont {G.}~\bibnamefont {Guo}}, \bibinfo {author} {\bibfnamefont {K.}~\bibnamefont {M{\o}lmer}},\ and\ \bibinfo {author} {\bibfnamefont {C.}~\bibnamefont {Adams}},\ }\bibfield  {title} {\bibinfo {title} {Enhanced metrology at the critical point of a many-body rydberg atomic system},\ }\href {https://doi.org/10.1038/s41567-022-01777-8} {\bibfield  {journal} {\bibinfo  {journal} {Nat. Phys.}\ }\textbf {\bibinfo {volume} {18}},\ \bibinfo {pages} {1447} (\bibinfo {year} {2022})}\BibitemShut {NoStop}%
\bibitem [{\citenamefont {B\"uttiker}\ \emph {et~al.}(1985)\citenamefont {B\"uttiker}, \citenamefont {Imry}, \citenamefont {Landauer},\ and\ \citenamefont {Pinhas}}]{buttiker1985generalized}%
  \BibitemOpen
  \bibfield  {author} {\bibinfo {author} {\bibfnamefont {M.}~\bibnamefont {B\"uttiker}}, \bibinfo {author} {\bibfnamefont {Y.}~\bibnamefont {Imry}}, \bibinfo {author} {\bibfnamefont {R.}~\bibnamefont {Landauer}},\ and\ \bibinfo {author} {\bibfnamefont {S.}~\bibnamefont {Pinhas}},\ }\bibfield  {title} {\bibinfo {title} {Generalized many-channel conductance formula with application to small rings},\ }\href {https://doi.org/10.1103/PhysRevB.31.6207} {\bibfield  {journal} {\bibinfo  {journal} {Phys. Rev. B}\ }\textbf {\bibinfo {volume} {31}},\ \bibinfo {pages} {6207} (\bibinfo {year} {1985})}\BibitemShut {NoStop}%
\bibitem [{\citenamefont {Lesovik}(1989)}]{lesovik}%
  \BibitemOpen
  \bibfield  {author} {\bibinfo {author} {\bibfnamefont {G.~B.}\ \bibnamefont {Lesovik}},\ }\bibfield  {title} {\bibinfo {title} {Excess quantum noise in 2d ballistic point contacts},\ }\href {http://jetpletters.ru/ps/0/article_16970.shtml} {\bibfield  {journal} {\bibinfo  {journal} {{JETP} {L}etters}\ }\textbf {\bibinfo {volume} {49}},\ \bibinfo {pages} {513} (\bibinfo {year} {1989})}\BibitemShut {NoStop}%
\bibitem [{\citenamefont {Meir}\ and\ \citenamefont {Wingreen}(1992)}]{meir1992landauer}%
  \BibitemOpen
  \bibfield  {author} {\bibinfo {author} {\bibfnamefont {Y.}~\bibnamefont {Meir}}\ and\ \bibinfo {author} {\bibfnamefont {N.~S.}\ \bibnamefont {Wingreen}},\ }\bibfield  {title} {\bibinfo {title} {Landauer formula for the current through an interacting electron region},\ }\href {https://doi.org/10.1103/PhysRevLett.68.2512} {\bibfield  {journal} {\bibinfo  {journal} {Phys. Rev. Lett.}\ }\textbf {\bibinfo {volume} {68}},\ \bibinfo {pages} {2512} (\bibinfo {year} {1992})}\BibitemShut {NoStop}%
\bibitem [{\citenamefont {Hartman}\ \emph {et~al.}(2018)\citenamefont {Hartman}, \citenamefont {Olsen}, \citenamefont {L{\"u}scher}, \citenamefont {Samani}, \citenamefont {Fallahi}, \citenamefont {Gardner}, \citenamefont {Manfra},\ and\ \citenamefont {Folk}}]{hartman2018direct}%
  \BibitemOpen
  \bibfield  {author} {\bibinfo {author} {\bibfnamefont {N.}~\bibnamefont {Hartman}}, \bibinfo {author} {\bibfnamefont {C.}~\bibnamefont {Olsen}}, \bibinfo {author} {\bibfnamefont {S.}~\bibnamefont {L{\"u}scher}}, \bibinfo {author} {\bibfnamefont {M.}~\bibnamefont {Samani}}, \bibinfo {author} {\bibfnamefont {S.}~\bibnamefont {Fallahi}}, \bibinfo {author} {\bibfnamefont {G.}~\bibnamefont {Gardner}}, \bibinfo {author} {\bibfnamefont {M.}~\bibnamefont {Manfra}},\ and\ \bibinfo {author} {\bibfnamefont {J.}~\bibnamefont {Folk}},\ }\bibfield  {title} {\bibinfo {title} {Direct entropy measurement in a mesoscopic quantum system},\ }\href {https://doi.org/10.1038/s41567-018-0250-5} {\bibfield  {journal} {\bibinfo  {journal} {Nat. Phys.}\ }\textbf {\bibinfo {volume} {14}},\ \bibinfo {pages} {1083} (\bibinfo {year} {2018})}\BibitemShut {NoStop}%
\bibitem [{\citenamefont {Han}\ \emph {et~al.}(2022)\citenamefont {Han}, \citenamefont {Iftikhar}, \citenamefont {Kleeorin}, \citenamefont {Anthore}, \citenamefont {Pierre}, \citenamefont {Meir}, \citenamefont {Mitchell},\ and\ \citenamefont {Sela}}]{han2022fractional}%
  \BibitemOpen
  \bibfield  {author} {\bibinfo {author} {\bibfnamefont {C.}~\bibnamefont {Han}}, \bibinfo {author} {\bibfnamefont {Z.}~\bibnamefont {Iftikhar}}, \bibinfo {author} {\bibfnamefont {Y.}~\bibnamefont {Kleeorin}}, \bibinfo {author} {\bibfnamefont {A.}~\bibnamefont {Anthore}}, \bibinfo {author} {\bibfnamefont {F.}~\bibnamefont {Pierre}}, \bibinfo {author} {\bibfnamefont {Y.}~\bibnamefont {Meir}}, \bibinfo {author} {\bibfnamefont {A.~K.}\ \bibnamefont {Mitchell}},\ and\ \bibinfo {author} {\bibfnamefont {E.}~\bibnamefont {Sela}},\ }\bibfield  {title} {\bibinfo {title} {Fractional entropy of multichannel kondo systems from conductance-charge relations},\ }\href {https://doi.org/10.1103/PhysRevLett.128.146803} {\bibfield  {journal} {\bibinfo  {journal} {Phys. Rev. Lett.}\ }\textbf {\bibinfo {volume} {128}},\ \bibinfo {pages} {146803} (\bibinfo {year} {2022})}\BibitemShut {NoStop}%
\bibitem [{\citenamefont {Child}\ \emph {et~al.}(2022)\citenamefont {Child}, \citenamefont {Sheekey}, \citenamefont {L\"uscher}, \citenamefont {Fallahi}, \citenamefont {Gardner}, \citenamefont {Manfra}, \citenamefont {Mitchell}, \citenamefont {Sela}, \citenamefont {Kleeorin}, \citenamefont {Meir},\ and\ \citenamefont {Folk}}]{child2022entropy}%
  \BibitemOpen
  \bibfield  {author} {\bibinfo {author} {\bibfnamefont {T.}~\bibnamefont {Child}}, \bibinfo {author} {\bibfnamefont {O.}~\bibnamefont {Sheekey}}, \bibinfo {author} {\bibfnamefont {S.}~\bibnamefont {L\"uscher}}, \bibinfo {author} {\bibfnamefont {S.}~\bibnamefont {Fallahi}}, \bibinfo {author} {\bibfnamefont {G.~C.}\ \bibnamefont {Gardner}}, \bibinfo {author} {\bibfnamefont {M.}~\bibnamefont {Manfra}}, \bibinfo {author} {\bibfnamefont {A.~K.}\ \bibnamefont {Mitchell}}, \bibinfo {author} {\bibfnamefont {E.}~\bibnamefont {Sela}}, \bibinfo {author} {\bibfnamefont {Y.}~\bibnamefont {Kleeorin}}, \bibinfo {author} {\bibfnamefont {Y.}~\bibnamefont {Meir}},\ and\ \bibinfo {author} {\bibfnamefont {J.}~\bibnamefont {Folk}},\ }\bibfield  {title} {\bibinfo {title} {Entropy measurement of a strongly coupled quantum dot},\ }\href {https://doi.org/10.1103/PhysRevLett.129.227702} {\bibfield  {journal} {\bibinfo  {journal} {Phys. Rev. Lett.}\ }\textbf {\bibinfo {volume} {129}},\ \bibinfo {pages} {227702} (\bibinfo {year}
  {2022})}\BibitemShut {NoStop}%
\bibitem [{\citenamefont {Campbell}\ \emph {et~al.}(2025)\citenamefont {Campbell}, \citenamefont {D'Amico}, \citenamefont {Ciampini}, \citenamefont {Anders}, \citenamefont {Ares}, \citenamefont {Artini}, \citenamefont {Auff{\`e}ves}, \citenamefont {Oftelie}, \citenamefont {Bettmann}, \citenamefont {Bonan{\c{c}}a} \emph {et~al.}}]{campbell2025roadmap}%
  \BibitemOpen
  \bibfield  {author} {\bibinfo {author} {\bibfnamefont {S.}~\bibnamefont {Campbell}}, \bibinfo {author} {\bibfnamefont {I.}~\bibnamefont {D'Amico}}, \bibinfo {author} {\bibfnamefont {M.~A.}\ \bibnamefont {Ciampini}}, \bibinfo {author} {\bibfnamefont {J.}~\bibnamefont {Anders}}, \bibinfo {author} {\bibfnamefont {N.}~\bibnamefont {Ares}}, \bibinfo {author} {\bibfnamefont {S.}~\bibnamefont {Artini}}, \bibinfo {author} {\bibfnamefont {A.}~\bibnamefont {Auff{\`e}ves}}, \bibinfo {author} {\bibfnamefont {L.~B.}\ \bibnamefont {Oftelie}}, \bibinfo {author} {\bibfnamefont {L.~P.}\ \bibnamefont {Bettmann}}, \bibinfo {author} {\bibfnamefont {M.~V.}\ \bibnamefont {Bonan{\c{c}}a}}, \emph {et~al.},\ }\bibfield  {title} {\bibinfo {title} {Roadmap on quantum thermodynamics},\ }\href@noop {} {\bibfield  {journal} {\bibinfo  {journal} {arXiv preprint arXiv:2504.20145}\ } (\bibinfo {year} {2025})}\BibitemShut {NoStop}%
\bibitem [{\citenamefont {Mitchell}\ \emph {et~al.}(2017)\citenamefont {Mitchell}, \citenamefont {Pedersen}, \citenamefont {Hedeg{\aa}rd},\ and\ \citenamefont {Paaske}}]{mitchell2017kondo}%
  \BibitemOpen
  \bibfield  {author} {\bibinfo {author} {\bibfnamefont {A.~K.}\ \bibnamefont {Mitchell}}, \bibinfo {author} {\bibfnamefont {K.~G.~L.}\ \bibnamefont {Pedersen}}, \bibinfo {author} {\bibfnamefont {P.}~\bibnamefont {Hedeg{\aa}rd}},\ and\ \bibinfo {author} {\bibfnamefont {J.}~\bibnamefont {Paaske}},\ }\bibfield  {title} {\bibinfo {title} {Kondo blockade due to quantum interference in single-molecule junctions},\ }\href {https://doi.org/10.1038/ncomms15210} {\bibfield  {journal} {\bibinfo  {journal} {Nat. Commun.}\ }\textbf {\bibinfo {volume} {8}},\ \bibinfo {pages} {15210} (\bibinfo {year} {2017})}\BibitemShut {NoStop}%
\bibitem [{\citenamefont {S.}\ and\ \citenamefont {Mitchell}(2024)}]{sen2023many}%
  \BibitemOpen
  \bibfield  {author} {\bibinfo {author} {\bibfnamefont {S.}~\bibnamefont {S.}}\ and\ \bibinfo {author} {\bibfnamefont {A.~K.}\ \bibnamefont {Mitchell}},\ }\bibfield  {title} {\bibinfo {title} {Many-body quantum interference route to the two-channel kondo effect: Inverse design for molecular junctions and quantum dot devices},\ }\href {https://doi.org/10.1103/PhysRevLett.133.076501} {\bibfield  {journal} {\bibinfo  {journal} {Phys. Rev. Lett.}\ }\textbf {\bibinfo {volume} {133}},\ \bibinfo {pages} {076501} (\bibinfo {year} {2024})}\BibitemShut {NoStop}%
\bibitem [{\citenamefont {Fr\'erot}\ and\ \citenamefont {Roscilde}(2018)}]{frerot2018quantum}%
  \BibitemOpen
  \bibfield  {author} {\bibinfo {author} {\bibfnamefont {I.}~\bibnamefont {Fr\'erot}}\ and\ \bibinfo {author} {\bibfnamefont {T.}~\bibnamefont {Roscilde}},\ }\bibfield  {title} {\bibinfo {title} {Quantum critical metrology},\ }\href {https://doi.org/10.1103/PhysRevLett.121.020402} {\bibfield  {journal} {\bibinfo  {journal} {Phys. Rev. Lett.}\ }\textbf {\bibinfo {volume} {121}},\ \bibinfo {pages} {020402} (\bibinfo {year} {2018})}\BibitemShut {NoStop}%
\bibitem [{\citenamefont {Mihailescu}\ \emph {et~al.}(2023)\citenamefont {Mihailescu}, \citenamefont {Campbell},\ and\ \citenamefont {Mitchell}}]{PhysRevA.107.042614}%
  \BibitemOpen
  \bibfield  {author} {\bibinfo {author} {\bibfnamefont {G.}~\bibnamefont {Mihailescu}}, \bibinfo {author} {\bibfnamefont {S.}~\bibnamefont {Campbell}},\ and\ \bibinfo {author} {\bibfnamefont {A.~K.}\ \bibnamefont {Mitchell}},\ }\bibfield  {title} {\bibinfo {title} {Thermometry of strongly correlated fermionic quantum systems using impurity probes},\ }\href {https://doi.org/10.1103/PhysRevA.107.042614} {\bibfield  {journal} {\bibinfo  {journal} {Phys. Rev. A}\ }\textbf {\bibinfo {volume} {107}},\ \bibinfo {pages} {042614} (\bibinfo {year} {2023})}\BibitemShut {NoStop}%
\bibitem [{\citenamefont {Mihailescu}\ \emph {et~al.}(2024)\citenamefont {Mihailescu}, \citenamefont {Bayat}, \citenamefont {Campbell},\ and\ \citenamefont {Mitchell}}]{mihailescu2023multiparameter}%
  \BibitemOpen
  \bibfield  {author} {\bibinfo {author} {\bibfnamefont {G.}~\bibnamefont {Mihailescu}}, \bibinfo {author} {\bibfnamefont {A.}~\bibnamefont {Bayat}}, \bibinfo {author} {\bibfnamefont {S.}~\bibnamefont {Campbell}},\ and\ \bibinfo {author} {\bibfnamefont {A.}~\bibnamefont {Mitchell}},\ }\bibfield  {title} {\bibinfo {title} {Multiparameter critical quantum metrology with impurity probes},\ }\href {https://doi.org/10.1088/2058-9565/ad438d} {\bibfield  {journal} {\bibinfo  {journal} {Quantum Sci. Technol.}\ }\textbf {\bibinfo {volume} {9}},\ \bibinfo {pages} {035033} (\bibinfo {year} {2024})}\BibitemShut {NoStop}%
\bibitem [{\citenamefont {Di~Fresco}\ \emph {et~al.}(2022)\citenamefont {Di~Fresco}, \citenamefont {Spagnolo}, \citenamefont {Valenti},\ and\ \citenamefont {Carollo}}]{SciPostPhys.13.4.077}%
  \BibitemOpen
  \bibfield  {author} {\bibinfo {author} {\bibfnamefont {G.}~\bibnamefont {Di~Fresco}}, \bibinfo {author} {\bibfnamefont {B.}~\bibnamefont {Spagnolo}}, \bibinfo {author} {\bibfnamefont {D.}~\bibnamefont {Valenti}},\ and\ \bibinfo {author} {\bibfnamefont {A.}~\bibnamefont {Carollo}},\ }\bibfield  {title} {\bibinfo {title} {{Multiparameter quantum critical metrology}},\ }\href {https://doi.org/10.21468/SciPostPhys.13.4.077} {\bibfield  {journal} {\bibinfo  {journal} {SciPost Phys.}\ }\textbf {\bibinfo {volume} {13}},\ \bibinfo {pages} {077} (\bibinfo {year} {2022})}\BibitemShut {NoStop}%
\bibitem [{\citenamefont {Laurell}\ \emph {et~al.}(2025)\citenamefont {Laurell}, \citenamefont {Scheie}, \citenamefont {Dagotto},\ and\ \citenamefont {Tennant}}]{laurell2024}%
  \BibitemOpen
  \bibfield  {author} {\bibinfo {author} {\bibfnamefont {P.}~\bibnamefont {Laurell}}, \bibinfo {author} {\bibfnamefont {A.}~\bibnamefont {Scheie}}, \bibinfo {author} {\bibfnamefont {E.}~\bibnamefont {Dagotto}},\ and\ \bibinfo {author} {\bibfnamefont {D.~A.}\ \bibnamefont {Tennant}},\ }\bibfield  {title} {\bibinfo {title} {Witnessing entanglement and quantum correlations in condensed matter: A review},\ }\href {https://doi.org/https://doi.org/10.1002/qute.202400196} {\bibfield  {journal} {\bibinfo  {journal} {Advanced Quantum Technologies}\ }\textbf {\bibinfo {volume} {8}},\ \bibinfo {pages} {2400196} (\bibinfo {year} {2025})}\BibitemShut {NoStop}%
\bibitem [{\citenamefont {Liu}\ \emph {et~al.}(2019)\citenamefont {Liu}, \citenamefont {Yuan}, \citenamefont {Lu},\ and\ \citenamefont {Wang}}]{Liu2020}%
  \BibitemOpen
  \bibfield  {author} {\bibinfo {author} {\bibfnamefont {J.}~\bibnamefont {Liu}}, \bibinfo {author} {\bibfnamefont {H.}~\bibnamefont {Yuan}}, \bibinfo {author} {\bibfnamefont {X.}~\bibnamefont {Lu}},\ and\ \bibinfo {author} {\bibfnamefont {X.}~\bibnamefont {Wang}},\ }\bibfield  {title} {\bibinfo {title} {Quantum fisher information matrix and multiparameter estimation},\ }\href {https://doi.org/10.1088/1751-8121/ab5d4d} {\bibfield  {journal} {\bibinfo  {journal} {J. Phys. A: Math. Theor.}\ }\textbf {\bibinfo {volume} {53}},\ \bibinfo {pages} {023001} (\bibinfo {year} {2019})}\BibitemShut {NoStop}%
\bibitem [{\citenamefont {Scandi}\ \emph {et~al.}(2025)\citenamefont {Scandi}, \citenamefont {Abiuso}, \citenamefont {Surace},\ and\ \citenamefont {De~Santis}}]{scandi2023}%
  \BibitemOpen
  \bibfield  {author} {\bibinfo {author} {\bibfnamefont {M.}~\bibnamefont {Scandi}}, \bibinfo {author} {\bibfnamefont {P.}~\bibnamefont {Abiuso}}, \bibinfo {author} {\bibfnamefont {J.}~\bibnamefont {Surace}},\ and\ \bibinfo {author} {\bibfnamefont {D.}~\bibnamefont {De~Santis}},\ }\bibfield  {title} {\bibinfo {title} {Quantum fisher information and its dynamical nature},\ }\href {https://doi.org/10.1088/1361-6633/ade453} {\bibfield  {journal} {\bibinfo  {journal} {Reports on Progress in Physics}\ }\textbf {\bibinfo {volume} {88}},\ \bibinfo {pages} {076001} (\bibinfo {year} {2025})}\BibitemShut {NoStop}%
\bibitem [{\citenamefont {Marvian}(2022)}]{marvian2022}%
  \BibitemOpen
  \bibfield  {author} {\bibinfo {author} {\bibfnamefont {I.}~\bibnamefont {Marvian}},\ }\bibfield  {title} {\bibinfo {title} {Operational interpretation of quantum fisher information in quantum thermodynamics},\ }\href {https://doi.org/10.1103/PhysRevLett.129.190502} {\bibfield  {journal} {\bibinfo  {journal} {Phys. Rev. Lett.}\ }\textbf {\bibinfo {volume} {129}},\ \bibinfo {pages} {190502} (\bibinfo {year} {2022})}\BibitemShut {NoStop}%
\bibitem [{\citenamefont {Poggi}\ \emph {et~al.}(2024)\citenamefont {Poggi}, \citenamefont {De~Chiara}, \citenamefont {Campbell},\ and\ \citenamefont {Kiely}}]{Poggi2024}%
  \BibitemOpen
  \bibfield  {author} {\bibinfo {author} {\bibfnamefont {P.~M.}\ \bibnamefont {Poggi}}, \bibinfo {author} {\bibfnamefont {G.}~\bibnamefont {De~Chiara}}, \bibinfo {author} {\bibfnamefont {S.}~\bibnamefont {Campbell}},\ and\ \bibinfo {author} {\bibfnamefont {A.}~\bibnamefont {Kiely}},\ }\bibfield  {title} {\bibinfo {title} {Universally robust quantum control},\ }\href {https://doi.org/10.1103/PhysRevLett.132.193801} {\bibfield  {journal} {\bibinfo  {journal} {Phys. Rev. Lett.}\ }\textbf {\bibinfo {volume} {132}},\ \bibinfo {pages} {193801} (\bibinfo {year} {2024})}\BibitemShut {NoStop}%
\bibitem [{\citenamefont {Deffner}\ and\ \citenamefont {Campbell}(2017)}]{Deffner2017}%
  \BibitemOpen
  \bibfield  {author} {\bibinfo {author} {\bibfnamefont {S.}~\bibnamefont {Deffner}}\ and\ \bibinfo {author} {\bibfnamefont {S.}~\bibnamefont {Campbell}},\ }\bibfield  {title} {\bibinfo {title} {Quantum speed limits: from heisenberg’s uncertainty principle to optimal quantum control},\ }\href {https://doi.org/10.1088/1751-8121/aa86c6} {\bibfield  {journal} {\bibinfo  {journal} {J. Phys. A: Math. Theor.}\ }\textbf {\bibinfo {volume} {50}},\ \bibinfo {pages} {453001} (\bibinfo {year} {2017})}\BibitemShut {NoStop}%
\bibitem [{\citenamefont {Meyer}(2021)}]{meyer2021}%
  \BibitemOpen
  \bibfield  {author} {\bibinfo {author} {\bibfnamefont {J.~J.}\ \bibnamefont {Meyer}},\ }\bibfield  {title} {\bibinfo {title} {Fisher information in noisy intermediate-scale quantum applications},\ }\href {https://doi.org/10.22331/q-2021-09-09-539} {\bibfield  {journal} {\bibinfo  {journal} {Quantum}\ }\textbf {\bibinfo {volume} {5}},\ \bibinfo {pages} {539} (\bibinfo {year} {2021})}\BibitemShut {NoStop}%
\bibitem [{\citenamefont {Gammelmark}\ and\ \citenamefont {M\o{}lmer}(2013)}]{Gammelmark2013}%
  \BibitemOpen
  \bibfield  {author} {\bibinfo {author} {\bibfnamefont {S.}~\bibnamefont {Gammelmark}}\ and\ \bibinfo {author} {\bibfnamefont {K.}~\bibnamefont {M\o{}lmer}},\ }\bibfield  {title} {\bibinfo {title} {Bayesian parameter inference from continuously monitored quantum systems},\ }\href {https://doi.org/10.1103/PhysRevA.87.032115} {\bibfield  {journal} {\bibinfo  {journal} {Phys. Rev. A}\ }\textbf {\bibinfo {volume} {87}},\ \bibinfo {pages} {032115} (\bibinfo {year} {2013})}\BibitemShut {NoStop}%
\bibitem [{\citenamefont {Gammelmark}\ and\ \citenamefont {M\o{}lmer}(2014)}]{Gammelmark2014}%
  \BibitemOpen
  \bibfield  {author} {\bibinfo {author} {\bibfnamefont {S.}~\bibnamefont {Gammelmark}}\ and\ \bibinfo {author} {\bibfnamefont {K.}~\bibnamefont {M\o{}lmer}},\ }\bibfield  {title} {\bibinfo {title} {Fisher information and the quantum cram\'er-rao sensitivity limit of continuous measurements},\ }\href {https://doi.org/10.1103/PhysRevLett.112.170401} {\bibfield  {journal} {\bibinfo  {journal} {Phys. Rev. Lett.}\ }\textbf {\bibinfo {volume} {112}},\ \bibinfo {pages} {170401} (\bibinfo {year} {2014})}\BibitemShut {NoStop}%
\bibitem [{\citenamefont {Radaelli}\ \emph {et~al.}(2024)\citenamefont {Radaelli}, \citenamefont {Smiga}, \citenamefont {Landi},\ and\ \citenamefont {Binder}}]{radaelli2024}%
  \BibitemOpen
  \bibfield  {author} {\bibinfo {author} {\bibfnamefont {M.}~\bibnamefont {Radaelli}}, \bibinfo {author} {\bibfnamefont {J.~A.}\ \bibnamefont {Smiga}}, \bibinfo {author} {\bibfnamefont {G.~T.}\ \bibnamefont {Landi}},\ and\ \bibinfo {author} {\bibfnamefont {F.~C.}\ \bibnamefont {Binder}},\ }\href {https://arxiv.org/abs/2402.06556} {\bibinfo {title} {Parameter estimation for quantum jump unraveling}} (\bibinfo {year} {2024}),\ \Eprint {https://arxiv.org/abs/2402.06556} {2402.06556} \BibitemShut {NoStop}%
\bibitem [{\citenamefont {Ilias}\ \emph {et~al.}(2022)\citenamefont {Ilias}, \citenamefont {Yang}, \citenamefont {Huelga},\ and\ \citenamefont {Plenio}}]{PRXQuantum.3.010354}%
  \BibitemOpen
  \bibfield  {author} {\bibinfo {author} {\bibfnamefont {T.}~\bibnamefont {Ilias}}, \bibinfo {author} {\bibfnamefont {D.}~\bibnamefont {Yang}}, \bibinfo {author} {\bibfnamefont {S.}~\bibnamefont {Huelga}},\ and\ \bibinfo {author} {\bibfnamefont {M.}~\bibnamefont {Plenio}},\ }\bibfield  {title} {\bibinfo {title} {Criticality-enhanced quantum sensing via continuous measurement},\ }\href {https://doi.org/10.1103/PRXQuantum.3.010354} {\bibfield  {journal} {\bibinfo  {journal} {PRX Quantum}\ }\textbf {\bibinfo {volume} {3}},\ \bibinfo {pages} {010354} (\bibinfo {year} {2022})}\BibitemShut {NoStop}%
\bibitem [{\citenamefont {Boeyens}\ \emph {et~al.}(2023)\citenamefont {Boeyens}, \citenamefont {Annby-Andersson}, \citenamefont {Bakhshinezhad}, \citenamefont {Perarnau-Llobet}, \citenamefont {Nimmrichter},\ and\ \citenamefont {Mehboudi}}]{boeyens2307probe}%
  \BibitemOpen
  \bibfield  {author} {\bibinfo {author} {\bibfnamefont {J.}~\bibnamefont {Boeyens}}, \bibinfo {author} {\bibfnamefont {B.}~\bibnamefont {Annby-Andersson}}, \bibinfo {author} {\bibfnamefont {G.}~\bibnamefont {Bakhshinezhad}, \bibfnamefont {P.and~Haack}}, \bibinfo {author} {\bibfnamefont {M.}~\bibnamefont {Perarnau-Llobet}}, \bibinfo {author} {\bibfnamefont {P.}~\bibnamefont {Nimmrichter}, \bibfnamefont {S.and~Potts}},\ and\ \bibinfo {author} {\bibfnamefont {M.}~\bibnamefont {Mehboudi}},\ }\bibfield  {title} {\bibinfo {title} {Probe thermometry with continuous measurements},\ }\href {https://doi.org/10.1088/1367-2630/ad0e8a} {\bibfield  {journal} {\bibinfo  {journal} {New J. Phys.}\ }\textbf {\bibinfo {volume} {25}},\ \bibinfo {pages} {123009} (\bibinfo {year} {2023})}\BibitemShut {NoStop}%
\bibitem [{\citenamefont {Khandelwal}\ \emph {et~al.}(2025)\citenamefont {Khandelwal}, \citenamefont {Landi}, \citenamefont {Haack},\ and\ \citenamefont {Mitchison}}]{mark-landi}%
  \BibitemOpen
  \bibfield  {author} {\bibinfo {author} {\bibfnamefont {S.}~\bibnamefont {Khandelwal}}, \bibinfo {author} {\bibfnamefont {G.~T.}\ \bibnamefont {Landi}}, \bibinfo {author} {\bibfnamefont {G.}~\bibnamefont {Haack}},\ and\ \bibinfo {author} {\bibfnamefont {M.~T.}\ \bibnamefont {Mitchison}},\ }\bibfield  {title} {\bibinfo {title} {Current-based metrology with two-terminal mesoscopic conductors},\ }\href@noop {} {\bibfield  {journal} {\bibinfo  {journal} {arXiv preprint arXiv:2507.12907}\ } (\bibinfo {year} {2025})}\BibitemShut {NoStop}%
\bibitem [{\citenamefont {Galpin}\ \emph {et~al.}(2014)\citenamefont {Galpin}, \citenamefont {Mitchell}, \citenamefont {Temaismithi}, \citenamefont {Logan}, \citenamefont {B\'eri},\ and\ \citenamefont {Cooper}}]{galpin2014conductance}%
  \BibitemOpen
  \bibfield  {author} {\bibinfo {author} {\bibfnamefont {M.~R.}\ \bibnamefont {Galpin}}, \bibinfo {author} {\bibfnamefont {A.~K.}\ \bibnamefont {Mitchell}}, \bibinfo {author} {\bibfnamefont {J.}~\bibnamefont {Temaismithi}}, \bibinfo {author} {\bibfnamefont {D.~E.}\ \bibnamefont {Logan}}, \bibinfo {author} {\bibfnamefont {B.}~\bibnamefont {B\'eri}},\ and\ \bibinfo {author} {\bibfnamefont {N.~R.}\ \bibnamefont {Cooper}},\ }\bibfield  {title} {\bibinfo {title} {Conductance fingerprint of majorana fermions in the topological kondo effect},\ }\href {https://doi.org/10.1103/PhysRevB.89.045143} {\bibfield  {journal} {\bibinfo  {journal} {Phys. Rev. B}\ }\textbf {\bibinfo {volume} {89}},\ \bibinfo {pages} {045143} (\bibinfo {year} {2014})}\BibitemShut {NoStop}%
\bibitem [{\citenamefont {Hanl}\ and\ \citenamefont {Weichselbaum}(2014)}]{hanl2014local}%
  \BibitemOpen
  \bibfield  {author} {\bibinfo {author} {\bibfnamefont {M.}~\bibnamefont {Hanl}}\ and\ \bibinfo {author} {\bibfnamefont {A.}~\bibnamefont {Weichselbaum}},\ }\bibfield  {title} {\bibinfo {title} {Local susceptibility and kondo scaling in the presence of finite bandwidth},\ }\href {https://doi.org/10.1103/PhysRevB.89.075130} {\bibfield  {journal} {\bibinfo  {journal} {Phys. Rev. B}\ }\textbf {\bibinfo {volume} {89}},\ \bibinfo {pages} {075130} (\bibinfo {year} {2014})}\BibitemShut {NoStop}%
\bibitem [{\citenamefont {Costi}(2000)}]{costi2000kondo}%
  \BibitemOpen
  \bibfield  {author} {\bibinfo {author} {\bibfnamefont {T.~A.}\ \bibnamefont {Costi}},\ }\bibfield  {title} {\bibinfo {title} {Kondo effect in a magnetic field and the magnetoresistivity of kondo alloys},\ }\href {https://doi.org/10.1103/PhysRevLett.85.1504} {\bibfield  {journal} {\bibinfo  {journal} {Phys. Rev. Lett.}\ }\textbf {\bibinfo {volume} {85}},\ \bibinfo {pages} {1504} (\bibinfo {year} {2000})}\BibitemShut {NoStop}%
\bibitem [{\citenamefont {Pezz\`e}\ \emph {et~al.}(2018)\citenamefont {Pezz\`e}, \citenamefont {Smerzi}, \citenamefont {Oberthaler}, \citenamefont {Schmied},\ and\ \citenamefont {Treutlein}}]{Pezze2018}%
  \BibitemOpen
  \bibfield  {author} {\bibinfo {author} {\bibfnamefont {L.}~\bibnamefont {Pezz\`e}}, \bibinfo {author} {\bibfnamefont {A.}~\bibnamefont {Smerzi}}, \bibinfo {author} {\bibfnamefont {M.~K.}\ \bibnamefont {Oberthaler}}, \bibinfo {author} {\bibfnamefont {R.}~\bibnamefont {Schmied}},\ and\ \bibinfo {author} {\bibfnamefont {P.}~\bibnamefont {Treutlein}},\ }\bibfield  {title} {\bibinfo {title} {Quantum metrology with nonclassical states of atomic ensembles},\ }\href {https://doi.org/10.1103/RevModPhys.90.035005} {\bibfield  {journal} {\bibinfo  {journal} {Rev. Mod. Phys.}\ }\textbf {\bibinfo {volume} {90}},\ \bibinfo {pages} {035005} (\bibinfo {year} {2018})}\BibitemShut {NoStop}%
\bibitem [{\citenamefont {Barry}\ \emph {et~al.}(2016)\citenamefont {Barry}, \citenamefont {Turner}, \citenamefont {Schloss}, \citenamefont {Glenn}, \citenamefont {Song}, \citenamefont {Lukin}, \citenamefont {Park},\ and\ \citenamefont {Walsworth}}]{barry2016optical}%
  \BibitemOpen
  \bibfield  {author} {\bibinfo {author} {\bibfnamefont {J.~F.}\ \bibnamefont {Barry}}, \bibinfo {author} {\bibfnamefont {M.~J.}\ \bibnamefont {Turner}}, \bibinfo {author} {\bibfnamefont {J.~M.}\ \bibnamefont {Schloss}}, \bibinfo {author} {\bibfnamefont {D.~R.}\ \bibnamefont {Glenn}}, \bibinfo {author} {\bibfnamefont {Y.}~\bibnamefont {Song}}, \bibinfo {author} {\bibfnamefont {M.~D.}\ \bibnamefont {Lukin}}, \bibinfo {author} {\bibfnamefont {H.}~\bibnamefont {Park}},\ and\ \bibinfo {author} {\bibfnamefont {R.~L.}\ \bibnamefont {Walsworth}},\ }\bibfield  {title} {\bibinfo {title} {Optical magnetic detection of single-neuron action potentials using quantum defects in diamond},\ }\href@noop {} {\bibfield  {journal} {\bibinfo  {journal} {Proceedings of the {N}ational {A}cademy of {S}ciences}\ }\textbf {\bibinfo {volume} {113}},\ \bibinfo {pages} {14133} (\bibinfo {year} {2016})}\BibitemShut {NoStop}%
\bibitem [{\citenamefont {Barry}\ \emph {et~al.}(2020)\citenamefont {Barry}, \citenamefont {Schloss}, \citenamefont {Bauch}, \citenamefont {Turner}, \citenamefont {Hart}, \citenamefont {Pham},\ and\ \citenamefont {Walsworth}}]{barry2020sensitivity}%
  \BibitemOpen
  \bibfield  {author} {\bibinfo {author} {\bibfnamefont {J.~F.}\ \bibnamefont {Barry}}, \bibinfo {author} {\bibfnamefont {J.~M.}\ \bibnamefont {Schloss}}, \bibinfo {author} {\bibfnamefont {E.}~\bibnamefont {Bauch}}, \bibinfo {author} {\bibfnamefont {M.~J.}\ \bibnamefont {Turner}}, \bibinfo {author} {\bibfnamefont {C.~A.}\ \bibnamefont {Hart}}, \bibinfo {author} {\bibfnamefont {L.~M.}\ \bibnamefont {Pham}},\ and\ \bibinfo {author} {\bibfnamefont {R.~L.}\ \bibnamefont {Walsworth}},\ }\bibfield  {title} {\bibinfo {title} {Sensitivity optimization for {NV}-diamond magnetometry},\ }\href {https://doi.org/10.1103/RevModPhys.92.015004} {\bibfield  {journal} {\bibinfo  {journal} {Reviews of {M}odern {P}hysics}\ }\textbf {\bibinfo {volume} {92}},\ \bibinfo {pages} {015004} (\bibinfo {year} {2020})}\BibitemShut {NoStop}%
\bibitem [{\citenamefont {Jensen}\ \emph {et~al.}(2016)\citenamefont {Jensen}, \citenamefont {Kehayias},\ and\ \citenamefont {Budker}}]{jensen2016magnetometry}%
  \BibitemOpen
  \bibfield  {author} {\bibinfo {author} {\bibfnamefont {K.}~\bibnamefont {Jensen}}, \bibinfo {author} {\bibfnamefont {P.}~\bibnamefont {Kehayias}},\ and\ \bibinfo {author} {\bibfnamefont {D.}~\bibnamefont {Budker}},\ }\bibfield  {title} {\bibinfo {title} {Magnetometry with nitrogen-vacancy centers in diamond},\ }in\ \href@noop {} {\emph {\bibinfo {booktitle} {High sensitivity magnetometers}}}\ (\bibinfo  {publisher} {Springer},\ \bibinfo {year} {2016})\ pp.\ \bibinfo {pages} {553--576}\BibitemShut {NoStop}%
\bibitem [{\citenamefont {Drung}\ \emph {et~al.}(2007)\citenamefont {Drung}, \citenamefont {Abmann}, \citenamefont {Beyer}, \citenamefont {Kirste}, \citenamefont {Peters}, \citenamefont {Ruede},\ and\ \citenamefont {Schurig}}]{drung2007highly}%
  \BibitemOpen
  \bibfield  {author} {\bibinfo {author} {\bibfnamefont {D.}~\bibnamefont {Drung}}, \bibinfo {author} {\bibfnamefont {C.}~\bibnamefont {Abmann}}, \bibinfo {author} {\bibfnamefont {J.}~\bibnamefont {Beyer}}, \bibinfo {author} {\bibfnamefont {A.}~\bibnamefont {Kirste}}, \bibinfo {author} {\bibfnamefont {M.}~\bibnamefont {Peters}}, \bibinfo {author} {\bibfnamefont {F.}~\bibnamefont {Ruede}},\ and\ \bibinfo {author} {\bibfnamefont {T.}~\bibnamefont {Schurig}},\ }\bibfield  {title} {\bibinfo {title} {Highly sensitive and easy-to-use {SQUID} sensors},\ }\href@noop {} {\bibfield  {journal} {\bibinfo  {journal} {{IEEE} {T}ransactions on {A}pplied {S}uperconductivity}\ }\textbf {\bibinfo {volume} {17}},\ \bibinfo {pages} {699} (\bibinfo {year} {2007})}\BibitemShut {NoStop}%
\end{thebibliography}

\end{document}

\title{\textbf{\underline{SUPPLEMENTARY MATERIAL}}\\Quantum Sensing with Nanoelectronics:\\Fisher Information for an \rev{Applied} Perturbation}

\author{George Mihailescu}
\author{Anthony Kiely}
\author{Andrew K. Mitchell}
\affiliation{School of Physics, University College Dublin, Belfield, Dublin 4, Ireland}
\affiliation{Centre for Quantum Engineering, Science, and Technology, University College Dublin, Ireland}

\maketitle
%%%%%%%%%%%%%%%%%%%%%%%%%%%%%%%%%%%%%%%%%%%%%%%%%%%%%%%%%%%%%%%%%%%%%%%
%%%%%%%%%%%%%%%%%%%%%%%%%%%%%%%%%%%%%%%%%%%%%%%%%%%%%%%%%%%%%%%%%%%%%%%

\rev{
\section{QFI for a finite-time ramp}
We consider in this work an arbitrary time-dependent Hamiltonian $\hat{H}_\gamma=\hat{H}_0 + \gamma(t) \hat{A}$, where we assume that $\hat{A}$ is a weak perturbation -- that is,  the resulting dynamics are in the linear response regime. The ramp from $\gamma(0)=0$ to $\gamma(\tau)=\lambda$ in a time $\tau$ encodes the small parameter $\lambda$ that we wish to estimate. We will quantify how well this can be done by the response of the system to this perturbation and the quantum Fisher information (QFI).

We take our quantum system to be initially in thermal equilibrium, described by the thermal density matrix $\hat{\rho}_0=\exp(-\beta \hat{H}_0)/Z_0$ at inverse temperature $\beta=1/T$, where $Z_0={\rm Tr}[\exp(-\beta \hat{H}_0)]$ is the partition function of $\hat{H}_0$. We set $k_{\rm B}\equiv 1$ and $\hbar\equiv 1$. In the energy eigenbasis we have $\hat{H}_0 |n_0\rangle = E^0_n |n_0\rangle$ and $\hat{\rho}_0 = \sum_n p_n^0 |n_0\rangle \langle n_0|$, where $p_n^0=\exp(-\beta E_n^0)/Z_0$ are the probabilities. 

This initial state then evolves as $\hat{\rho}_\lambda = \hat{U}_\lambda^{\phantom{\dagger}} \:\hat{\rho}_0\: \hat{U}_\lambda^\dagger$ where the unitary time evolution operator $\hat{U}_\lambda$ encodes the relevant parameter $\lambda$. The QFI for such a unitary encoding is given by, \cite{toth}
\begin{eqnarray}
\label{eq:qfi_gen}
F_Q[\lambda] &=&  2\sum_{n \ne m}\frac{(p_n^0-p_m^0)^2}{p_n^0+p_m^0} |\bra{m_0} \hat{U}_\lambda^\dagger (\partial_\lambda \hat{U}_\lambda) \ket{n_0} |^2 \;.
\end{eqnarray}

%############

\subsection{Perturbation Theory}
To be more explicit, the full time evolution operator from a time $t_1$ to a later time $t_2$ can be denoted as $\hat{U}_\lambda(t_2,t_1)$. We make the approximation $\hat{U}_\lambda \approx \hat{U}_0 + \lambda \hat{U}_1$ where $\hat{U}_0(t_2,t_1)=e^{-i \hat{H}_0(t_2-t_1)}$ is the unperturbed evolution (of a time-independent Hamiltonian $\hat{H}_0$) and,
\begin{eqnarray}
\hat{U}_1(t_2,t_1) =  - i \int_{t_1}^{t_2} ds \, \eta(s) \hat{U}_0(t_2,s)  \hat{A} \hat{U}_0(s,t_1) \;,
\end{eqnarray}
is the first order correction and $\eta(s)=\gamma(s)/\lambda$ is just a rescaling of the ramp function $\gamma(s)$ such that $\eta(\tau)=1$.

In general, $F_Q[\lambda]$ will have a non-trivial dependence on $\lambda$.  However we will focus solely on the zeroth order term, which allows us to approximate the key part of $F_Q$ as,
\begin{eqnarray}
\bra{m_0} \hat{U}_\lambda^\dagger (\partial_\lambda \hat{U}_\lambda) \ket{n_0}&\approx&  \bra{m_0} \hat{U}_0^\dagger(\tau,0) \hat{U}_1(\tau,0) \ket{n_0} \\
&=& -i \int_0^{\tau} ds \bra{m_0} \hat{U}_0^\dagger (s,0) \eta(s) \hat{A} \hat{U}_0 (s,0) \ket{n_0} \;.
\end{eqnarray}

Putting this all together, the QFI then reads,
\begin{eqnarray}
F_Q[\lambda] &\approx& 2 \sum_{m \neq n} \frac{(p_n^0-p_m^0)^2}{p_n^0+p_m^0} \left| \int_0^{\tau} ds \bra{m_0} \hat{U}_0^\dagger (s,0) \eta(s) \hat{A} \hat{U}_0 (s,0) \ket{n_0}  \right|^2 \\
&=& 2 \sum_{m \neq n} \frac{(p_n^0-p_m^0)^2}{p_n^0+p_m^0} \left| \bra{m_0}\hat{A}\ket{n_0} \right|^2 \left| \int_0^{\tau} ds \,    \eta(s) e^{-i (E_n^0-E_m^0) s}  \right|^2 \\
&=& 2 \sum_{m \neq n} \frac{(p_n^0-p_m^0)^2}{p_n^0+p_m^0} \left| \bra{m_0}\hat{A}\ket{n_0} \right|^2 \times Q_{\tau}(E_n^0-E_m^0)
\end{eqnarray}
where details of the ramp appear only in the function,
\begin{eqnarray}
Q_{\tau}(\omega) &=& \left| \int_0^{\tau} ds \,    \eta(s) e^{-i \omega s}  \right|^2 \\
&=& \left| \frac{1}{- i  \omega }  \int_0^{\tau} ds \,    \eta(s) \frac{\partial}{\partial s} e^{-i \omega s} \right|^2  = \frac{1}{\omega^2} \left|   1 - \int_0^{\tau} ds \, \dot{\eta}(s) e^{i \omega  (\tau-s)}    \right|^2\;,
\end{eqnarray}
where on the second line we applied integration by parts and used $\eta(0)=0$ and $\eta(\tau)=1$.

%###########

\subsection{Adiabatic Limit}
We now consider the long-time limit of slow driving, $\tau \to \infty$. In this limit we expect the QFI to become independent of the particular ramp profile used. Within the adiabatic approximation (where the rate of the ramp $\dot{\eta}$ is slow relative to the timescale set by the inverse gaps), we indeed find that $Q(\omega) \rightarrow \omega^{-2}$. This recovers Eq.~(1) of the main text. 

%##############

\subsection{QFI from susceptibilities}
\noindent Our result can be cast in the alternative form of a susceptibility, similar to the  result of Ref.~\onlinecite{zoller}, by using the identity,
\begin{equation}
\int d\omega~\frac{\tanh(\omega/2T)}{\omega^2}\delta(\omega-E_m^0+E_n^0) = \frac{p_n^0-p_m^0}{p_n^0+p_m^0}\times\frac{1}{(E_n^0-E_m^0)^2} \;.
\end{equation}
Together with the the definition of the correlation function,
\begin{equation}\label{eq:Kspec}
    {\rm Im}K (\omega) = \pi\sum_{n,m} (p_n^0-p_m^0)|\langle n_0|\hat{A}|m_0\rangle|^2\:\delta(\omega-E_m^0+E_n^0)  \;,
\end{equation}
we can now express the QFI as,
\begin{equation}\label{eq:qfi_Kbar}
\begin{split}
F_Q[\lambda] &= \frac{2}{\pi}\int d\omega~ \tanh(\omega/2T) \:{\rm Im} K(\omega)\times  Q_{\tau}(\omega) \;.
\end{split}
\end{equation}
Here, $K(\omega) \equiv \langle\langle \hat{A};\hat{A}\rangle\rangle =\int dt~ \exp(i\omega t) K(t)$ is the Fourier transform of the retarded, real-time correlator $K(t)=-i\theta(t)\langle [\hat{A}(0),\hat{A}(t)]\rangle_0$, where $\hat{\Omega}(t)=e^{i\hat{H}_0t} \:\hat{\Omega}\: e^{-i\hat{H}_0t}$. 
$K(\omega)$ is evaluated in $\hat{H}_0$ and its Lehmann representation is given by Eq.~\ref{eq:Kspec}. Thus, the QFI for an applied perturbation is a physical observable, since the susceptibility is a physical observable, and experimentally measurable in principle. This holds for a drive in finite-time $\tau$ as well as in the adiabatic limit $\tau \to \infty$ where $Q_{\tau}(\omega)\to \omega^{-2}$ in Eq.~\ref{eq:qfi_Kbar}. The latter yields Eq.~(6) of the main text.

We note the difference between the adiabatic limit of Eq.~\ref{eq:qfi_Kbar} and the result in Ref.~\onlinecite{zoller}, which expresses the QFI for an operator without connecting it to any physical perturbation. A family of related expressions was considered in Ref.~\onlinecite{PhysRevA.94.062316}. By recognizing that the generator of the quasistatic evolution is the adiabatic gauge potential and not the perturbation $\hat{A}$ itself, we get an additional squared excitation energy denominator in our expression for the QFI. This has important consequences. However, it does not affect the results or conclusions of Refs.~\onlinecite{zoller} or \onlinecite{PhysRevA.94.062316}, which were obtained in a different context and did not address quantum parameter estimation. 

We also note that Ref.~\onlinecite{mehboudi2018fluctuation} derives a similar result for perturbed systems that have relaxed to thermal equilibrium (rather than following adiabatic evolution). In this case the QFI picks up additional corrections relative to Eq.~\ref{eq:qfi_Kbar}.

It may be convenient to express Eq.~\ref{eq:qfi_Kbar} in terms of a related correlation function.  We therefore introduce the retarded current-current correlation function $\bar{K}(\omega)\equiv \langle\langle \dot{A} ; \dot{A} \rangle \rangle$, 
where $\dot{A}=\tfrac{d}{dt} \hat{A}$. Here $\bar{K}(\omega)$ is the Fourier transform of $\bar{K}(t)=-i\theta(t)\langle [\dot{A}(0),\dot{A}(t)]\rangle_0$, again evaluated  in $\hat{H}_0$. Using bosonic Green's function equations of motion,\cite{zubarev1960double}
\begin{eqnarray*}
\omega\langle \langle \hat{X} ; \hat{Y} \rangle \rangle -\langle [\hat{X},\hat{Y}] \rangle  =  \langle\langle \hat{X} ;[\hat{H},\hat{Y}] \rangle \rangle  = \langle\langle [\hat{X},\hat{H}] ;\hat{Y} \rangle \rangle
\end{eqnarray*}
with $\dot{\Omega}\equiv \tfrac{d}{dt}\hat{\Omega}=i[\hat{H},\hat{\Omega}]$, we find that $\langle\langle \dot{A} ; \dot{A} \rangle\rangle = \omega^2\langle\langle \hat{A} ; \hat{A} \rangle\rangle+\omega\langle [\hat{A},[\hat{A},\hat{H}_0]]\rangle$. Since $\hat{A}$ and $\hat{H}_0$ are both Hermitian, $\langle [\hat{A},[\hat{A},\hat{H}_0]]\rangle$ is real, 
and therefore ${\rm Im}\bar{K}(\omega)=-\omega^2 {\rm Im}K(\omega)$. Thus we can write, 
\begin{equation}\label{eq:qfi_K}
\begin{split}
F_Q[\lambda] &= -\frac{2}{\pi}\int d\omega~\frac{\tanh(\omega/2T)}{\omega^2}\: {\rm Im} \bar{K}(\omega) \times  Q_{\tau}(\omega)\;,
\end{split}
\end{equation}
which expresses the QFI for a perturbation $\lambda$ in terms of the correlation function for the associated currents $\dot{A}$.

%##################

\subsection{QFI from transport coefficients}
As shown above, the QFI for a perturbation can be computed from correlation functions evaluated in the unperturbed system at equilibrium. The linear response of a system to a weak perturbation can also be expressed in terms of equilibrium correlation functions via the Kubo formula. As we show below, it is possible to relate the QFI for a perturbation to dynamical (ac) linear response transport coefficients -- even though the perturbation for which we compute the QFI is dc steady-state. 

Generalizing to an ac perturbation, we define,
\begin{equation}
    \hat{H}'_1(t)=\lambda \cos(\omega t) \hat{A} \;,
\end{equation}
where $\omega$ is the ac frequency. 
Taking the perturbation $\lambda$ to be small, we can look at the response of the system to the perturbation in terms of the induced currents that flow, to linear order in $\lambda$. Within linear response, it follows that,
\begin{equation}
    \langle \dot{A} \rangle = \chi(\omega) \lambda
\end{equation}
where $\chi(\omega)=\tfrac{d}{d\lambda}\langle \dot{A}\rangle|_{\lambda\to 0} $ is the ac-frequency-dependent linear-response transport coefficient, and $\dot{A}=\tfrac{d}{dt} \hat{A}$.

The Kubo formula \cite{kubo1957statistical,minarelli2022linear} allows us to calculate the transport coefficient in terms of \emph{equilibrium} properties of the unperturbed system $\hat{H}_0$,
\begin{equation}\label{eq:chi_ac}
    \chi(\omega)=\frac{{\rm Im} \bar{K}(\omega)}{\omega}
\end{equation}
where $\bar{K}(\omega)\equiv \langle\langle \dot{A} ; \dot{A} \rangle \rangle$ as before. We have set $\hbar\equiv 1$. Equivalently, this may be expressed in terms of $K(\omega)$ in Eq.~\ref{eq:Kspec},
\begin{equation}\label{eq:chiw}
    \chi(\omega)=-\omega{\rm Im} K(\omega) \;.
\end{equation}

The dc limit corresponds to taking $\omega \to 0$, such that $\hat{H}_1(t) \to \hat{H}_1=\lambda \hat{A}$ and $\chi(\omega) \to \chi_{\rm dc}$ where, 
\begin{equation}
    \chi_{\rm dc} = -\lim_{\omega \to 0} [\omega {\rm Im} K(\omega)]=-\lim_{\omega \to 0} [{\rm Im} \bar{K}(\omega)/\omega] \;.
\end{equation}
When the dc transport coefficient $\chi_{\rm dc}$ is finite, the low-frequency behaviour of the correlation functions is therefore ${\rm Im}K(\omega)\sim 1/\omega$ and ${\rm Im} \bar{K}(\omega)\sim \omega$.

It is now clear that the QFI for a small dc perturbation $\lambda$ can be understood in terms of the ac linear-response transport coefficient $\chi(\omega)$ corresponding to the currents $\langle \dot{A}\rangle$ induced by the operator $\hat{A}$ coupling to $\lambda$,
\begin{equation}\label{eq:qfi_chi}
\begin{split}
F_Q[\lambda] 
& = -\frac{2}{\pi}\int d\omega~\frac{\tanh(\omega/2T)}{\omega}\:\chi(\omega) \times  Q_{\tau}(\omega)  \;.
\end{split}
\end{equation}
Note that in the adiabatic limit $\tau \to \infty$, this reduces to Eq.~(7) of the main text since $Q_{\tau}(\omega)\to \omega^{-2}$. An immediate consequence is that $F_Q[\lambda]$ is divergent unless $\chi(\omega)$ vanishes faster than $\omega$.
}

%###############
%###############

\rev{\section{QFI for simple models}

Here we provide the QFI for adiabatic perturbations in some standard simple models, where one can obtain an exact formula for the QFI. This will provide some intuitive insight into the sensing of adiabatically-encoded parameters.

%#####
\subsection{Example 1: Landau-Zener model}

As a first example,  we assume a Landau-Zener model $\hat{H}_0= \Delta \sigma_x + g \sigma_z$ with an adiabatic perturbation in the field strength $\hat{H}_1=\omega \sigma_z$ such that $\hat{A}\equiv \sigma_z$ and $\lambda\equiv\omega$. Computing the QFI gives,
\begin{eqnarray}
F_Q[\omega] = \frac{4 \omega ^2 \tanh ^2\left(\beta  \sqrt{\Delta
   ^2+g^2}\right)}{\Delta ^2+g^2}.
\end{eqnarray}
This formula confirms the intuition that the highest sensitivity is found at the smallest energy gap $g=0$ (for a fixed positive $\Delta$). There is also increasing sensitivity for lower temperatures.  The maximum QFI is $4 \omega^2/ \Delta^2$.

%#####
\subsection{Example 2: Quantum harmonic oscillator}
Next,  we consider a quantum harmonic oscillator $\hat{H}_0= \nu a^\dagger a$ which is perturbed by a linear potential $\hat{H}_1=\lambda \hat{X}$ (such that $\hat{A}\equiv \hat{X}$). Using the spectrum is this case is cumbersome since it requires a double infinite sum.  Instead we will directly compute the response function $\chi(\omega)$.  We start by noting that,
\begin{eqnarray}
A(t) &= & e^{i \nu a^\dagger a t} \hat{X} e^{-i \nu a^\dagger a t} \\
&=& \frac{1}{\sqrt{2}} \left[ e^{- i \nu t} a + e^{i \nu t} a^\dagger \right].
\end{eqnarray}
From this we can compute the commutator as,
\begin{eqnarray}
[A(0),A(t)] &=& \frac{1}{2} \left[a+ a^\dagger,  e^{- i \nu t} a + e^{i \nu t} a^\dagger \right] \\
&=& i \sin(\nu t),
\end{eqnarray}
where we have used $[a,a^\dagger]=1$.  Putting this altogether, we have,
\begin{eqnarray}
K (\omega) &=&   \int dt \, e^{i \omega t} \theta(t) \sin(\nu t) \\
&=& \frac{1}{2} i \pi  \delta (\omega -\nu )-\frac{1}{2} i
   \pi  \delta (\nu +\omega )+\frac{\nu }{\nu ^2-\omega
   ^2}.
\end{eqnarray}
From this we get that
\begin{eqnarray}
\chi(\omega) &=& - \omega \, {\rm Im} K (\omega) \\
&=& \frac{\omega \pi}{2} \left[ \delta (\nu +\omega ) - \delta (\omega -\nu ) \right],
\end{eqnarray}
which inserted into our expression for the QFI gives (for $\nu>0$),
\begin{eqnarray}
F_Q[\lambda]   = \nu^{-2} \tanh \left(\beta \nu/2 \right).
\end{eqnarray}
This gives an enhanced QFI at lower frequencies $\nu$ because the perturbation $\hat{X}$ generates a translation in momentum space.  The thermal state is broader in momentum space for increasing frequency $\nu$ which in turn makes the perturbation more difficult to infer from measurements on the system.

%###########
\subsection{Example 3: Jaynes-Cummings Model}

Consider the Hamiltonian $\hat{H}_0=\frac{\omega_q}{2} \sigma_z + \omega_r a^\dagger a$ where $\omega_q$ and $\omega_r$ are the natural frequencies of a qubit and a resonator, respectively. The thermal state in this case is a product state $\hat{\rho}_0=\hat{\rho}_q \otimes \hat{\rho}_r$ where,
\begin{eqnarray}
    \hat{\rho}_q &=& e^{-\beta \omega_q \sigma_z/2}/Z_q, \\
    \hat{\rho}_r &=& e^{-\beta \omega_r a^\dagger a}/Z_r.
\end{eqnarray}
The partition functions can be simply computed as,
\begin{eqnarray}
    Z_q &=& 2 \cosh \left(\beta  \omega_q/2\right), \\
    Z_r &=& \frac{e^{\beta  \omega_r }}{e^{\beta  \omega_r }-1}.
\end{eqnarray}
We assume a perturbation which couples the qubit and resonator, $\hat{H}_1=g(\sigma_+ a+ \sigma_- a^\dagger) \equiv \lambda\hat{A}$. The time evolution is
\begin{eqnarray}
    A(t) = g \left( e^{-i (\omega_r - \omega_q)t} \sigma_+ a + e^{-i (\omega_q - \omega_r)t} \sigma_- a^\dagger \right),
\end{eqnarray}
which gives a commutator,
\begin{eqnarray}
    [A(0),A(t)] &=& g^2 \left[\sigma_+ a+ \sigma_- a^\dagger,  e^{-i (\omega_r - \omega_q)t} \sigma_+ a + e^{-i (\omega_q - \omega_r)t} \sigma_- a^\dagger \right] \\
    &=& g^2 \left\{ e^{-i (\omega_q - \omega_r)t} [\sigma_+ a, \sigma_- a^\dagger]+  e^{-i (\omega_r - \omega_q)t} [\sigma_- a^\dagger,\sigma_+ a]  \right\} \\
    &=& g^2  \left\{ e^{-i (\omega_q - \omega_r)t}-  e^{-i (\omega_r - \omega_q)t} \right\} [\sigma_+ a, \sigma_- a^\dagger] \\
    &=& -2 i g^2  \sin[(\omega_q-\omega_r)t] [\sigma_+ a, \sigma_- a^\dagger]
\end{eqnarray}
The only thing left unknown is the thermal expectation value constant,
\begin{eqnarray}
C &=&\avg{[\sigma_+ a, \sigma_- a^\dagger]}_0 \\
&=& \avg{\sigma_+ \sigma_-}_q \avg{a a^\dagger}_r- \avg{\sigma_- \sigma_+}_q \avg{a^\dagger a}_r \\
&=&  \frac{1}{e^{\beta  \omega_q}+1}   \frac{e^{2 \beta  \omega_r }}{\left(e^{\beta  \omega_r
   }-1\right)^2} - \frac{e^{\beta \omega_q}}{e^{\beta  \omega_q}+1} \frac{e^{\beta  \omega_r }}{\left(e^{\beta  \omega_r
   }-1\right)^2} \\
&=& \frac{e^{\beta \omega_r} \left[ e^{\beta  \omega_r } - e^{\beta \omega_q}\right]}{(e^{\beta  \omega_q}+1)\left(e^{\beta  \omega_r
   }-1\right)^2}   ,
\end{eqnarray}
where $\avg{\cdot}_x={\rm Tr}(\cdot \hat{\rho}_x)$. Up to some factors, we arrive at the same integral as the previous case apart from some factors,
\begin{eqnarray}
    K (\omega) &=&  -2 g^2 C \int dt \, e^{i \omega t} \theta(t) \sin(\Delta t),
\end{eqnarray}
where we have defined $\Delta=\omega_q-\omega_r$. The susceptibility is then
\begin{eqnarray}
\chi(\omega) &=&  g^2 C \omega \pi \left[ \delta (\Delta +\omega ) - \delta (\omega -\Delta ) \right].
\end{eqnarray}
The QFI finally follows as,
\begin{eqnarray}
   F_Q[g] =  4 \Delta^{-2} C g^2
   \tanh \left(\beta \Delta/2\right).
\end{eqnarray}

}

%##########
\rev{
\section{Optimal Observable}
The results in the main text show how there can be a large discrepancy between the exponential scaling predicted by the QFI and the actual precision obtained from the canonical choice of observable in nanoelectronics, namely the electric current. 
This begs the question: in general for an adiabatically-encoded parameter, what \textit{is} the optimal observable which would saturate the CRB? Fortunately, this optimal observable is known to be,\cite{Paris_Quantum_Estimation}
\begin{eqnarray}
X \approx  \frac{L_0}{F_Q[\lambda]},
\end{eqnarray}
where $L_\lambda$ is the symmetric logarithmic derivative (SLD) for a parameter value $\lambda$.  As in the main text,  we approximate this by the nearby value $\lambda=0$,  since the exact value of the parameter is unknown. This is valid in the perturbative (linear response) regime considered. We can compute the SLD directly as,
\begin{eqnarray}
L_\lambda &=& 2 \sum_{n \neq m} \frac{p_n^0-p_m^0}{p_n^0+p_m^0} \braket{m_\lambda}{\partial_\lambda n_\lambda} \ketbra{m_\lambda}{n_\lambda} \\
&=&  2 \sum_{n \neq m} \tanh \left( \frac{E_n^0-E_m^0}{2 T} \right) i \bra{m_\lambda} \hat{\mathcal{O}} \ket{n_\lambda} \ketbra{m_\lambda}{n_\lambda}.
\end{eqnarray}
Choosing the operator $\hat{\mathcal{O}}$ which corresponds to a weak adiabatic perturbation of $\hat{A}$ as before and evaluating at $\lambda=0$, the optimal observable can be explicitly written as,
\begin{eqnarray}
X = - \frac{2}{F_Q[\lambda]} \sum_{n \neq m} \tanh \left( \frac{E_n^0-E_m^0}{2 T} \right) \frac{\langle m_0|\hat{A}|n_0\rangle}{E_n^0-E_m^0} \ketbra{m_0}{n_0}.
\end{eqnarray}
This is in general a highly complex nonlocal operator which is not experimentally  accessible. However there are several methods which have been used to approximate (using only local operators) the adiabatic gauge potential \cite{sels2017minimizing,claeys2019floquet,morawetz2024} which could potentially be adapted to this setting.
}

\vfill

\section{QFI for the Anderson Impurity Model} 
\noindent A nontrivial application is to quantum dot (QD) nanoelectronics devices. Here we define $\hat{H}_0$ as the QD nanostructure coupled to source and drain leads at equilibrium. The leads are held at the same chemical potential and have the same temperature; everything is at zero field. In general, the QD-leads coupling can be strong,\cite{pustilnik2004kondo} and electronic interactions cannot typically be neglected because quantum confinement in any nano-scale device ubiquitously results in Coulomb blockade physics.\cite{meirav1990single} The ensuing macroscopic entanglement between the QD and leads at low temperatures can produce highly non-markovian dynamics, such as the Kondo effect.\cite{hewson1997kondo}

We model our equilibrium nanodevice as a generalized quantum impurity model\cite{bulla2008numerical} of the form $\hat{H}_0=\hat{H}_{\rm leads}+\hat{H}_{\rm nano} + \hat{H}_{\rm hyb}$. In the quantum transport setup, the nanostructure is tunnel-coupled to infinite source and drain leads, which are taken to be continuum reservoirs of non-interacting conduction electrons, in the thermodynamic limit.

The approach we develop can be straightforwardly generalized to more complex nanostructures but for the sake of a simple demonstration, we restrict ourselves here to a single QD orbital, tunnel-coupled symmetrically to source and drain leads. With local Coulomb interactions on the QD, this is the celebrated Anderson impurity model (AIM):\cite{hewson1997kondo}
\begin{equation}\label{eq:aim}
    \hat{H}_0=t\sum_{j=1}^L \sum_{\alpha,\sigma} \left (c_{\alpha j \sigma}^{\dagger} c_{\alpha j+1 \sigma}^{\phantom{\dagger}} + {\rm H.c.} \right ) + \epsilon_d^{\phantom{\dagger}} \sum_{\sigma} \Big (d_{\sigma}^{\dagger}d_{\sigma}^{\phantom{\dagger}}\Big ) + U_d^{\phantom{\dagger}} \left (d_{\uparrow}^{\dagger}d_{\uparrow}^{\phantom{\dagger}}d_{\downarrow}^{\dagger}d_{\downarrow}^{\phantom{\dagger}} \right )+ V\sum_{\alpha,\sigma}  \left ( d_{\sigma}^{\dagger}c_{\alpha 1 \sigma}^{\phantom{\dagger}} + {\rm H.c.}  \right ) \;, 
\end{equation}
where $\alpha=s,d$ for source and drain leads, $\sigma=\uparrow,\downarrow$ for up and down spin, $c_{\alpha j \sigma}^{(\dagger)}$ are annihilation (creation) operators for the conduction electrons, and $d_{\sigma}^{(\dagger)}$ are annihilation (creation) operators for the QD. Here we have given the Hamiltonian for the each of the leads in the form of a 1d nanowire, comprising $L$ sites. This allows us to study the scaling of the QFI with system size. The thermodynamic limit corresponds to $L\to \infty$.
The noninteracting ($U_d=0$) limit of this model is the resonant level model (RLM). 

One useful simplification arising in the case of the AIM or RLM is the `proportionate coupling' QD-lead hybridization geometry.\cite{meir1992landauer} This means that we may introduce even and odd lead orbital combinations: $c_{ej\sigma}=\tfrac{1}{\sqrt{2}}(c_{sj\sigma}+c_{dj\sigma})$ and $c_{oj\sigma}=\tfrac{1}{\sqrt{2}}(c_{sj\sigma}-c_{dj\sigma})$. The QD then only couples to the even lead orbitals in $\hat{H}_0$ and the odd lead orbitals formally decouple. The Hamiltonian for the leads takes exactly the same form as in Eq.~\ref{eq:aim} but now $\alpha=e,o$. The hybridization part of the Hamiltonian becomes simply $H_{\rm hyb}=\sqrt{2}V\sum_{\sigma}(d_{\sigma}^{\dagger}c_{e 1 \sigma}^{\phantom{\dagger}} + {\rm H.c.})$. This effective one-channel description is not essential to the following, but does permit an elegant description.

To this model we add a perturbation $\hat{H}_1$, switched on adiabatically, and study the resulting QFI for this perturbation. In the following we consider, separately, a voltage bias perturbation and a magnetic field applied to the QD.

%##############

\subsection{Bias Voltage Perturbation}
First we consider applying an ac bias voltage,
\begin{equation}\label{eq:ac_bias}
\hat{H}_1(t) = - e V_b \cos(\omega t)\;\tfrac{1}{2}\Big [\hat{N}_s - \hat{N}_d \Big ] \;,    
\end{equation}
where $V_b$ is the bias voltage, taken to be split equally across source and drain leads. Here $\hat{N}_{\alpha}=\sum_{j \sigma}^L c_{\alpha j \sigma}^{\dagger} c_{\alpha j \sigma}^{\phantom{\dagger}}$ is the total number operator for lead $\alpha$. This perturbation can be defined for the model with finite or infinite $L$.

In this setup $\lambda=-eV_b$ and $\hat{A}=\tfrac{1}{2}(\hat{N}_s-\hat{N}_d)$. For infinite leads in the quantum transport context, the natural observable is the (average) electrical current into the drain lead, $\langle \hat{I} \rangle$ with current operator $\hat{I}=-e\dot{N}_d$ and $\dot{N}_{\alpha}=\tfrac{d}{dt}\hat{N}_{\alpha}$. For infinite leads, due to current conservation we have $\langle \dot{N}_s\rangle = -\langle \dot{N}_d\rangle$. In linear response we may write $\langle \hat{I} \rangle = \chi(\omega) V_b$ with $\chi(\omega)$ the ac conductance. The dc conductance is simply $\chi^{dc}=\chi(\omega \to 0)$, and is typically finite. The conductance is given in units of $e^2/h$. From the Kubo formula,\cite{kubo1957statistical,minarelli2022linear} $\chi(\omega)=\left(\tfrac{e^2}{h}\right)\times \tfrac{2\pi\omega}{4}\:{\rm Im} \langle\langle \hat{N}_s-\hat{N}_d ; \hat{N}_s -\hat{N}_d\rangle\rangle$, which is evaluated at equilibrium in $\hat{H}_0$.
The QFI for estimation of the bias voltage in the dc limit is therefore,
\begin{equation}\label{eq:qfi_cond}
F_Q[V_b] =  \frac{2}{\pi}\int d\omega~\frac{\tanh(\omega/2T)}{\omega^3} \chi(\omega) \; ,
\end{equation}
where here and in the following we set $\hbar\equiv 1$, $e\equiv 1$ and $k_{\rm B}\equiv 1$. 

With infinite leads as thermal reservoirs, $\chi^{dc}$ is typically finite and so the integral is divergent, meaning an infinite voltage QFI. However, an interesting question addressed in the following is \textit{how} the QFI diverges with system size? For finite $L$ it is no longer meaningful to talk of quantum transport and steady state currents. Therefore we express the voltage QFI directly via Eq.~\ref{eq:qfi_K} in terms of the correlation function $K(\omega)$.

%##############

\subsubsection{Green's function formulation}
For the following discussion, it will be useful to find an efficient formulation for the QFI in terms of the QD retarded Green's function. The key object is the correlator $\bar{K}(\omega)=\tfrac{1}{4}\langle\langle\dot{N}_s-\dot{N}_d ; \dot{N}_s-\dot{N}_d\rangle\rangle$, where $\hat{N}_{\alpha}=\sum_{j=1}^L\sum_{\sigma} c_{\alpha j \sigma}^{\dagger} c_{\alpha j \sigma}$ and $\dot{N}_{\alpha}=\tfrac{d}{dt}\hat{N}_{\alpha}  =i[\hat{H},\hat{N}_{\alpha}]$. Thus we find,
\begin{equation}
    \bar{K}(\omega)=-\tfrac{1}{2}V^2\sum_{\sigma,\sigma'} \langle\langle ( d^{\dagger}_{\sigma}c_{o1\sigma} - {\rm H.c.} ); (d^{\dagger}_{\sigma'}c_{o1\sigma'} - {\rm H.c.} )\rangle\rangle = \tfrac{1}{2}V^2\sum_{\sigma} \left [ 
    \langle\langle d^{\dagger}_{\sigma}c_{o1\sigma} ; 
 c_{o1\sigma}^{\dagger} d_{\sigma}\rangle\rangle + \langle\langle c_{o1\sigma}^{\dagger} d_{\sigma} ; d^{\dagger}_{\sigma}c_{o1\sigma} \rangle\rangle 
    \right ] \;,
\end{equation}
where the second equality follows from the fact that the odd conduction electron sector is decoupled and is therefore subject to separate charge and spin conservation. Furthermore, the retarded correlators can be decomposed into independent contributions corresponding to the decoupled even and odd sectors. After some manipulations we obtain,
\begin{equation}\label{eq:ImKcurrent}
    {\rm Im}\; \bar{K}(\omega) = \pi V^2 \int d\omega' \;A_{QD}(\omega') \times  \left [ \rho_0(\omega'-\omega) \{ f_{\rm eq}(\omega'-\omega) - f_{\rm eq}(\omega')\} - \rho_0(\omega'+\omega)\{ f_{\rm eq}(\omega'+\omega) - f_{\rm eq}(\omega')\} \right] \;,
\end{equation}
where $A_{QD}(\omega) = -\tfrac{1}{\pi}{\rm Im}\:G_{QD}(\omega)$ is the QD spectral function, and where $G_{QD}(\omega)\equiv \langle\langle d_{\sigma} ; d_{\sigma}^{\dagger}\rangle\rangle$ is the full lead-coupled QD retarded Green's function, evaluated in $\hat{H}_0$ (which is independent of spin $\sigma$). Also, we have used $\rho_0(\omega)=-\tfrac{1}{\pi}\:{\rm Im}\:\mathcal{G}_{\rm lead}(\omega)$ as the density of states of the free, uncoupled, leads at the QD position. It is given in terms of the free leads Green's function $\mathcal{G}_{\rm lead}(\omega) \equiv \langle\langle c_{\alpha 1 \sigma} ; c_{\alpha 1 \sigma}^{\dagger}\rangle\rangle_0$, which is taken to be independent of channel $\alpha$ and spin $\sigma$. As we see later, it is crucial in certain circumstances to keep track of the finite lead bandwidth, and so here we retain the full lead density of states as an arbitrary function for generality. The equilibrium Fermi-Dirac distribution is denoted $f_{\rm eq}(\omega)=[e^{\omega/T}+1]^{-1}$. We note that Eq.~\ref{eq:ImKcurrent} is exact, holds for interacting or noninteracting models, and for finite lead length $L$ as well as in the thermodynamic limit $L\to \infty$.\\

With a knowledge of the lead density of states $\rho_0(\omega)$ and the QD spectral function $A_{QD}(\omega)$ we can evaluate $\bar{K}(\omega)$ and hence the voltage QFI,
\begin{equation}\label{eq:FqVb}
\begin{split}
    F_Q[V_b] &= 4V^2 \int d\omega\int d\omega'~\frac{\tanh(\frac{\omega-\omega'}{2T})}{(\omega-\omega')^4} \left [  f_{\rm eq}(\omega') - f_{\rm eq}(\omega) \right] \times \rho_0(\omega)  A_{QD}(\omega') \;.
    \end{split}
\end{equation}

For $L\to \infty$ we have $t\mathcal{G}_{\rm lead}(\omega)=\omega/2t-i\sqrt{1-(\omega/2t)^2}$ such that $\rho_0(\omega)=\tfrac{1}{\pi t}\sqrt{1-(\omega/2t)^2}\Theta(1-|\omega/2t|)$ is finite within a band of halfwidth $D=2t$. For the noninteracting RLM, the lead-coupled QD Green's function reads,
\begin{equation}\label{eq:Grlm}
    G_{QD}^{RLM}(\omega) = \frac{1}{\omega - \epsilon_d -2V^2\mathcal{G}_{\rm lead}(\omega) } \;.
\end{equation}
$A_{QD}(\omega)$ is then approximately Lorentzian, with a peak centered on $\omega\sim \epsilon_d$ of width $\sim V^2/t$. Combining these results, we indeed find that the dc electrical conductance $\chi_c^{dc}=\lim_{\omega\to 0} {\rm Im} \bar{K}(\omega)/\omega$ is finite, and hence the voltage QFI $F_{Q}[V_b]$ from Eq.~\ref{eq:qfi_chi} or Eq.~\ref{eq:FqVb} diverges. 

For the interacting AIM, the Green's function has a non-trivial self-energy correction,
\begin{equation}
    G_{QD}^{AIM}(\omega) = \frac{1}{\omega - \epsilon_d -2V^2\mathcal{G}_{\rm lead}(\omega) -\Sigma(\omega) } \;,
\end{equation}
but the dc conductance is again finite, and the QFI again diverges.\\

For finite $L$, both the free lead density of states $\rho_0(\omega)$ and the lead-coupled QD Green's function $G_{QD}(\omega)$ consist of a finite sum of poles, and the QFI integrals collapse to sums over pole contributions. For the free lead,
\begin{equation}
    \rho_0(\omega)=\sum_{k=1}^L |a_k|^2 \delta(\omega-\xi_k) \;,
\end{equation}
which has $L$ poles for a nanowire lead of length $L$. In the diagonal basis of lead $\alpha$, we can write $\hat{H}_{\rm lead}^{\alpha} = \sum_{k\sigma} \xi_k c_{\alpha k \sigma}^{\dagger} c_{\alpha k \sigma}$ and $a_k$ is the weight of free lead state $k$ at the site $j=1$ coupled to the QD.

On the other hand, for the non-interacting RLM, we can write,
\begin{equation}\label{eq:Gdd}
    A_{QD}(\omega)= \sum_p |b_p|^2 \delta(\omega-\epsilon_p)
\end{equation} 
where the sum over $p$ runs over all $2L+1$ poles of the full lead-QD-lead composite system. In the diagonal representation of the full (equilibrium) Hamiltonian, $\hat{H}_0=\sum_{p\sigma} \epsilon_p f_{p\sigma}^{\dagger}f_{p\sigma}$ with single-particle energies $\epsilon_p$. Here $b_p$ is the weight of eigenstate $p$ on the QD orbital.

In the case of the interacting AIM with finite $L$, we may still write $A_{QD}(\omega)$ as a sum of poles as per Eq.~\ref{eq:Gdd} but the pole weights and positions must be obtained from the Lehmann representation in the many-particle basis.\cite{weichselbaum2007sum} The sum over $p$ includes exponentially-many terms, corresponding to the proliferation of many-particle excitations.\\

For the finite-sized RLM or AIM, we therefore find that the QFI can be expressed as,
\begin{equation}
\begin{split}
    F_Q[V_b] &= 4V^2 \sum_{k,p} \frac{\tanh \left (\frac{\xi_k-\epsilon_p}{2T}\right )}{(\xi_k-\epsilon_p)^4}\left [f_{\rm eq}(\epsilon_p)  - f_{\rm eq}(\xi_k)  \right ] \times |a_k \: b_p|^2 \;.
    \end{split}
\end{equation}
In general this expression for the QFI yields a finite result, although clearly $F_Q[V_b]$ blows up as the energy gap between single-particle excitation poles closes. 

Our numerical results for the noninteracting RLM obtained by diagonalizing the single-particle Hamiltonian $\hat{H}_0$ shown in Fig.~2(a,b) of the main text demonstrate a scaling ${\rm max}(F_Q[V_b]) \sim L^2$. This is the result of entanglement of all $2L+1$ sites in the system in this setup. However, note that we are ultimately dealing with the single-particle physics of independent electrons: for $2L+1$ sites in our full system, we have $2L+1$ poles of the single-particle QD Green's function on which the QFI depends. The energy gap between excitations is then typically $\sim 1/L$.

For the interacting AIM we obtain the QD Green's function of the coupled system using exact diagonalizaion (ED) in the many-particle basis (exploiting charge and spin symmetries), which we can do up to $L=8$. We find very strong scaling with system size -- roughly exponential, ${\rm max}(F_Q[V_b])\sim (4^{L})^2$ -- see Fig.~2(c,d). This is consistent with a typical minimum excitation energy gap $\sim 1/(4^L)$ as discussed in the main text.

%##############

\subsection{Magnetic Field Perturbation}
\noindent We now consider adding instead a perturbation to $\hat{H}_0$ corresponding to a magnetic field $B$ on the QD,
\begin{equation}
    \hat{H}_1 = B\hat{S}_d^z \;,
\end{equation}
with $B$ a Zeeman field along $z$ and $\hat{S}_d^z=\tfrac{1}{2} [d_{\uparrow}^{\dagger}d_{\uparrow}^{\phantom{\dagger}} -d_{\downarrow}^{\dagger}d_{\downarrow}^{\phantom{\dagger}}]$ the spin projection of the QD orbital. Due to SU(2) spin symmetry the direction of $B$ is arbitrary.  
In this case the QFI for estimation of the field strength $B$ is,
\begin{equation}\label{eq:qfi_B}
F_Q[B] =  \frac{2}{\pi}\int d\omega~\frac{\tanh(\omega/2T)}{\omega^2} {\rm Im}K(\omega) \;,
\end{equation}
where $K(\omega)=\langle \langle S_d^z ; S_d^z\rangle\rangle$ is the dynamical spin susceptibility evaluated in $\hat{H}_0$. In a Fermi liquid phase\cite{hewson1997kondo} we typically have ${\rm Im}K(\omega) \sim \omega$. Thus the integral is convergent and $F_Q[B]$ is expected to be finite. Since the perturbation is \emph{local}, the correlation function $K(\omega)$ also involves local operators. However, it does of course still encode correlations in the full systems that affect the QD.

The RLM has no spin dynamics since $\sigma=\uparrow$ and $\sigma=\downarrow$ sectors are decoupled. Therefore in the case of magnetometry we must consider the interacting AIM. In Fig.~4 of the main text we compare ED calculations for finite-size systems with numerical renormalization group (NRG) results\cite{bulla2008numerical,weichselbaum2007sum} for the system in the thermodynamic limit. Although we see a modest increase in $F_Q[B]$ with $L$ from ED up to $L=8$, the QFI does not diverge with $L$. NRG results for $L\to \infty$ show a finite magnetometry QFI for any finite $T$, although the QFI does diverge logarithmically as the temperature $T$ is decreased. 

To understand the numerical results for the QFI we note that NRG results for ${\rm Im}~K(\omega)$ are consistent with the ansatz
\begin{equation}
    T_K \:{\rm Im}\:K(\omega) \sim \frac{\omega/T_K}{1+(T/T_K)^2+(\omega/T_K)^2} \;, 
\end{equation}
where $T_K$ is the Kondo temperature, defined here as the $T=0$ peak in ${\rm Im}~K(\omega)$. The universal scaling result is approximate but found to be  rather accurate for all $\omega$ and $T$ considered, up to log corrections. Using our ansatz in Eq.~\ref{eq:qfi_B} for the QFI yields directly the scaling prediction that $T_K^2 F_Q[B]$ is a universal scaling function of $T/T_K$, as confirmed by the full NRG results in Fig.~4(b). Furthermore, we extract from the ansatz the QFI asymptotes $T_K^2 F_Q[B]\sim (T_K/T)^2$ for $T\gg T_K$ and  $\sim \log[T_K/T]$ for $T\ll T_K$. We expect similar low-$T$ results for magnetometry in any Fermi liquid, since by the Korringa-Shiba relation\cite{hewson1997kondo} ${\rm Im}\:K(\omega)\sim \omega$ for such a system on the lowest energy scales.

%###############################
%###############################

\section{Fisher information for a current measurement}
\noindent In the nanoelectronics context, the natural experimental observable is of course the electrical current between source and drain leads passing through the QD. We imagine switching on a bias voltage $V_b$ adiabatically and measuring the current, to estimate some parameter $\lambda$. Note that $\lambda$ here can be a parameter of $\hat{H}_0$ or $\hat{H}_1$, or indeed temperature $T$, and need not be small. To make contact with the QFI results of the previous sections, here we consider explicitly once again voltometry or magnetometry (estimation of $V_b$ or $B$), but this time we use the current specifically, rather than the optimal measurement. We focus here on an instantaneous current measurement $\langle \hat{I}\rangle$, although we consider the integrated current (collected charge) in the next section. To characterize the precision of our system to estimation of $V_b$ or $B$, we use the error-propagation formula to define the quantity,\cite{toth}
\begin{equation}\label{eq:FIlambda}
F_I[\lambda]=\frac{(\partial_{\lambda} \langle \hat{I}\rangle)^2}{{\rm Var}(I)} \;,    
\end{equation}
where ${\rm Var}(I)=\langle \delta \hat{I}^2 \rangle$ and $\delta \hat{I}=\hat{I}-\langle\hat{I}\rangle$ in terms of the current operator $\hat{I}=-e\dot{N}_d$ as before. In general, $F_I[\lambda]\le \mathcal{F}_I[\lambda]$, where $\mathcal{F}_I[\lambda]$ is the true (classical) Fisher information (FI) for a current measurement. The precision $F_I[\lambda]$ is equal to the FI $\mathcal{F}_I[\lambda]$ when measurement outcomes are Gaussian distributed.\cite{ding2022enhanced} Thus we have a chain of inequalities, $F_Q[\lambda]\ge \mathcal{F}_I[\lambda]\ge F_I[\lambda]$, and from the CRB ${\rm Var}(\lambda)\ge 1/F_Q[\lambda]$ (one-shot case).\cite{toth}

The variance of the instantaneous current ${\rm Var}(I)$ appearing in Eq.~\ref{eq:FIlambda} can be obtained from the current autocorrelator $S(t)=\langle \delta\hat{I}(0)\delta\hat{I}(t)\rangle$ since $S(t=0)=\langle \delta \hat{I}^2 \rangle$. Therefore, ${\rm Var}(I)$ can also be calculated by integrating the noise spectrum $S(\omega)=\int dt\: e^{i\omega t}\:S(t)$, viz:
\begin{equation}\label{eq:varI}
   {\rm Var}(I)=S(t=0)=\frac{1}{2\pi}\int d\omega\: S(\omega).
\end{equation}
We conclude that if we know the current and its noise spectrum, we can calculate the precision for an instantaneous current measurement. In the following we therefore focus on finding expressions for $\langle\hat{I}\rangle$ and $S(\omega)$ for our nanoelectronics models of interest. Note that the electrical current has units $e/h$, the conductance is in units $e^2/h$, and the noise has units $e^2/h^2$. Unless otherwise stated, we take $e\equiv 1$, $k_B\equiv 1$ and $\hbar\equiv 1$.

For parameter estimation of $V_b$ or $B$ we therefore have,
\begin{equation}\label{eq:FI}
    F_I[V_b] = \frac{(G^{dc})^2}{S(t=0)} \qquad\qquad ; \qquad\qquad  F_I[B] = \frac{(\partial_B \langle \hat{I}\rangle)^2}{S(t=0)} 
\end{equation}
where $G^{dc}=d\langle \hat{I}\rangle/dV_b$ is the dc differential conductance, which can be evaluated at any $V_b$.

%############

\subsection{Linear Response but interacting}
In linear response we can apply a small ac bias voltage $V_b$ according to Eq.~\ref{eq:ac_bias}. The (average) current $\langle \hat{I}\rangle$ is then proportional to the voltage $V_b$. The proportionality factor is the linear response ac conductance $\chi(\omega)$. For $\omega\to 0$ this becomes the dc linear response conductance $\chi^{dc}$, which is the differential conductance $G^{dc}$ appearing in Eq.~\ref{eq:FI} evaluated in the zero-bias limit $V_b\to 0$ and we have $\langle\hat{I}\rangle=\chi^{dc} V_b$. In the general ac transport setup, with $\omega$ the ac bias frequency, the Kubo formula gives an expression\cite{kubo1957statistical,minarelli2022linear} for the linear response conductance $\chi(\omega)={\rm Im}\:\langle\langle \hat{I};\hat{I}\rangle\rangle/\omega$. Importantly, the current-current correlation function is evaluated at equilibrium in $\hat{H}_0$. Since in $\hat{H}_0$ at zero bias no net current flows, $\langle \hat{I}\rangle_0=0$ and so $\delta \hat{I}\equiv \hat{I}-\langle\hat{I}\rangle = \hat{I}$ at equilibrium. Thus,
\begin{equation}\label{eq:LRcond}
    \chi(\omega) = {\rm Im}\:\langle\langle \delta\hat{I} ; \delta\hat{I}\rangle\rangle / \omega \qquad ;\qquad \lim_{V_b \to 0} G^{dc}\equiv \chi^{dc}=\lim_{\omega\to 0} \chi(\omega) 
\end{equation}
Furthermore, the fluctuation-dissipation relation tells us that the retarded and lesser Green's functions in equilibrium are connected. Therefore we can write,
\begin{equation}\label{eq:fd}
   \pi S(\omega) = n_B(\omega)\:{\rm Im}\:\langle\langle \delta\hat{I} ; \delta\hat{I}\rangle\rangle  \equiv \omega n_B(\omega) \chi(\omega) \;,
\end{equation}
where $n_B(\omega)=[e^{\omega/T}-1]^{-1}$ is the Bose-Einstein distribution.  For $F_I(\lambda)$ we need $S(t=0)$, which we can obtain from the ac linear response conductance $\chi(\omega)$, or the current-current correlation function $\langle\langle \delta\hat{I} ; \delta\hat{I}\rangle\rangle$.

The above formulation is restricted to linear response, but is otherwise completely general: it makes no assumptions about the form of the nanostructure, which can be single or multi-orbital in any geometry, interacting or non-interacting, and connected to the leads in any geometry. The central object needed is the current-current correlation function evaluation at equilibrium, which can be calculated even for complex interacting systems by e.g.~NRG.\cite{bulla2008numerical,weichselbaum2007sum}

%########

\subsubsection{Single QD case}
If we specialize now to the single-QD case of the interacting AIM (or non-interacting RLM) then we can in fact make further simplifications. Up to factors of the electric charge $e$, $\hbar$ and $k_B$ (which we have set to unity), the current-current correlator $\langle\langle \delta\hat{I} ; \delta\hat{I}\rangle\rangle = \bar{K}(\omega)$ is simply the correlator given in Eq.~\ref{eq:ImKcurrent}.
The dc conductance in linear response for the AIM (or RLM) can then be obtained by inserting Eq.~\ref{eq:ImKcurrent} into Eq.~\ref{eq:LRcond}. After some manipulations we recover the famous Meir-Wingreen formula,\cite{meir1992landauer}
\begin{equation}\label{eq:mw}
    \chi^{dc} = \lim_{\omega\to 0}[{\rm Im}\:\bar{K}(\omega)/\omega] = 2\pi V^2\int d\omega\:[-\partial_{\omega} f_{\rm eq}(\omega)]\:\rho_0(\omega)A_{QD}(\omega)
\end{equation}
Similarly, inserting Eq.~\ref{eq:ImKcurrent} into Eq.~\ref{eq:fd} yields an explicit expression for the equilibrium noise spectrum,
\begin{equation}\label{eq:eqS}%CHECK FACTORS of 2pi !!!!
\begin{split}
S(\omega) &= V^2  \int d\omega' \:A_{QD}(\omega') \times n_B(\omega)  \left [ \rho_0(\omega'-\omega) \{ f_{\rm eq}(\omega'-\omega) - f_{\rm eq}(\omega')\} - \rho_0(\omega'+\omega)\{ f_{\rm eq}(\omega'+\omega) - f_{\rm eq}(\omega')\} \right] \;, \\
&\equiv  V^2\int d\omega'\:A_{QD}(\omega') \times  \left [ \rho_0(\omega'-\omega)f_{\rm eq}(\omega')\Big (1-f_{\rm eq}(\omega'-\omega)\Big ) + \rho_0(\omega'+\omega) f_{\rm eq}(\omega'+\omega)\Big (1-f_{\rm eq}(\omega')\Big )  \right]
\end{split}
\end{equation}
This can be integrated to find the variance of the instantaneous current,
\begin{equation}
\begin{split}
   {\rm Var}(I) &= \frac{1}{2\pi}\int d\omega\:S(\omega) = \frac{1}{2\pi^2}\int d\omega\; n_B(\omega) \:{\rm Im}\; \bar{K}(\omega) \\
   &=  \frac{V^2}{2\pi} \int\int d\omega\:d\omega' \;n_B(\omega) A_{QD}(\omega') \times  \left [ \rho_0(\omega'-\omega) \{ f_{\rm eq}(\omega'-\omega) - f_{\rm eq}(\omega')\} - \rho_0(\omega'+\omega)\{ f_{\rm eq}(\omega'+\omega) - f_{\rm eq}(\omega')\} \right] \\
   &=\frac{V^2}{4\pi}  \left [ \int d\omega'\:A_{QD}(\omega')\times \int d\omega\:\rho_0(\omega) - \int d\omega'\:\tanh(\omega'/2T)A_{QD}(\omega')\times \int d\omega \:\tanh(\omega'/2T)\rho_0(\omega)\right ]
   \end{split}
\end{equation}
Due to spectral normalization, and for the standard case of a particle-hole symmetric free lead density of states $\rho_0(\omega)=\rho_0(-\omega)$, we therefore have the remarkably simple result,
\begin{equation}\label{eq:varIeq}
    {\rm Var}(I) = \frac{V^2}{4\pi} \equiv \frac{D\Gamma}{2\pi^2}  \;,
\end{equation}
which is as such a constant trivial factor that plays a spectating role in the calculation of $F_I(\lambda)$ for single QDs (noninteracting RLM and interacting AIM) in linear response. In the second equality we used $\Gamma=\pi\rho_0 V^2$ and the flat-band result $\rho_0=1/(2D)$. We note that in the naive calculation in which the finite bandwidth is neglected is UV divergent.

%##################

\subsection{Beyond linear response but noninteracting}
We now consider the full nonequilibrium transport problem at finite voltage bias. In the presence of strong electron interactions, however, this is largely still a hard open problem. Therefore in this section we restrict ourselves to the simpler case of non-interacting models, where a Landauer-B\"uttiker type approach holds.\cite{buttiker1985generalized} This greatly simplifies the formulation. We note however that although the current calculation is straightforward, the full nonequilibrium noise spectrum is more subtle. In particular, we emphasize that it is crucial to consider a realistic (microscopic) model for the QD circuit since nonuniversal high-energy scales like the hybridization and the conduction electron bandwidth play a role. One must therefore consider from the outset a model with a finite conduction electron bandwidth -- the commonly-used wide-band limit will not suffice, as shown below. As such, and for concreteness, in the following we take the metallic free lead density of states to be flat within a band of half-width $D$, such that $\rho_0(\omega)=\rho_0\Theta(D-|\omega|)$ and $\rho_0=1/2D$. In practice we set $D=1$.

For non-interacting impurity-type models, the quantum transport problem is controlled by the transmission function $\mathcal{T}(\omega)$, which is independent of both temperature and bias voltage.\cite{buttiker1985generalized,minarelli2022linear} In general, we can write $\mathcal{T}(\omega)=4\Gamma_s\Gamma_d\sum_{\sigma}|G_{sd;\sigma}(\omega)|^2$ where $\Gamma_{\alpha}=\pi\rho_0 V_{\alpha}^2$ is the hybridization of the nanostructure frontier orbital $d_{r_{\alpha};\sigma}$ to lead $\alpha=s,d$ and $G_{sd;\sigma}(\omega)=\langle\langle d_{r_s;\sigma} ; d^{\dagger}_{r_d;\sigma}\rangle\rangle$ is the full lead-coupled retarded nanostructure Green's function connecting source and drain leads. For clarity, we restore units of $e$ and $h$ in the following.\\

Taking care to keep track of the finite bandwidth, we now express the current in Landauer-B\"uttiker form,\cite{buttiker1985generalized}
\begin{equation}\label{eq:LBI}
    \langle \hat{I}\rangle = \frac{e}{h} \int d\omega\: \mathcal{T}(\omega)\times [f_s(\omega)-f_d(\omega)] \;,
\end{equation}
where $f_{\alpha}(\omega)=f_{\rm eq}(\omega-\mu_{\alpha})\Theta(D-|\omega-\mu_\alpha|)$ in terms of the chemical potential $\mu_{\alpha}$ for lead $\alpha$, and we are keeping track of the finite bandwidth with the step function. 
The bias voltage is $eV_b=\mu_s-\mu_d$, which one can assume is split equally across source and drain leads. 

Generalizing the results of Lesovik,\cite{lesovik} we can similarly formulate the nonequilibrium noise spectrum,
\begin{equation}\label{eq:noneqS}
\begin{split}
S(\omega) = \frac{e^2}{h^2}\int d\omega'\:\bigg [ &f_s(\omega'+\omega)\bar{f}_s(\omega') \mathcal{T}(\omega')\mathcal{T}(\omega'+\omega) \:+\: f_d(\omega'+\omega)\bar{f}_d(\omega') \mathcal{T}(\omega')\mathcal{T}(\omega'+\omega) \\
+& f_s(\omega'+\omega)\bar{f}_d(\omega') \Big (1-\mathcal{T}(\omega')\Big )\mathcal{T}(\omega'+\omega) \:+\: f_d(\omega'+\omega)\bar{f}_s(\omega') \mathcal{T}(\omega')\Big (1-\mathcal{T}(\omega'+\omega)\Big )\bigg] \;,
\end{split}
\end{equation}
where we have defined $\bar{f}_{\alpha}(\omega)=[1-f_{\rm eq}(\omega-\mu_{\alpha})]\Theta(D-|\omega-\mu_\alpha|)$ for notational simplicity.

The current variance can be obtained by integrating the noise spectrum as before, but now we use the nonequilibrium formula for noninteracting fermions. After some manipulation we obtain,
\begin{equation}
\begin{split}
{\rm Var}(I) &\equiv \langle \hat{I}^2\rangle - \langle \hat{I}\rangle^2 = \frac{1}{2\pi}\int d\omega\:S(\omega) \\
&=D\Delta_{V_b} + \langle \hat{I}\rangle\delta_{V_b} - \langle \hat{I}\rangle^2 \;,
\end{split}
\end{equation}
where $D$ is the lead bandwidth as before, $\Delta_{V_b}=\tfrac{e^2}{h^2}(2\pi)^{-1}\int d\omega\:\mathcal{T}(\omega)\Theta(D-|\omega-\mu_s|)$ is an effective hybridization, and $\delta_{V_b}=\tfrac{e}{h}(2\pi)^{-1}\int d\omega\:\mathcal{T}(\omega)[\Theta(D-|\omega-\mu_s|)-\Theta(D-|\omega-\mu_d|)]$ describes particle-hole asymmetry in the transmission function at the band edges in a window of width $V_b$. For a symmetric transmission function $\mathcal{T}(\omega)=\mathcal{T}(-\omega)$ this factor vanishes, $\delta_{V_b}=0$. The current $\langle \hat{I}\rangle$ is given by Eq.~\ref{eq:LBI} as usual.

%#############

\subsubsection{Single QD case}
Here we consider the single QD case of the RLM (with spin). 
For simplicity we assume equal coupling to source and drain leads, $V_s=V_d\equiv V$ such that $\Gamma_s=\Gamma_d\equiv \Gamma=\pi \rho_0 V^2$. Then the transmission function reads, \begin{equation}
    \mathcal{T}(\omega)=8\Gamma^2|G_{QD}(\omega)|^2\equiv 4\pi\Gamma A_{QD}(\omega) = \frac{2\Theta(D-|\omega|)}{1+(\omega-\epsilon_d)^2/4\Gamma^2},
\end{equation}
where we have used the exact form of the RLM Green's function in Eq.~\ref{eq:Grlm} and approximate $\mathcal{G}_{\rm lead}(\omega)=-i\pi \rho_0(\omega)$. We take care of the finite lead bandwidth, but neglect the small Lamb shift coming from the real part of the lead Green's function.

%###############
%###############

\section{Integrated current measurements}
In the previous section we considered instantaneous current measurements. However, the formalism can be straightforwardly extended to \textit{integrated} current measurements.\cite{mark}

We denote the charge collected in the drain lead over some time interval $\Delta t$ as $C(\Delta t)$. For some time-dependent current trace $I(t)$ we define $C(\Delta t)=\int_0^{\Delta t} dt\:I(t)$. 
In the steady state, and taking $\Delta t$ large, the integrated current becomes $C\simeq \Delta t \langle \hat{I}\rangle$ while the variance of the integrated current is ${\rm Var}(C)\simeq \langle (\Delta t \delta \hat{I})^2\rangle$. Since more information is gained about the parameter of interest from the collected charge the longer we wait, the relevant quantity is really the precision \textit{rate} (or time average) which we denote $\bar{F}_I[\lambda]$.  Thus we define,\cite{mark}
\begin{equation}
\bar{F}_I[\lambda]=\frac{(\partial_{\lambda} \langle \hat{I}\rangle)^2}{\tfrac{d}{dt}{\rm Var}(C)} \;.    
\end{equation}
It can be shown that $\tfrac{d}{dt}{\rm Var}(C)$ can also be obtained by integrating the current autocorrelation function $S(t)$, or equivalently in terms of the zero-frequency value of the noise spectrum,
\begin{equation}
   \tfrac{d}{dt}{\rm Var}(C)= \int dt\: S(t) = S(\omega=0).
\end{equation}

In the interacting but linear-response regime we can take the $\omega\to 0$ (dc) limit of Eq.~\ref{eq:fd} to obtain the completely general relation,
\begin{equation}\label{eq:S0lr}
    S(\omega=0)=\left(\frac{T}{\pi\hbar}\right)\chi^{dc} \equiv  2T \chi^{dc} \:h^{-1}\;. 
\end{equation}
Thus we may calculate $\bar{F}_I(\lambda)$ purely from a knowledge of the dc linear response conductance $\chi^{dc}$.\\ 

For the noninteracting but nonequilibrium case, we take the $\omega\to 0$ (dc) limit of Eq.~\ref{eq:noneqS} instead. We obtain,
\begin{equation}\label{eq:noneqS0}
\begin{split}
S(\omega=0) = \frac{e^2}{h^2}\int d\omega'\:\bigg [ &  \mathcal{T}(\omega')\Big( f_s(\omega')\bar{f}_d(\omega') + f_d(\omega')\bar{f}_s(\omega') \Big ) - [\mathcal{T}(\omega')]^2 \Big ( f_s(\omega') - f_d(\omega') \Big )^2 \bigg] \;,
\end{split}
\end{equation}
which is a generalization of the well-known Levitov formula,\cite{lesovik} in which we keep track of the finite lead bandwidth.

%######################

\subsection{Voltometry via current measurements}
\begin{figure}[t!]
    \centering \includegraphics[width=0.9\linewidth]{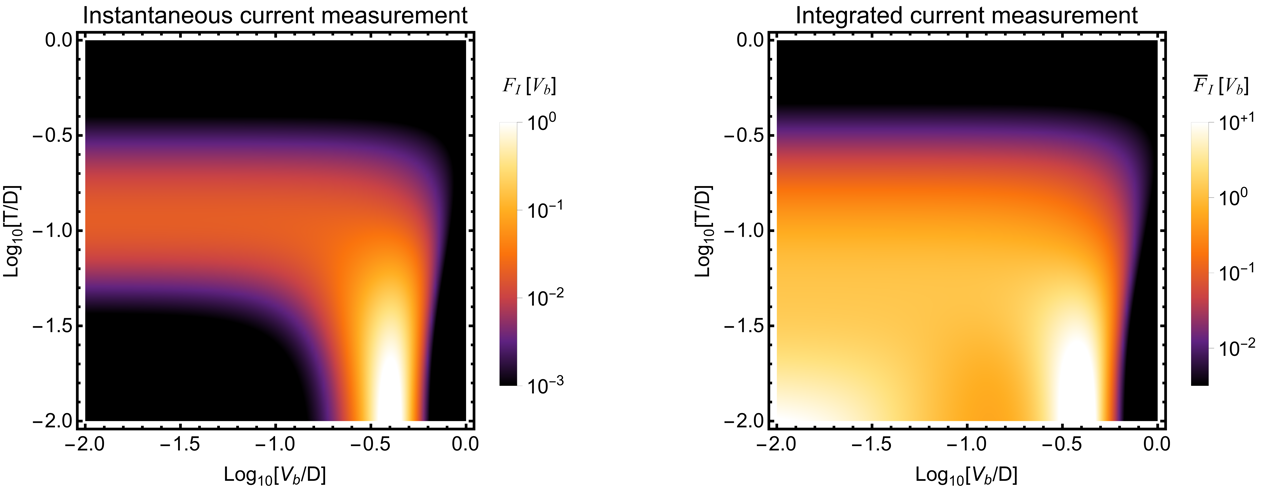}
\caption{Voltometry precision for estimation of the bias voltage $V_b$ in the non-interacting single-QD resonant level model, as a function of temperature and bias. Left panel shows $F_I[V_b]$ for an instantaneous current measurement, whereas the right panel shows the rate $\bar{F}_I[V_b]$ for an integrated current measurement. Exact results plotted for representative values $\epsilon_d=-0.2$, $\Gamma=0.05$, and finite lead bandwidth $D=1$. Recall that the corresponding voltometric QFI is infinite for this model.}
    \label{fig:FIVbRLM}
\end{figure}

We briefly compare estimation of the bias voltage using either an instantaneous current measurement $F_I[V_b]$, or an integrated current measurement, $\bar{F}_I[V_b]$. In Fig.~\ref{fig:FIVbRLM} we consider both quantities for the non-interacting RLM in the non-linear regime, as a function of temperature and bias. There are qualitative similarities between the two measures, but the major difference is that the instantaneous current measurement precision has vanishing sensitivity at low $T$ and $V_b$ when the transmission function is small at low energies (as with these parameters); while the integrated current precision rate is relatively enhanced in this region.

\begin{figure}[t!]
    \centering
\includegraphics[width=0.95\linewidth]{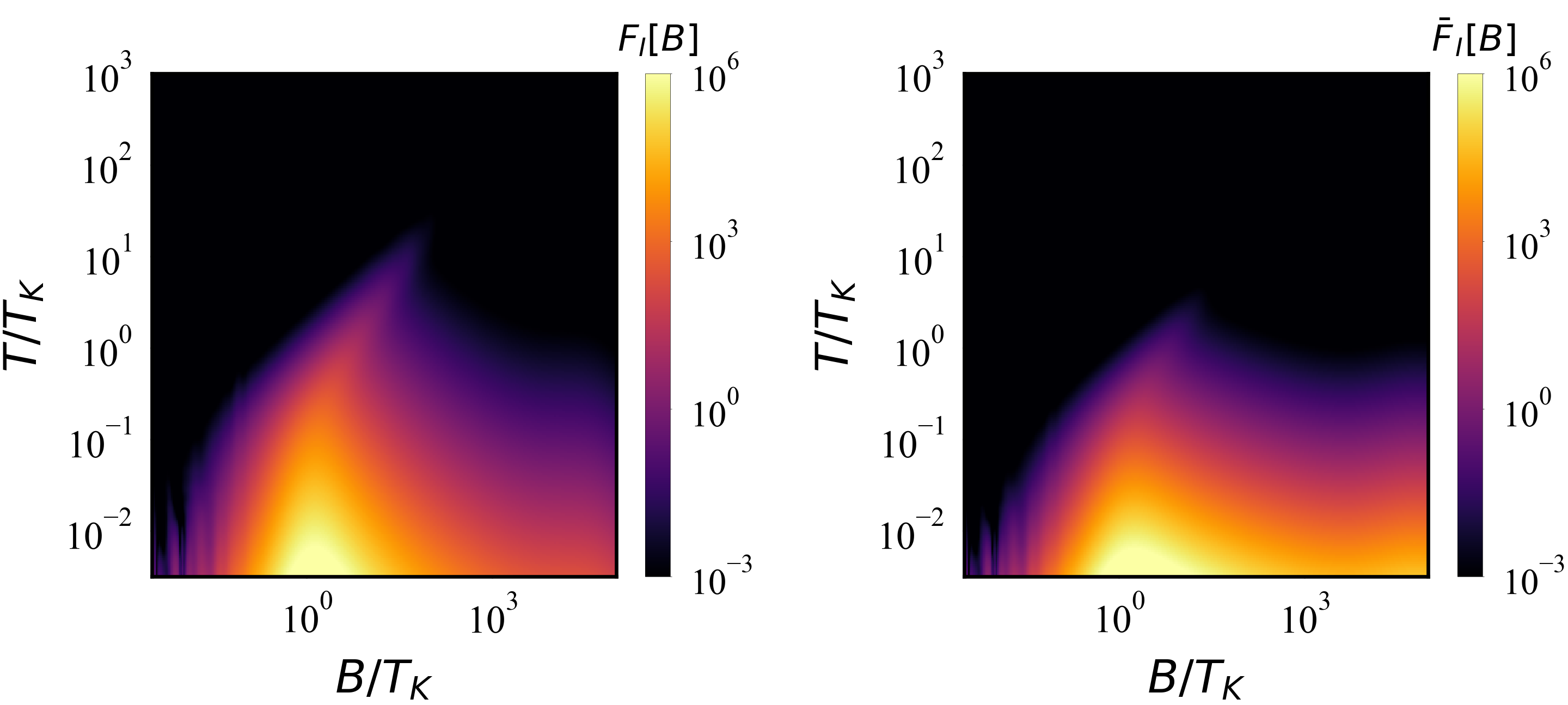}
\caption{Magnetometry precision for estimation of an applied magnetic field $B$ using a current measurement in linear response, for the \textit{interacting} Anderson impurity model. Exact results obtained using NRG, plotted as a function of temperature and field strength, rescaled in terms of the Kondo temperature $T_K$. Left panel shows $F_I[B]$ for an instantaneous current measurement, whereas right panel shows the rate $\bar{F}_I[B]$
for an integrated current measurement. Shown for representative values $\epsilon_d=-0.15$, $U_d=0.3$, $V=0.09$, $D=1$.}
    \label{fig:FImag}
\end{figure}

%########################

\subsection{Magnetometry via current measurements}
In Fig.~\ref{fig:FImag} we present numerical results for the precision for estimation of a magnetic field applied to the QD. The left panel shows the precision for an instantaneous current measurement $F_I(B)$, while in the right panel we show the precision rate $\bar{F}_I(B)$  for an integrated current measurement. This is computed for the interacting AIM, since the field couples to the QD spin, and the non-interacting RLM has no spin dynamics. Therefore we use NRG to compute the linear response conductance, from which the precision is obtained. We are thereby restricted to linear response in the bias voltage and current; the precision is proportional to $V_b^2$. We plot universal results in the Kondo regime. Fig.~\ref{fig:FImag} shows nontrivial behaviour for both, plotted here as a function of rescaled temperature $T/T_K$ and field strength $B/T_K$, with peak sensitivity in both cases at low $T$ around $B\sim T_K$.

%###################

\section{NRG calculations}
For many-body calculations of the interacting AIM in the thermodynamic limit, we use Wilson's NRG,\cite{bulla2008numerical} with real-frequency dynamical correlation functions obtained using the full-density-matrix method.\cite{weichselbaum2007sum} We exploit conserved charge and spin-projection in the iterative block-diagonalization procedure. A logarithmic discretization parameter $\Lambda=2.5$ was used and $N_s=3000$ states retained at each NRG step.

%###############################
%###############################

\section{Validity of the Adiabatic approximation}
Our general results on the QFI for an adiabatic perturbation (Eq.~1 of the main text) were illustrated with the example of the AIM and RLM (Figs.~2 and 3). Care must be taken in defining the adiabatic process in such cases, especially when taking the thermodynamic limit of system size $L\to \infty$.

The standard adiabatic approximation requires that the instantaneous spectral decomposition of the time-evolving Hamiltonian does not contain accidental degeneracies. The AIM and RLM do feature degenerate eigenstates, but typically these are symmetry protected and cannot mix. Specifically, we may label eigenstates by conserved total charge $Q$ and total $S^z$ spin projection; the voltage perturbation $V_b$ and QD magnetic field $B$ preserve these symmetries and so adiabatic evolution occurs independently in these block diagonal symmetry subspaces. 

The switch-on protocol $\gamma(t)$ for the perturbation $\hat{H}_1(t)=\gamma(t)\hat{A}$ must be \textit{slow} in an adiabatic process. In particular, the ramp time $\Delta t$ must be long compared with the inverse minimum spectral gap $\hbar/\Delta E$ to satisfy\cite{Comparat2009} %$\hbar \dot{\gamma}(t) |\langle n_\gamma |\hat{A}|m_\gamma\rangle|\ll |E_n^\gamma - E_m^\gamma |^2$ for $n\ne m$. 
$\sum_{m\ne n} |\hbar \langle m_{\gamma}|\dot{n}_{\gamma}\rangle/(E_n^\gamma - E_m^\gamma )|\ll 1$. 
This condition can be checked by analyzing the spectrum of the instantaneous Hamiltonian $H_\gamma$. Thus as the gap closes ($\Delta E \to 0$) the adiabatic switch-on time diverges ($\Delta t\to \infty$). For the AIM and RLM, the system becomes \textit{gapless} as $L\to \infty$. As discussed in Ref.~\onlinecite{polkovnikov2008breakdown}, the adiabatic limit exists in gapless systems when the limits $L\to \infty$ and $\Delta t\to \infty$ commute -- that is, we can take $\Delta t\to \infty$ first before $L\to \infty$. Real many-body systems have this property when 
weak coupling to the environment or relaxation scattering processes cut off incipient infrared divergences of the energy coming from low-energy modes.\cite{polkovnikov2008breakdown} We therefore expect our results to hold for the AIM and RLM with finite $L$ for a slow ramp of finite duration. 
For the noninteracting RLM the characteristic spectral gap is $\Delta E \sim D/L$ due to the single-particle states, whereas for the interacting AIM we have $\Delta E \sim D/(4^L)$ due to strong correlation effects. In the thermodynamic limit $L\to \infty$ where both RLM and AIM are gapless, the adiabatic switch-on of the perturbation in principle therefore requires an infinite ramp time $\Delta t\to \infty$.

On a technical level, the calculation of the QFI in the thermodynamic limit is handled automatically using Eq.~6 and Eq.~7 through the formulation involving integrals over the correlation functions $\chi(\omega)$ or $K(\omega)$, which simply become continuous functions of $\omega$ within a finite bandwidth $D$ as $L\to \infty$. For finite $L$ note that $\chi(\omega)$ and $K(\omega)$ consist of a discrete sum of poles, obtainable from their Lehmann representation. This is used in Figs.~2 and 4 for the finite-size scaling analysis. Controlled finite-size approximations that converge to the thermodynamic limit can be naturally achieved by broadening the discrete poles by e.g.~convolution of the spectra with a Gaussian kernel.

In the second part of the paper we consider instead the metrological precision for a current measurement in nanoelectronics devices in the thermodynamic limit $L\to \infty$, rather than the QFI for the optimal many-body measurement. Here we employ the standard assumption that the voltage bias $V_b$ is switched on adiabatically in the distant past (but the parameter $\lambda$ can be a parameter of $\hat{H}_0$, $\hat{H}_1$ or even $T$). 
The adiabatic approximation has a long history in the quantum transport context.\cite{kubo1957statistical,henheik2021justifying} In experiment, where the nanostructure is contacted with a macroscopic external electronic circuit, the voltage bias is obviously switched on in a finite time, yet the measured linear-response transport coefficients are accurately given by the predictions of the Kubo formula which relies on the linear response assumption.\cite{kubo1957statistical,goldhaber1998kondo,*van2000kondo,pustilnik2004kondo} Indeed, detailed universal scaling predictions using the Kubo formula for the electrical conductance in strongly correlated critical systems have been quantitatively confirmed experimentally.\cite{iftikhar2018tunable,mitchell2016universality,pouse2023quantum,*karki2023z} We conclude that the adiabatic switch-on of the voltage bias in nanoelectronics systems in the thermodynamic limit is entirely practical.

\begin{figure}[t!]
    \centering
\includegraphics[width=0.9\linewidth]{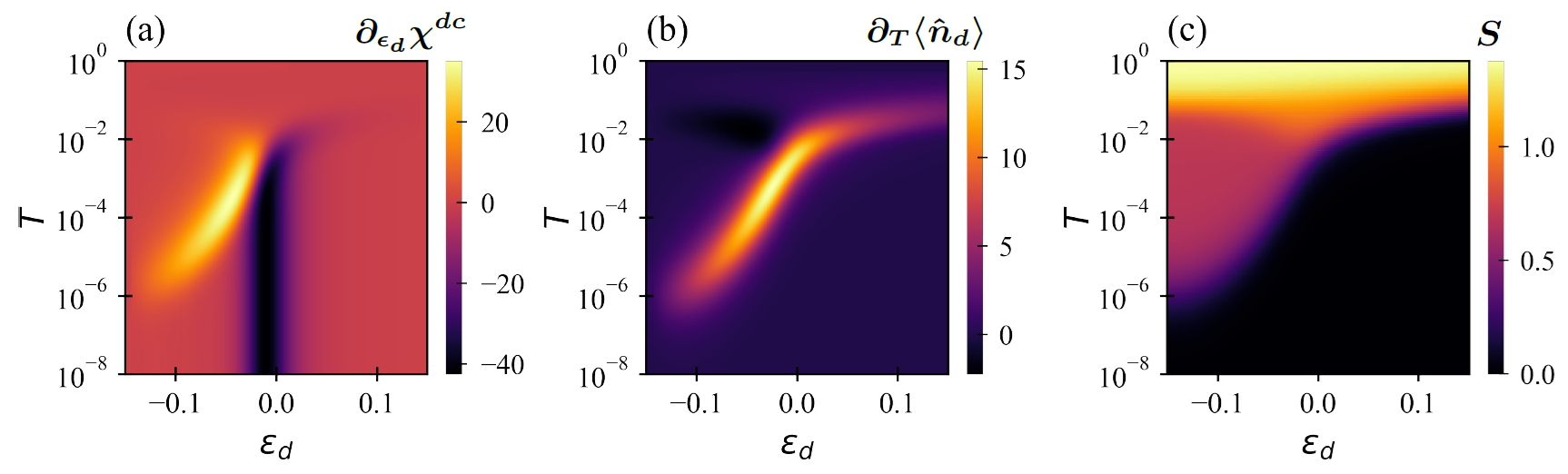}
\caption{Precision of thermodynamic entropy estimation from current measurements in QD systems modelled by the AIM in the Kondo regime. (a) Derivative of the experimentally-measurable linear-response electrical conductance $\chi^{dc}$ with respect to the QD potential $\epsilon_d$ as a function of $T$ and $\epsilon_d$. (b) Temperature derivative of the experimentally-measurable QD charge $\langle \hat{n}_d\rangle$ as a function of $T$ and $\epsilon_d$. (c) Thermodynamic entropy $S$ of the QD as a function of $T$ and $\epsilon_d$. Shown for representative values of AIM model parameters $U_d=0.3$, $V=0.09$ ($\Gamma=0.013$) and $B=0$ with a flat conduction electron band of half-width $D=1$.}
    \label{fig:SMent}
\end{figure}

\section{Entropy Estimation}
\noindent In Fig.~3d of the main text we provide a non-standard application of the parameter estimation paradigm, to thermodynamic quantities. We focus on the thermodynamic entropy of the QD in the AIM, defined $S=S_{\rm tot}-S_0$ where $S_{\rm tot}=-\partial_T \mathcal{F}$ is the equilibrium entropy of the full lead-coupled QD system $\hat{H}_0$, and $S_0$ is the entropy of a trivial reference state in which the QD is unoccupied (in practice obtained for $\epsilon_d\to \infty$). Here $\mathcal{F}=-T\ln(Z_0)$ is the free energy of the system ($Z_0$ is the partition function of $\hat{H}_0$ as before). Recently,\cite{child2022entropy} it was pointed out that since $\partial_T\partial_{\epsilon_d}\mathcal{F}=\partial_{\epsilon_d}\partial_T\mathcal{F}$, one can write down a local Maxwell relation,
\begin{equation}
S(\epsilon_d,T) = -\int_{\epsilon_d}^{\infty} d\bar{\epsilon}~ \partial_T\langle\hat{n}_d\rangle_{\bar{\epsilon},T} \;,
\end{equation}
for the entropy $S$ in terms of the experimentally measurable QD charge $\langle\hat{n}_d\rangle=\partial_{\epsilon_d}\mathcal{F}$, where $\hat{n}_d=\sum_{\sigma}d_{\sigma}^{\dagger}d_{\sigma}^{\phantom{\dagger}}$. Indeed, the entropy of a single QD modelled by the AIM in the strongly correlated regime was extracted experimentally in Ref.~\onlinecite{child2022entropy} in this way.

Here we consider the precision of estimation of $S$ from a current measurement, $F_I[S]=|\partial_{S} \langle \hat{I}\rangle|^2/{\rm Var}(I)$, as defined from Eq.~10 of the main text. We wish to express the precision in terms of the experimentally-measurable quantities $\langle\hat{n}_d\rangle$ and $\chi^{dc}=\partial_{V_b}\langle \hat{I}\rangle$. We do this by writing $\partial_S \langle \hat{I}\rangle = \partial_{\epsilon_d} \langle \hat{I}\rangle / \partial_{\epsilon_d} S$ and then exploiting the Maxwell relation $\partial_{\epsilon_d} S=-\partial_T \langle \hat{n}_d\rangle$ as used above. This leads to,
\begin{equation}\label{eq:sm_ent}
    F_I[S]=\frac{|\partial_{\epsilon_d} \langle \hat{I}\rangle/\partial_T \langle \hat{n}_d\rangle|^2}{{\rm Var}(I)} \;.
\end{equation}
Furthermore, in linear response $\partial_{\epsilon_d} \langle \hat{I}\rangle=V_b\partial_{\epsilon_d} \chi^{dc}$ in terms of the dc electrical conductance $\chi^{dc}$.

For the AIM, we compute both $\chi^{dc}$ and $\langle\hat{n}_d\rangle$ as a full function of $\epsilon_d$ and $T$ using NRG. From these we easily obtain their numerical derivatives, which are plotted in Figs.~\ref{fig:SMent}(a,b). For reference, we show the entropy $S$ itself in panel (c). Evaluation of Eq.~\ref{eq:sm_ent} using the data in Figs.~\ref{fig:SMent}(a,b) yields the plot in Fig.~4d of the main text.

We note that the entropy in Fig.~\ref{fig:SMent}(c) shows characteristic regions corresponding to the renormalization group fixed points of the AIM.\cite{hewson1997kondo} In particular we have $S=\ln(4)$ at high temperatures for the free-orbital fixed point corresponding to quasi-degenerate empty, singly-occupied, and doubly-occupied QD states. We also see a local-moment regime of degenerate QD spin states giving $S=\ln(2)$ (but charge degrees of freedom frozen out), and $S=0$ fermi liquid regimes. For $\epsilon_d<0$ and $T\ll T_K$ the quenched entropy has a nontrivial origin in the Kondo effect.\cite{hewson1997kondo} Note that the metallic AIM does not support any critical points; the complex behavior of Fig.~\ref{fig:SMent}(c) arises from crossovers, rather than transitions, between fixed points. The precision diagram Fig.~4d can now be understood: we see enhanced sensitivity to estimation of the entropy using a current measurement when the entropy itself changes rapidly with $\epsilon_d$ and $T$. This occurs along the crossovers between the renormalization group fixed points described above.

%###################
%###################

%\bibliography{bibo}